# Highly polar groups and supramolecular order enable charge injection and long-range conductivity in organic materials without extended π-systems

H. Mager[1], A. A. Butkevich[1], S. Klubertz[1], S. V. Haridas[2], O. Shyshov[2], V. C. Wakchaure[2], J. Borstelmann[3], I. Michalsky[3], M. García-Iglesias[4], V. Rodríguez[4,5], D. González-Rodríguez[5], A. R. A. Palmans[6], M. Kivala[3], M. von Delius[*,2] and M. Kemerink[*,1]

[1] Institute for Molecular Systems Engineering and Advanced Materials, Heidelberg University, Im Neuenheimer Feld 225, 69120 Heidelberg, Germany

[2] Institute of Organic Chemistry, Ulm University, Albert-Einstein-Allee 11, 89081 Ulm, Germany

[3] Institute of Organic Chemistry, Heidelberg University, Im Neuenheimer Feld 270, 69120 Heidelberg, Germany

[4] QUIPRE Department, Nanomedicine-IDIVAL, Universidad de Cantabria, Avd. de Los Castros, 46, 39005 Santander, Spain.

[5] Nanostructured Molecular Systems and Materials Group, Organic Chemistry Department, Universidad Autónoma de Madrid, 28049 Madrid, Spain

[6] Laboratory of Macromolecular and Organic Chemistry, Institute for Complex Molecular Systems, Eindhoven University of Technology, P.O. Box 513, 5600 MB Eindhoven, The Netherlands.

*Corresponding authors: max.vondelius@uni-ulm.de; martijn.kemerink@uni-heidelberg.de

## Abstract

Electronic conductivity in organic materials is well-established. Both semiconductive and metallic behavior is observed in (quasi) 0-, 1-, 2- and 3-dimensional carbon-based materials and is applied in a wide range of commercial devices. Despite their large structural variety, these materials commonly have an extended π-system, formed by a double-digit number of conjugated $sp^2$-hybridized carbon atoms, which is responsible for the conductivity. Here, we present a class of organic molecular materials that, despite the absence of an extended π-system, show a distinct direct current conductivity in quasi-1D supramolecular stacks of small organic molecules. Long-range conductivity takes place by removal of an electron from the highest occupied molecular orbital, i.e. oxidation, followed by charge transfer between neighboring molecules. Kelvin probe force microscopy (KPFM) on thin-film devices shows that the resulting transport band only becomes energetically accessible for charge injection from the contacts thanks to interfacial dipoles that shift the relevant molecular orbital levels by up to 2 eV. Long-range order in the form of fibrillar supramolecular polymers with lengths exceeding several micrometers enhances the macroscopic conductivity but is not essential. While current densities are moderate, this work provides a compelling explanation for the existing body of knowledge on non-π-conjugated conductivity and suggests a new way to endow organic materials with conductivity.

## 1 Introduction

Carbon allotropes are a well-known and well-understood class of electronically active materials of varying dimensionality. The latter range from (quasi) zero-dimensional (0D) for fullerenes[1], via 1D for carbon nanotubes[2] and 2D for graphene[3] to 3D for diamond.[4] In contrast to the other allotropes, the C atoms in diamond are $sp^3$ hybridized, which, amongst others, leads to its wide bandgap of 5.47 eV. The other carbon allotropes share the $sp^2$ hybridization of extended arrays of carbon atoms, which enables the formation of spatially extended π-systems that give rise to the highest occupied (HOMO) and lowest unoccupied (LUMO) molecular orbitals that are involved in optoelectronic functionality like charge transport and the absorption and emission of light. For these orbital levels to be energetically accessible, that is, enable charge injection from/to common metals, which have work functions ~3-5 eV below vacuum, and to allow optical transitions in the visible regime, the π-system must be sufficiently large. Although the latter is hard to quantify exactly, it must be significantly larger than a single benzene ring that has HOMO and LUMO energies around -7.1 and -0.3 eV, respectively.[5] Although formally a wide-gap semiconductor with an optical absorption in the ultraviolet region, charge transport in benzene has not been observed. In contrast, pentacene with its extended π-system comprising 22 carbon atoms in five condensed benzene rings and HOMO and LUMO energies of -4.9 eV and -3.0 eV is the prototypical organic semiconductor.[6–8] The existence of a sufficient wavefunction overlap between carrier sites is generally considered a prerequisite for long-range charge transport in organic matter that is usually fulfilled by the presence of an extended π-system as this facilitates maintaining reasonably large transfer integrals in the presence of structural disorder. The latter is a common denominator in general organic (semi)conductors that find application in light emitting diodes, solar cells, transistors, sensors and more[9–13].

Here, we present the experimental discovery of stable direct current (DC) electrical conductivity in a class of organic molecular materials that do not have a chemical structure that allows significant HOMO or LUMO delocalization, and in some cases do not have a π-system altogether. Per conventional wisdom, and especially in comparison to organic semiconductors with an extended π-system, no significant conductivity would be expected in these compounds. Instead of an extended π-system, these materials share the presence of a nitrogen-containing moiety that can be reversibly oxidized and show a strong tendency for unidimensional supramolecular polymerization through intermolecular hydrogen bonding. In addition, they contain one or more strong dipolar units such as a pentafluorocyclohexane and/or a (thio)carboxamide group. The three functionalities, providing an oxidizable center, hydrogen bonding and a dipole, can be integrated in a single moiety like an amide group, but this does not seem to be necessary. In such materials, hole transport is possible by oxidation of the molecule, that is, by the removal of an electron from the HOMO, followed by charge transfer from/to neighboring molecules. Injection into the HOMO level from the metal contacts in thin-film metal/active layer/metal devices is enabled by interfacial dipoles that have been aligned by the applied field, whereas long-range connectivity is improved by the supramolecular polymerization.

The work presented herein, addressing the long-range transport on a micrometer scale of, in this case, electronic excitations in supramolecular polymers is reminiscent of earlier remarkable findings for such materials. For example, Faramarzi *et al.* described metallic-type transport in supramolecular organic nanowires, self-assembled from a $\pi$-conjugated triarylamine derivative, between two metallic electrodes[14], which was later theoretically studied by Akande *et al*. via density functional theory (DFT) and Monte-Carlo methods.[15] Despite these remarkable findings, the authors observed a collapse of the conductivity for device lengths above a few hundred nanometers,[14] limited by self-assembly properties, which underlines the difficulty to achieve truly long-range transport in supramolecular polymers.[14]No such limitation is found for the compounds investigated here, as charge transport over distances exceeding the length of the supramolecular polymers remains finite.

Along a very different line of research, Gorbunov *et al.* and later Casellas *et al.* showed a nonlinear coupling between the ferroelectric polarization and the charge transport in semiconducting molecular compounds equipped with dipolar moieties that show a strong tendency for quasi-1D supramolecular organization, giving rise to a bistable switchable conductivity.[16,17] Finally, a parallel can be drawn between the conductivity mechanism discussed herein and the conductive properties of certain bioelectronic materials that lack explicit π-stacking.[18] For example, electron transport over distances up to ~10 nm in α-helical peptide self-assembled monolayers, presumably by hopping along the amide-groups, was observed by Arikuma and coworkers.[19,20] Ing *et al.* reported long-range charge transport with a band-like temperature dependence in supramolecular $\alpha$-helical nanofibers of a synthetic peptide.[21] In both cases charge transport is occurring independent of extended $\pi$-conjugation in highly ordered hydrogen-bonded assemblies, even though the precise mechanism remained unclear. The work presented herein is distinct from these earlier reports by its focus on a broad variety of organic compounds and the proposal of a concrete mechanism for charge injection and (long-range) transport.

## 2 Results and Discussion

### 2.1 Observation of unexpected conductivity

We started our experiments with the all-*cis* fluorinated cyclohexanes, a relatively new[22] class of facially polarized small organic molecules with dipole moments up to 6.2 Debye.[23–29] The structures of these molecules are shown in Fig. 1. Three molecules, named **FCH-Cn-A** (n = 2, 3, 4), have an all-*cis* pentafluoro cyclohexane unit which is connected to a saturated spacer of differing length (2, 3 or 4 $sp^3$-hybridized carbon atoms) via an ester motif. The spacer connects to an amide group which is linked to a gallic acid unit decorated with long alkyl chains that guarantee high solubility in organic solvents and therefore enable solution-processing. Compound **FCH-E** on the other hand lacks both the spacer and the amide group, whereas compound **FCH-C3-TA** closely resembles compound **FCH-C3-A**, with the sole difference being a replacement of the amide-oxygen atom with a sulfur atom. Full chemical names and synthetic details are given in Section 1 of the Supplementary Information (SI). Previously, **FCH-C3-A** was shown to demonstrate supramolecular polymerization[30] in solution, leading to the formation of supramolecular double helices that could be transferred to solid state by simple spin coating on a substrate.[31,32] Apart from the phenyl ring to which the solubilizing linear alkyl chains are attached, the molecule contains two strong dipolar units: an amide (3.7 Debye) and, connected through a $(CH_2)_3$ spacer, an all-*cis* 1,2,3,4,5,6-hexafluorocyclohexane unit (6.2 Debye). Note that the estimated dipole moments are for isolated units; for head-to-tail stacked units, the net dipole moment will depend on the number of units stacked and might be somewhat larger due to cooperativity .[33,34]

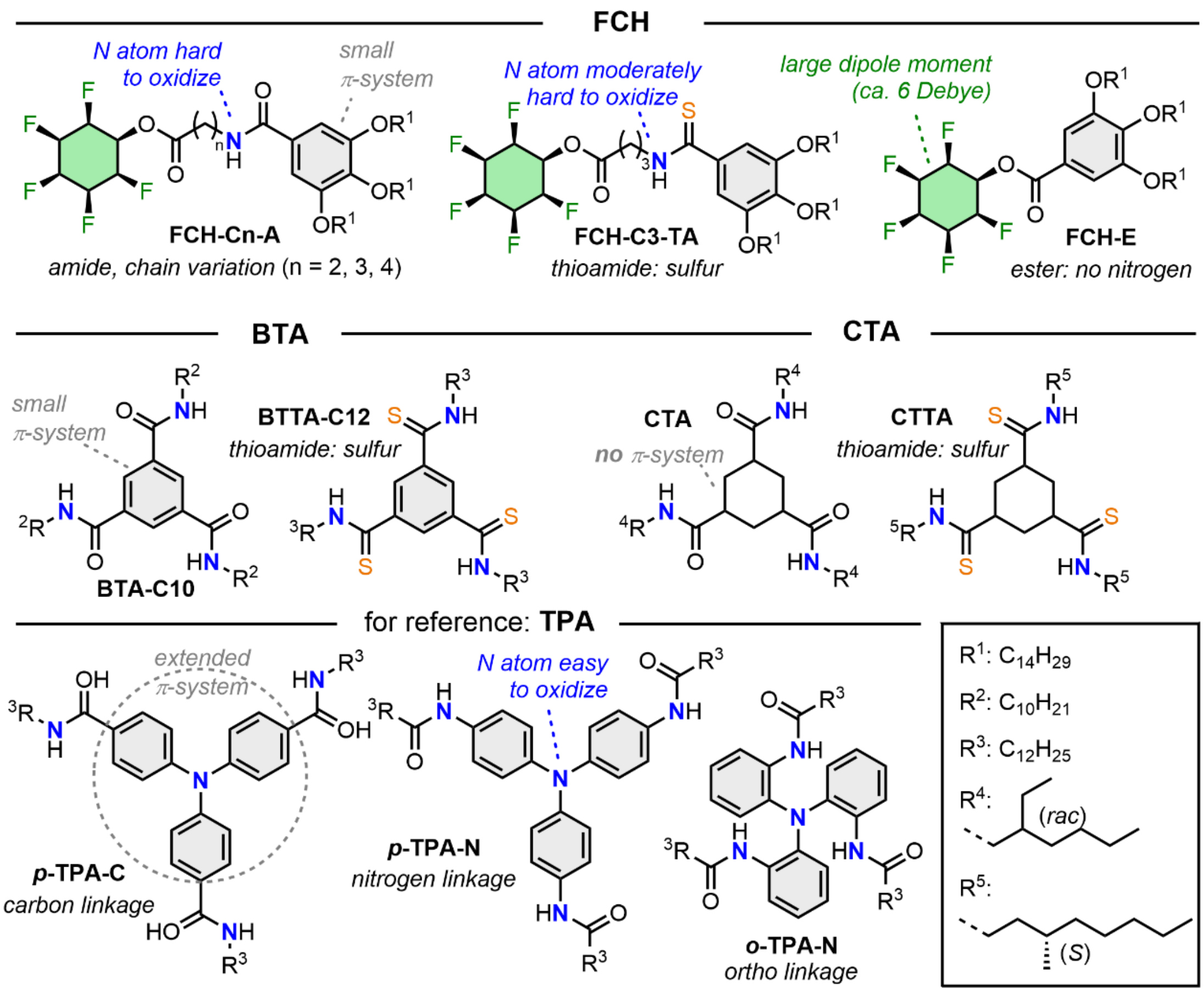

**Figure 1**: Molecular structures of the compounds investigated in this article, including research rationale. Molecular families - **FCH**: all-*cis* fluorinated cyclohexanes; **BTA**: benzene-1,3,5-tricarboxamides; **CTA**: cyclohexane-1,3,5-tricarboxamides; **TPA**: triphenylamines.

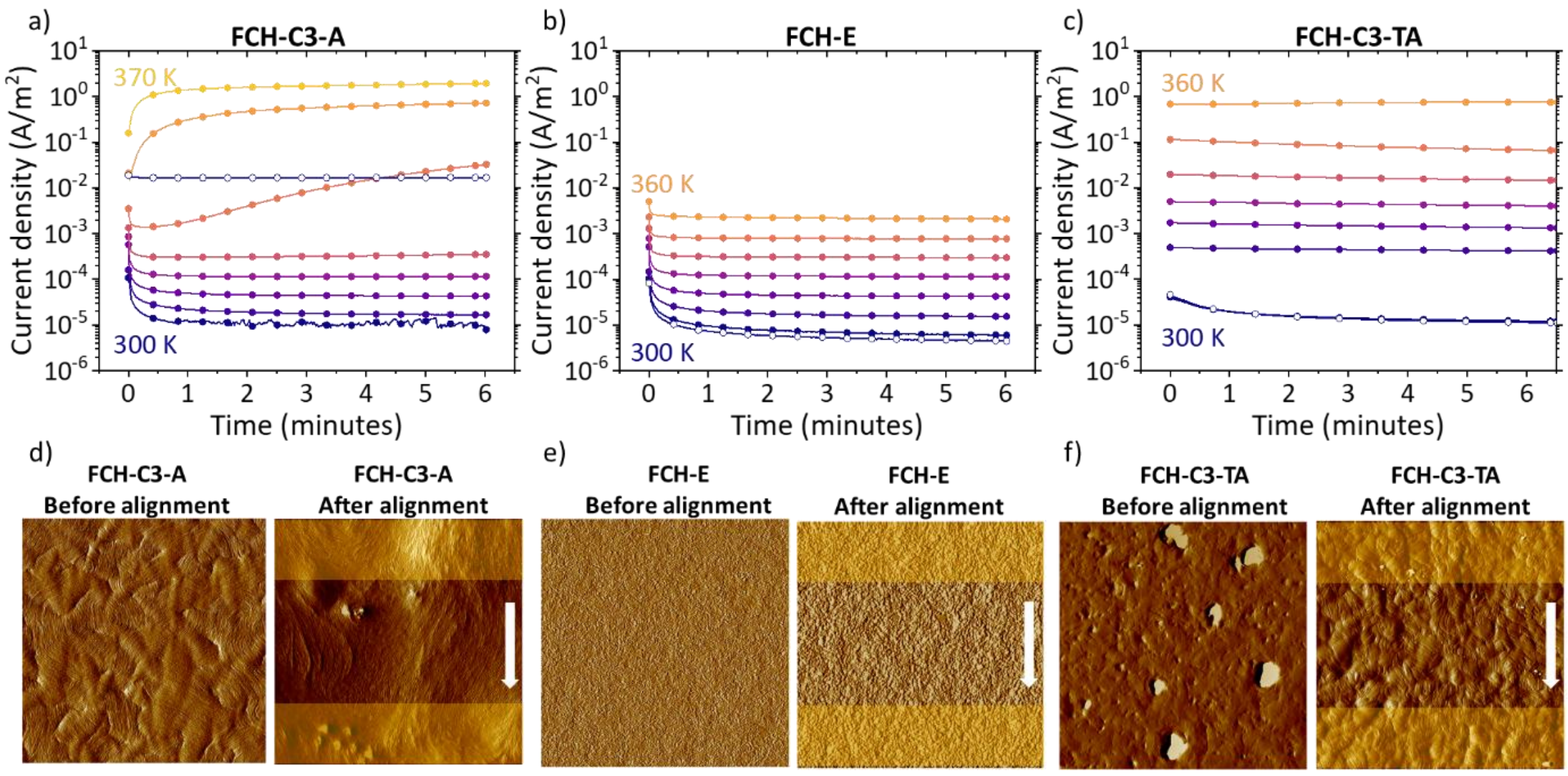

**Figure 2**: *a-c*) Measured current densities at constant voltage (100 V) plotted against measurement time for different temperatures, from 300 K up to ~20 K below the melting temperature in 10 K steps. Full symbols show data from heating up, empty symbols from cooling down. *d-f*) AFM amplitude images for thin films before and after the field-annealing procedure. The films were field-annealed for one hour under an applied field of 20 V/µm, at respective temperatures of 100, 90 and 90 °C. Electrodes are schematically indicated by lighter bars and the field direction by white arrows. Scan size is 10×10 µm$^2$, vertical scale is in arbitrary units. Corresponding topography images are given in SI Fig. S6.

Since several of the investigated compounds are prone to form rough and incompletely closed layers, precluding the use of out-of-plane diode-type devices, and in order to have access to the surface for topographical and structural characterization prior and after measurement, all electrical characterization was done on in-plane devices, using interdigitated electrodes (IDE) with a 5 µm gap. Images of the used IDE can be found in SI Fig. S5. Thin films were deposited by simple drop-casting and measured in vacuum. For the same reasons, converting measured currents to current densities leads to relatively large uncertainties of about a factor 2 or less, which is, however, sufficient for our purposes. Details on sample layout, fabrication and characterization can be found in SI Section 2.

Continuously measuring the conductivity of a fresh **FCH-C3-A** device, while stepwise increasing the temperature upward from room temperature, reveals a rich phenomenology, see Fig. 2a. At low temperatures, we observe a short transient decay, which we attribute to ionic motion, towards a plateau in the (sub-)nA-range that is similar to, but somewhat above the background current from an empty IDE at the same temperature, see SI Fig. S5. At higher temperatures, the decay is followed, and later overwhelmed by, a pronounced increase that speeds up with temperature, reaching values in the µA-range. Comparing the behavior of the compounds in the **FCH-Cn-A** series (SI Fig. S7) shows that the temperature where the upswing sets in correlates with the melting temperature but otherwise is qualitatively identical. We therefore attribute the upswing to a macroscopic rearrangement of the material during field annealing. This attribution is confirmed by the three orders of magnitude enhanced conductivity when cooling back to 300 K (cf. closed and open blue symbols in Fig. 2a) and the different surface morphologies prior and after the temperature sweep. Comparing the atomic force microscopy (AFM) images in Fig. 2d shows (bundles of) supramolecular polymer chains draped more or less randomly on the surface prior to annealing (left), while lying mostly aligned in the field direction afterwards (right). After annealing, the length of the fibers reach and partially exceed 6 µm, which is roughly double that of the pristine film, and larger than the gap width. Fig. S8 shows high-resolution images, allowing to identify individual (supramolecular) polymers with a 2-3 nm width, confirming that drop-casting the material **FCH-Cn-A** leads to the formation of quasi-1D chains in the solid state.

Stable alignment of supramolecular polymers by electrical field and elevated temperatures below the melting point has been demonstrated previously, e.g. for the discotic organic ferroelectric trialkyl-benzene-1,3,5-tricarboxamide (**BTA**).[35] Grazing incidence wide-angle X-ray scattering (GIWAXS, SI Fig. S9) confirms the presence of the corresponding short-range order with a molecular stacking distance of ~4 Å and an inter-columnar packing distance of ~20 Å.

## 2.2 Temperature-dependent studies

The observed conductivity is completely unexpected as the **FCH-C3-A** does not have an extended $\pi$-system, and DFT calculations, see SI Section 5, estimate HOMO and LUMO energy levels at -6.47 eV and -1.31 eV, respectively. Experimental data of the frontier molecular orbital energy levels of **FCH-C3-A** and other compounds obtained by cyclic voltammetry and UV-Vis spectroscopy is shown in the SI in Table S1. The experimental HOMO levels are similar or slightly less deep than those from the DFT calculations and as such support the line of arguments following below. With a bandgap similar to diamond, the material should be an insulator. We carefully ruled out spurious effects like material degradation, ion, proton or impurity conduction, or double layer formation as potential 'trivial' explanations as extensively discussed in SI Section 3. In particular, the conductivity is time-stable over hours (in ambient) or days (in $N_2$ and vacuum) and can be reversibly changed between high- and low-conductivity states by melting and field-annealing the active layer, cf. Fig. S10. The temperature-dependent conductivity of all five fluorinated cyclohexane materials shown in Fig. 1 is summarized in Fig. 3a.

Temperature-dependent conductivity measurements of **FCH-E** (see Fig. 1), which possesses a structure similar to the **FCH-Cn-A** series, with the significant difference of not containing an amide group, are shown in Fig. 2b. An increase in conductivity by roughly two orders of magnitude with temperature is observed, which is similar to an empty IDE (Fig. S5) and is to be compared to the five orders of magnitude increase for the **FCH-Cn-A** materials. Moreover, no transient increase of the current is observed, and the conductivity falls back to its original magnitude after cooling down back to 300 K. For these reasons, we classify the measured current as not significantly conductive. Although GIWAXS (SI Fig. S9) shows a more pronounced crystallization than for the **FCH-C3-A** that gets enhanced by field annealing, AFM did not show any signs of fiber formation, neither prior to nor after field-annealing.

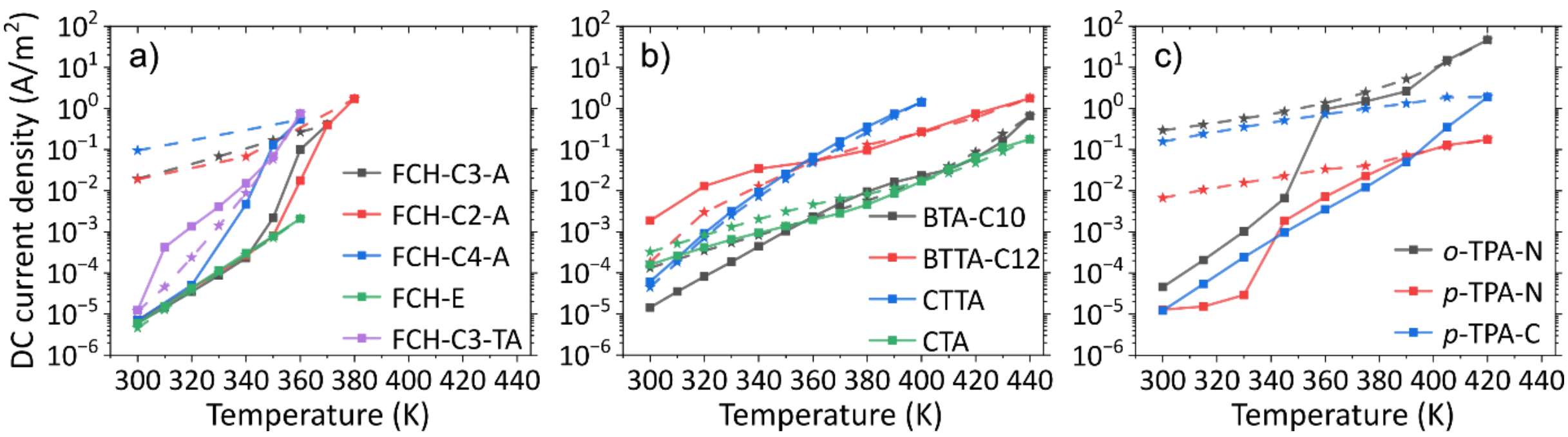

**Figure 3**: DC-current densities vs. temperature for increasing (solid lines, squares) and decreasing (dashed lines, stars) temperatures, for (a) **FCH-C3-A** derivatives, (b) **BT(T)A** and **CT(T)A** molecular discotics and (c) **TPA** compounds. Current values are measured at 100 V, in quasi-steady state after 6 minutes, cf. Fig. 2a-c.

The **FCH-E** results rule out the $\pi$-system in the phenyl group being solely responsible for the charge transport in **FCH-Cn-As**; instead, they point to the amide group playing a crucial role. Hence, we synthesized the thioamide analog to **FCH-C3-A**, cf. Fig. 1. Specifically, substitution of the oxygen atom by a less electronegative sulfur atom significantly lifts the HOMO level of the molecule (-5.71 eV compared to -6.47 eV as calculated by DFT), facilitating oxidation. Indeed, upon first annealing, the **FCH-C3-TA** compound shows a similar, five orders of magnitude conductivity increase with temperature as **FCH-C3-A**, but with significantly higher currents at lower $T$, see Fig. 2a and c and Fig. 3a. However, when cooling back down, the conductivity returns to its initial value at 300 K. The latter is consistent with the limited changes in long-range order seen in Fig. 2f; although up to ~1 µm long fibers are seen to appear after annealing, these are not as pronounced as for **FCH-C3-A** and occur in isolated domains that are significantly smaller than the electrode gap width, an observation that is in line with the general notion that hydrogen bonding between thioamides is weaker than between amides.[36] We therefore attribute the lack of permanent conductivity increases for the **FCH-C3-TA** compound to, first, the persistent presence of unordered domain boundaries in the transport path, and, second, to the lesser packing order in **FCH-C3-TA** than **FCH-C3-A** after alignment, as shown in the XRD data in SI Fig. S9.

### 2.3 Proposed mechanism for charge injection and transport

From the measurement series **FCH-Cn-A**, **FCH-E** and **FCH-C3-TA**, we conclude that the (thio)amide group is essential for the electronic transport since this is the only missing structural element in material **FCH-E**, which shows no unusual conductivity. Long-range supramolecular organization in the form of macroscopic fibers seems to enhance the conductivity but is likely not essential. In view of the latter, differences in rigidity between **FCH-Cn-A** and **FCH-C3-TA** on the one hand and **FCH-E** on the other are less likely to be essential for long-range transport. We propose that the electronic species involved in the long-range transport are holes, formed by the removal of an electron (oxidation) from the HOMO. DFT calculations in SI Fig. S3 indicate that the HOMO is at least partially localized on the nitrogen atom in all compounds that show enhanced conductivity. This oxidation could either be induced by a sufficiently strong (and unknown) accepting species or, as suggested by space-charge-limited current behavior shown later, by electron transfer to the metal contact. In both cases, hole transport takes place by inter-molecular charge transfer between adjacent molecules. Fig. 4 schematically illustrates the proposed mechanism. This scenario would not only explain the absence of significant conductivity in **FCH-E**, but also the initially higher currents in the easier-to-oxidize **FCH-C3-TA** that, at higher temperatures, get compensated by the better long-range order induced by field annealing in the **FCH-C3-A**.

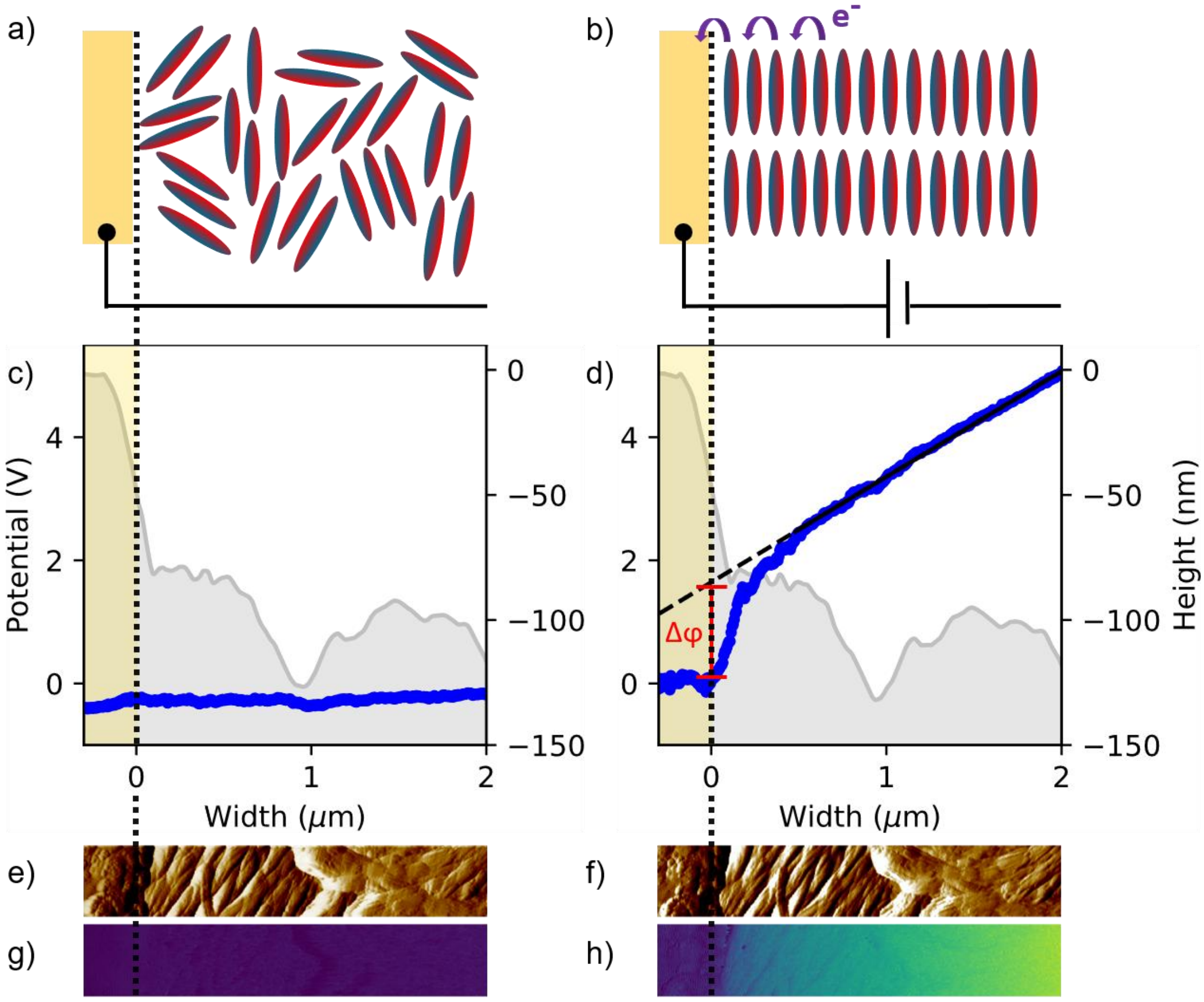

**Figure 4**: Proposed mechanism for electron injection into (hole injection from) a metal electrode. a) and b) schematically depict the positive electrode and part of the active layer in real space, in absence (a) and presence (b) of an external bias. In c) and d) the corresponding measured electrostatic potential profiles (mean over 70 lines) are plotted against the distance from the organic-electrode interface for applied voltages of 0 V and 10 V, respectively, showing a distinct potential step $\Delta\varphi$ caused by the interfacial dipolar density that facilitates charge injection (oxidation). The height profile of electrode and channel is shown in grey. e), f) show AFM amplitude images and g), h) KPFM potential maps corresponding to c) and d). The boundary between electrode and channel is marked by a black dashed line. Measurement area is 3.0×0.4 $\mu m^2$. The device was not poled prior to measurements, showing that interface dipole alignment occurs irrespective of long-range alignment.

It is not immediately evident how the initial oxidation of the first molecule in a supramolecular stack or fiber that sits next to the metal contact, i.e. charge injection, can energetically happen since the HOMO energy level of some of the materials is at -6.5 eV to -7.1 eV, while the work function of the IDE Au electrodes in ambient is around -4.5 eV.[37] DFT calculations shown in SI Section 5 indicate a HOMO level at -6.47 eV for **FCH-C3-A** and -5.71 eV for **FCH-C3-TA**, while experimental cyclic voltammetry data shows HOMO levels of -5.73 eV and -5.6 eV, respectively. In absence of interfacial dipoles, the resulting $\varphi \approx$ 1.1-2 eV injection barrier should be unsurmountable around room temperature, even if image charges and energetic disorder might lower the effective barrier by a few hundreds of meV. However, all the investigated compounds share strong dipolar moieties, which in related materials were shown to give rise to substantial interfacial energy shifts.[38] In SI Section 4, we estimate the dipolar shift due to interfacial alignment of the molecular dipoles to be around $\Delta\varphi \approx$ 1.9 eV, meaning that at the applied fields and temperatures, the oxidation level might well be accessible for what is actually best described as hole injection from the Au electrode. This scenario predicts a lowering of the injected current when an electrode material with a lower work function is used. Indeed, using Al instead of Au IDEs was found to significantly reduce the device current as shown in SI Fig. S11.

## 2.4 Response to external electric fields

To first show that the dipolar moieties in our materials are indeed able to respond to external electrical fields, we performed capacitance-voltage (CV) and double-wave method (DWM) measurements, which are typical measurements used to identify ferroelectricity. Further details are given in SI Section 2. For our current purposes, the question whether any of the compounds investigated is truly ferroelectric is irrelevant and the data in Fig. 5 and SI Fig. S12 only serve to prove the presence of field-responsive dipoles. Panel (a) shows for the case of **FCH-C3-A**, the butterfly shape that is characteristic of dipoles becoming free to rotate at the coercive field, which, for the used parameters, sits around 10 V/µm.[39] Panel (b) confirms the presence of switchable bistable polarization by showing near-rectangular polarization hysteresis curves. The slightly different amplitudes in (b) result from a minor asymmetry in positive and negative branches of the DWM response.

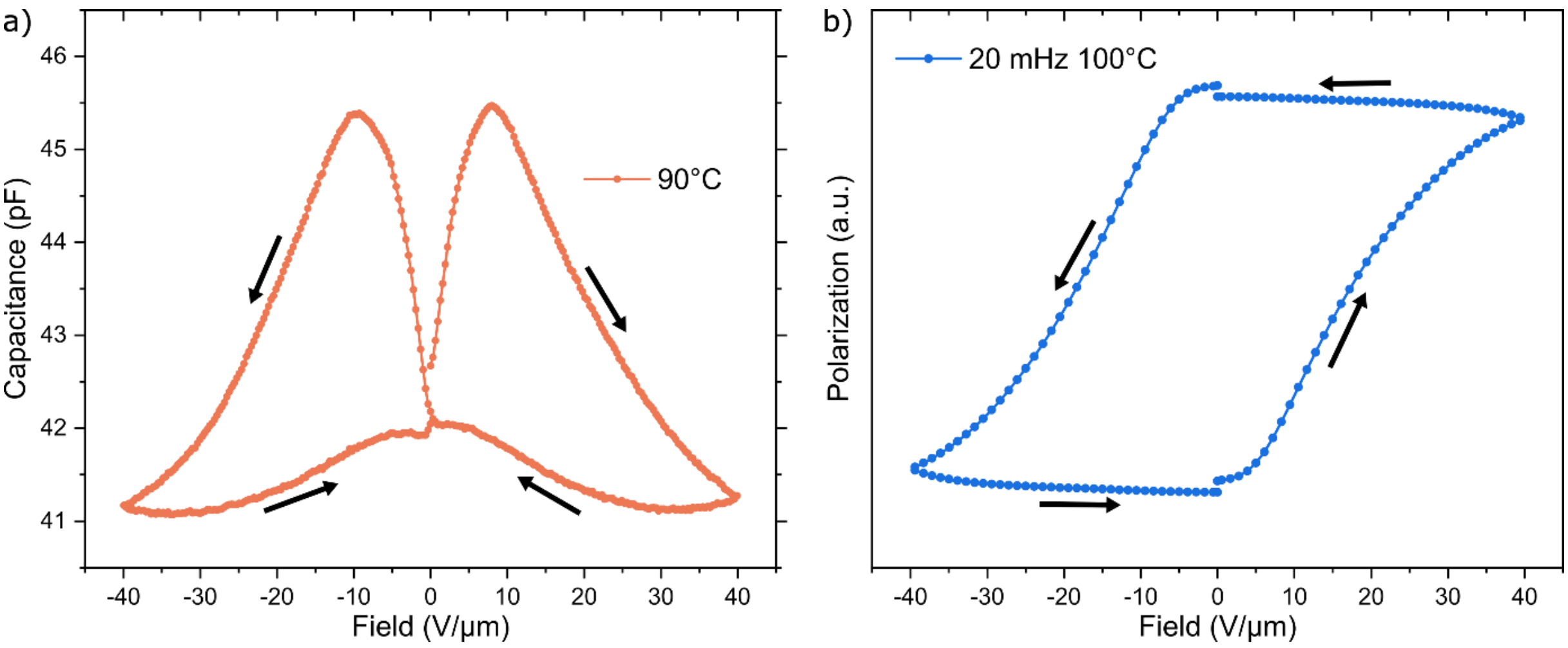

**Figure 5**: Capacitance-voltage (a) and polarization hysteresis measurements (b) for **FCH-C3-A**. Arrows denote the sweep direction. Normalized polarization is plotted in view of the ill-defined layer thickness and inhomogeneous field profiles in IDEs. The corresponding measurements for **FCH-E** and **FCH-C3-TA** are given in SI Fig. S12.

Given the larger rigidity of **FCH-E** as compared to **FCH-C3-(T)A**, one might, alternatively,

attribute the conductivity differences to an inability to align interfacial dipoles for the former material. However, the CV and DWM measurements for **FCH-E** in SI Fig. S12 show that the dipole on the pentafluorocyclohexane ring is still mobile in this compound, ruling out that scenario.

2.5 Modulation of charge injection barrier and generality

Direct proof of the hypothesized injection barrier modulation by field-aligned interfacial dipoles is provided by in-operando Kelvin probe force microscopy (KPFM) of the channel region (see Fig. 4c and d) that shows a potential step of $\Delta\varphi \approx 1$ eV at the organic-electrode interface that changes polarity with bias. The measured barrier height reduction is on the lower end of the predicted range of interface barrier heights. This is expected, not only because of the modest (instrument-limited) voltages applied, but also since the effects of image charges and energetic disorder lower the barrier height reduction that is needed for ohmic charge injection; once a contact is ohmic, further barrier lowering is prevented by the injected charges. The process is self-limiting.

To further test the universality of our hypothesis, we investigated other compounds that are known, or can be expected, to meet all three criteria that we propose to give rise to conductance without extended $\pi$-systems: (i) an oxidation center, (ii) H bonding to enable supramolecular polymerization, (iii) strong dipoles. The first class of materials we (re)visited are the molecular discotics trialkylbenzene-1,3,5-tricarboxamide (**BTA**) and trialkylbenzene-1,3,5-triscarbothioamide (**BTTA**) (Fig. 1), which show a rich self-organization behavior by hydrogen bonding into supramolecular stacks.[40] In the solid state, the molecular stacks build a typically hexagonal lattice that enters a liquid crystalline phase around or slightly above room temperature. A field annealing procedure that is similar to what is used here aligns the supramolecular stacks parallel to the applied electric field, leading to stable ferroelectric switching with a molecular dipole moment of ~12 Debye per molecule.[39] Interestingly, the thioamide-analog of **BTA** was previously found to show a pronounced 'leakage' current at elevated temperatures when sandwiched between noble metal electrodes, which was tentatively attributed to electrochemical reactions induced by sulfur in the presence of noble metals.[41]

When explicitly tested for DC conductivity, both **BTA-C10** and **BTTA-C12** showed a DC-conductivity at elevated temperature, as shown in SI Fig. S13 (raw data) and Fig. 3b. However, the relative conductivity increase with temperature for the **BTA-C10** is significantly less than for **FCH-C3-A**. This can be readily explained by the deeper lying HOMO energy level (-7.13 eV vs. -6.47 eV) and the smaller dipolar density and correspondingly smaller dipolar shift (~1 eV vs. ~1.9 eV) due interfacial alignment of the amide dipoles, resulting in a higher net injection barrier (~1.6 eV vs. ~0.1 eV) at the organic-metal interface. As for **FCH-C3-A** and **FCH-C3-TA**, the thio-analog of **BTA** (HOMO of -5.88 eV) shows an increased conductivity already at lower temperatures than the amide-analog, which is consistent with the much lower net injection barrier, estimated at ~0.4 eV. We stress that the phenomenology (changes in optical spectra, formation of gas bubbles) that was previously observed for pyrrole-substituted **BTA** that also showed a finite conductivity due to (intended) electropolymerization, is not seen in any of the compounds investigated herein.[42] In further contrast to the **BTA-C10-pyrrole** investigated in Ref. [42], for which the current starts to drop on the time scale of seconds, the current in **BTA-C10** and **BTTA-C12** is time-stable after initial transients have settled.

With a similar molecular discotic structure, and similar self-organization properties to **BT(T)A**s, but with a cyclohexane instead of a benzene core, cyclohexane-tris(amide) (**CTA**) and its thio-equivalent **CTTA** is an excellent candidate to investigate how the complete absence of a π-system affects the conductivity[43]. As for **BTA-C10**, the **CTA** compound shows only a modest conductivity increase with increasing temperature, whereas the **CTTA**, like the **BTTA-C12**, exhibits higher conductivity already at lower temperatures and a significant relative increase in conductivity over the whole temperature range, see Fig. 3b and raw data in SI Fig. S14. In view of the very similar HOMO levels of the **CTA**-compounds to those of the **BTA**s, the differences between **CTA** and **CTTA** can readily be explained in terms of (net) injection barriers. More importantly, the highly similar behavior of **BTA** and **CTA** derivatives is a strong indicator that not even a small π-system is required for long-range charge transport in organic materials.

Both **BTA-10** and **BTTA-C12** are known to exhibit a large degree of long-range order due to supramolecular polymerization that can be oriented and improved by field-annealing, as further confirmed by the AFM and GIWAXS images in SI Fig. S15 and SI Fig. S16.[35,39,41] Also, for **CTA**, AFM (SI Fig. S17) and GIWAXS images (SI Fig. S16) point towards increased structural order and long-range alignment, whereas **CTTA** shows less clear trends, with the film becoming very smooth in combination with a reduction in observed GIWAXS reflexes, which might point to an onset of degradation. In this context, it is slightly surprising that for neither the **BT(T)A** nor the **CT(T)A** molecules a permanently enhanced conductivity at lower temperatures is observed. We interpret this as being mainly the result of the currents becoming limited by the remaining charge injection barriers, which is consistent with the conductivity not being space-charge (bulk) limited at these temperatures, as further discussed below.

To put the observed behavior qualitatively and quantitatively in perspective, we synthesized a series of triphenylamine (**TPA**) derivatives decorated with peripheral amide units as depicted in Fig. 1. The **TPA** motif is a well-established unit, especially in hole-conducting polymeric and small molecular materials that allows for the wavefunction to extend over the entire molecule, i.e. the aromatic lobes and the connecting N-atom[14,15]. In spite of the phenyl rings oriented in a propeller-like fashion, the HOMO wavefunction is almost equally distributed over the entire **TPA** scaffold including the central nitrogen with its electron lone pair, see SI Section S5[44]. However, the HOMO energy levels obtained experimentally and by DFT calculations are comparable to those of the **FCH-C3-TA, BTTA** and **CTTA**.

Despite extensive investigation, we could not establish any dipolar activity in the **TPA** compounds, which is in line with previous results for similar non-planar C3-symmetric compounds.[45] The fact that we nevertheless see a very similar development of the electrical conductivity in all **TPA** compounds is consistent with the notion that the dipolar moieties are only needed to make deeper-lying energy levels accessible for charge transfer with the contacts, but not per se for the charge transport itself. The very similar behavior of the three **TPA**s, which only show quantitative differences, is consistent with the lack of differences found in the AFM and GIWAXS analyses, cf. SI Fig. S18 and SI Fig. S19. For all compounds, minor changes in surface topography upon field annealing are observed by AFM, in combination with minor improvements in crystallinity. More importantly, the three **TPA**s show an enhanced low-temperature conductivity after field-annealing. Fig. 3c shows that both in terms of qualitative behavior and magnitude, the conductivities in the **TPA** compounds are comparable to **FCH-C3-A**.

2.6 Quantitative model

In the preceding discussions, it was assumed that charge transport occurs by charges (holes) that are first injected from the contacts, that is, the material itself is intrinsic. The absence of a significant number of free charges at zero applied field is expected on basis of the simple fact that doping a material with a HOMO level around or below -6 eV is far from trivial and unlikely to occur spontaneously by, e.g. an impurity, as further discussed in the SI, Section 3. Strong evidence for this notion is found in the IV-curves after cooling back to 300 K shown in Fig. 6a. In the absence of effective doping, charge transport in an (intrinsic) semiconductor occurs by charges that are injected from the contacts. As these necessarily build up a finite space charge in the system, the resulting space charge limited current (SCLC) follows a characteristic quadratic bias dependence. For a system with ohmic contacts, this can be described according to Murgatroyd[46]:

$$J=\frac{9}{8}\epsilon_r\epsilon_0\mu_0\frac{(V\text{-}V_{\mathrm{bi}})^2}{L^3}\exp\left(0.891\gamma\sqrt{\frac{V\text{-}V_{\mathrm{bi}}}{L}}\right) \quad (1)$$

where $\epsilon_r\epsilon_0$ is the dielectric constant and $\mu_0$ the low-field mobility. The exponential term phenomenologically captures the, for organic materials typical, field- and/or carrier-density-dependence of the mobility, with $\gamma$ as an empirical factor.[47] In passing, we note that for systems showing SCLC behavior, the conductivity is inherently voltage-dependent and as such not meaningful, and one should accordingly focus on the (low-field) mobility for material comparison.

The black lines in Fig. 6a are fits to Eq. 1; the fit parameters are given in SI Table S2. Only a subset of materials is shown, namely representative examples of material groups that show SCLC behavior around room temperature. E.g. members of the BTA family of materials only show SCLC behavior at elevated temperatures, or in some cases not at all, as expected on basis of the estimated injection barriers. Corresponding IV-curves for the discotic molecules at elevated temperatures and a brief analysis are given at SI Fig. S20. For the material groups for which a negligible net injection barrier is expected, indeed SCLC behavior is found, as illustrated for **FCH-C3-A** and ***p/o*-TPA-N**; minor deviations from Eq. 1 at low fields are expected to occur due to, e.g., leakage currents stemming from the substrate[47]. We attribute the current offset at low bias observed for ***p*-TPA-N** to motion of (unknown) ionic species that give rise to a small battery-like background current that becomes visible due to the relatively low mobility of this compound. More importantly, the low-field mobility of the compounds showing the novel conductivity mechanism, **FCH-C3/4-A**, **BTTA-C12** and **CTA** (SI Table S2) lies in the range $10^{-8}$ to $10^{-6}$ $cm^2/Vs$, which is very much comparable to the values found for the established π-conjugated semiconductor motif **TPA,** cf. Fig. 3c and SI Table S2, and to other conventional disordered organic semiconducting polymers.[47–50] Note also that the seemingly high voltages required to reach appreciable current densities are mostly due to the large inter-electrode gap of 5 µm; from the $V^2/L^3$-scaling of Eq. 1, the same current densities as here at 100 V would be reached below 0.3 V for an out-of-plane device with a common active layer thickness of 100 nm.

For conventional organic semiconductors, that is, materials where the HOMO (and LUMO) levels are located on an extended π-system, charge transport is commonly described by the Gaussian disorder model (GDM)[51]. In the GDM, transport is assumed to take place by hopping in the low energy tail of a Gaussian-shaped density of localized states of width $\sigma_{\mathrm{DOS}}$, giving rise to a temperature-dependent mobility of the form

$$\mu(T) \propto \exp\left(\text{-}C\left(\frac{\sigma_{DOS}}{k_{\mathrm{B}}T}\right)^2\right) (2)$$

with $C$ as a constant of order unity and Boltzmann's constant $k_{\mathrm{B}}$. As expected, the ***p***- and ***o*-TPA-N** compounds follow the GDM at lower temperatures, see Fig. 6b, with a modest energetic disorder around 60 meV (SI Table S2). As application of the GDM model requires a sufficiently time-stable behavior, only few of the novel compounds could be analyzed using Eq. 2. Nevertheless, the fact that the **FCH-C3-A** compound shows an equally good fit to Eq. 2 over the full range of measured temperatures, with a somewhat higher but still modest disorder of ~76 meV, further illustrates the strong similarity between, and potentially the universality of charge transport in organic materials with and without an (extended) π-system. In all cases, deviations from a perfect fit to Eq. 2 are well within range of what is commonly observed for intrinsic organic semiconductors.[47]

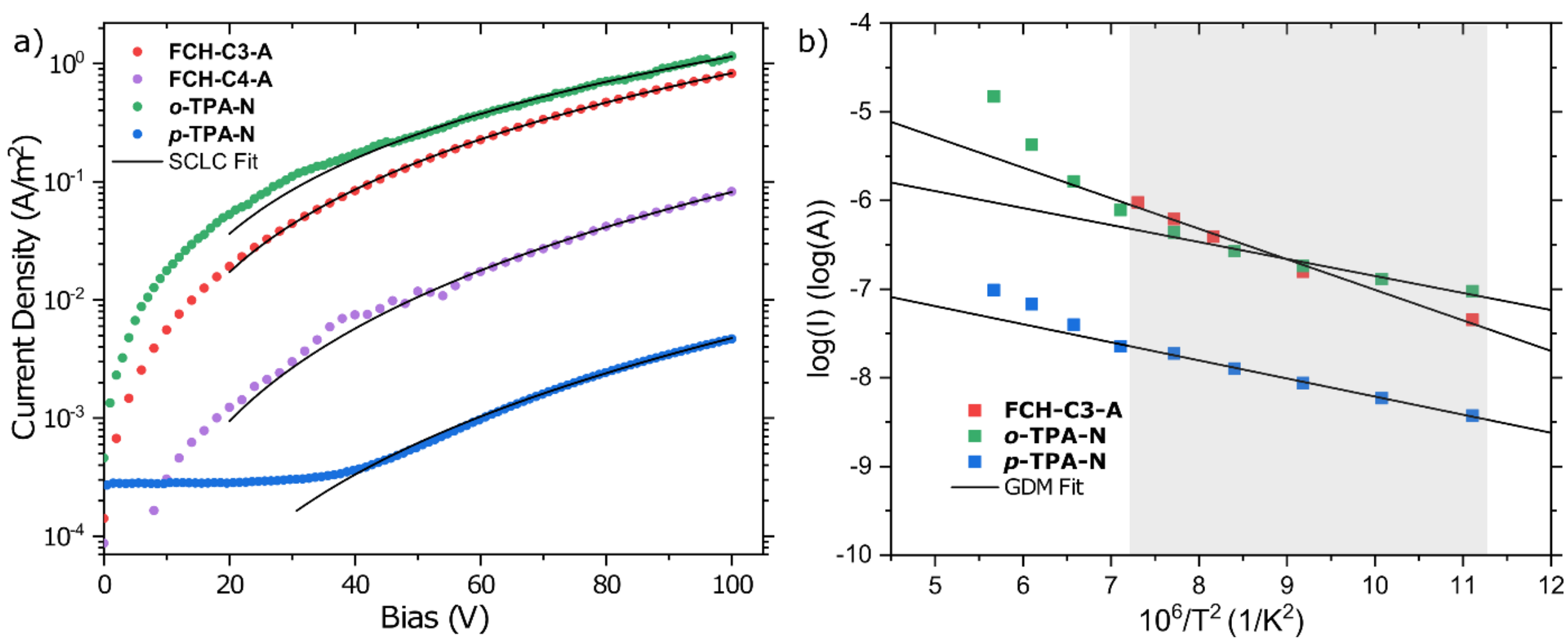

**Figure 6:** a) Current-voltage characteristics for selected materials at 300 K after field annealing, fitted with the SCLC expression Eq. 1. b) Steady-state current vs temperature (cooling) for selected materials showing SCLC behavior over a sufficiently wide temperature range, fitted with the GDM expression Eq. 2. The fitting range is marked in grey, higher temperature points are left out to avoid (transient) annealing effects. The corresponding fitting parameters are given in the SI Table S2.

## 3 Conclusion

In this work, we studied the charge transport in two families of small-molecule organic materials that share a strong tendency for supramolecular polymerization, forming quasi-1D molecular fibers. In contrast to conventional organic semiconductors, all materials contain dipolar moieties and either contain only a small π-system, typically a single benzene ring, or no π-system at all. An unexpected electrical conductivity is observed that shows qualitatively and quantitatively the same characteristics as established (π-conjugated) small-molecule semiconductors. In particular, the conductivity can become space-charge-limited, following the Gaussian disorder model for hopping of localized charges. We rationalize these observations as follows: charge transport occurs as a two-step process, in which an electron is drawn from the HOMO level to a contact electrode (oxidation), followed by charge transfer between neighboring molecules. The HOMO level becomes energetically accessible to oxidation only due to the (field-driven) alignment of the dipolar moieties at the metal-organic interface, which was directly visualized by in-operando SKPM. The long-range charge transport is facilitated by, but does not strictly require, long-range order due to molecular self-organization under the effect of an external field.

This two-fold explanation – (i) charge injection thanks to dipoles, (ii) hole transport thanks to supramolecular organization – may not only apply to the compounds presented herein, but also for scattered reports in the literature on conductivity in organic molecules lacking an extended π-system, including peptides[19–21] and radical polymer glasses[52]. Since the absolute values for charge carrier mobility and energetic disorder are roughly at the level of conventional disordered organic semiconductors, materials following the design rules laid out above greatly, and realistically so, expand the catalogue of semiconducting organics. The absence of the unsaturated bonds that are needed to form an extended π-system may be anticipated to lead to an enhanced environmental stability of such materials.

**Acknowledgements**

We thank the Deutsche Forschungsgemeinschaft (DFG, German Research Foundation) for support of this work (project numbers 281029004 (SFB 1249 and DE-1830/6-1). M.Ke. thanks the Carl Zeiss Foundation for financial support. The authors acknowledge support by the state of Baden-Württemberg through bwHPC and the German Research Foundation (DFG) through grant no INST 40/575-1 FUGG (JUSTUS 2 cluster). O. S. thanks the DAAD for a doctoral scholarship. M.G.I. and D.G.-R. are grateful to projects Santander Talent Attraction Research (STAR2), PID2020-116921GB-I00, PID2021-125429NA-I00, CNS2022-135129 and TED2021-132602B-I00, funded by MCIN/AEI/10.13039/ 501100011033 and NextGeneration EU/PRTR. We thank E.W. Meijer for providing materials and stimulating discussions and Andreas Gruber and Kerstin Leopold for elemental analysis via TXRF.

**Conflict of Interest**

The authors declare no conflict of interest.

**Data Availability Statement**

The data that support the findings of this study are available in the supplementary material of this article.

Supplementary Information to

# Highly polar groups and supramolecular order enable charge injection and long-range conductivity in organic materials without extended π-systems

H. Mager[1], A. A. Butkevich[1], S. Klubertz[1], S. V. Haridas[2], O. Shyshov[2], V. C. Wakchaure[2], J. Borstelmann[3], I. Michalsky[3], M. García-Iglesias[4], V. Rodríguez[4,5], D. González-Rodríguez[5], A. R. A. Palmans[6], M. Kivala[3], M. von Delius[*,2] and M. Kemerink[*,1]

[1] Institute for Molecular Systems Engineering and Advanced Materials, Heidelberg University, Im Neuenheimer Feld 225, 69120 Heidelberg, Germany

[2] Institute of Organic Chemistry, Ulm University, Albert-Einstein-Allee 11, 89081 Ulm, Germany

[3] Institute of Organic Chemistry, Heidelberg University, Im Neuenheimer Feld 270, 69120 Heidelberg, Germany

[4] QUIPRE Department, Nanomedicine-IDIVAL, Universidad de Cantabria, Avd. de Los Castros, 46, 39005 Santander, Spain.

[5] Nanostructured Molecular Systems and Materials Group, Organic Chemistry Department, Universidad Autónoma de Madrid, 28049 Madrid, Spain

[6] Laboratory of Macromolecular and Organic Chemistry, Institute for Complex Molecular Systems, Eindhoven University of Technology, P.O. Box 513, 5600 MB Eindhoven, The Netherlands.

*Corresponding authors: max.vondelius@uni-ulm.de; martijn.kemerink@uni-heidelberg.de

## 1 – Materials

### 1.1 Synthesis of FCH-Cn-A, FCH-E and FCH-C3-TA:

All commercially available reagents and solvents were purchased from Merck- Sigma Aldrich, ABCR, and TCI Germany and were used without further purification. Hydrogenation reactions were carried out in a 150 mL Roth High pressure steel reactor. NMR spectra were recorded on a BrukerAvance 600 ($^{1}$H: 600 MHz; $^{13}$C: 151 MHz) and BrukerAvance 400 ($^{1}$H: 400 MHz; $^{13}$C: 126 MHz, $^{19}$F: 376 MHz) spectrometers at 298 K and referenced to the residual solvent peak ($^{1}$H: chloroform-*d*, 7.26 ppm; acetone-$d_6$, 2.05 ppm; dichloromethane-$d_2$, 5.32 ppm; methanol-$d_4$, 3.31 ppm; THF-$d_8$, 1.72 ppm; $^{13}$C: chloroform-*d*, 77.16 ppm; acetone-$d_6$, 206.26 ppm; methanol-$d_4$, 49 ppm; dichloromethane-$d_2$, 53.84 ppm; THF-$d_8$, 25.31 ppm). Coupling constants (*J*) are denoted in Hz and chemical shifts ($\delta$) in ppm. Multiplicities are denoted as follows: s = singlet, d = doublet, t = triplet, m = multiplet, br = broad. High-resolution mass spectrometry (HRMS) was performed using a Fourier Transform Ion Cyclotron Resonance (FT-ICR) mass spectrometer solariX (Bruker Daltonik GmbH, Bremen, Germany) equipped with a 7.0 T superconducting magnet and interfaced to an Apollo II Dual ESI/MALDI source (using *trans*-2[3-(4-*tert*-butylphenyl)-2methyl-2-propenylidene]malononitrile (DCTB) as matrix for MALDI), which can be switched from ESI/APCI to MALDI operation almost instantaneously. Analytical thin layer chromatography was used to monitor reactions on pre-coated aluminum plates and visualized by dipping in $KMnO_4$ stain followed by heating.

**Scheme S1**. Synthesis of all-*cis* fluorinated cyclohexane based monomers: **FCH-Cn-A**, **FCH-C3-A-TA** and **FCH-E**.

The synthesis and characterization of **FCH-C3-A** have described previously[1]. **FCH-C2-A**, **FCH-C4-A** were synthesized following the same **general procedure**:

Pentafluorophenyl ester (**S2**) (4 mmol, 2 equiv.), 5 mL of DMF and 3 mL of DIPEA were added to the amine (**S1**, 2 mmol, 1 equiv.) in dry DCM (10 mL) and the reaction mixture was stirred at room temperature overnight. After removal of solvents under reduced pressure the crude material was purified by flash column chromatography ($CH_2Cl_2$ to $CH_2Cl_2$/acetone 8:2) to yield the title compound.

Comments on reaction conditions, side products and purification:

1. DMF is needed to completely dissolve starting materials containing all-cis $C_6H_6F_5$ group and to prevent gelation which sometimes can happen if DCM alone is used as a solvent.

2. To completely remove pentafluorophenol from the reaction mixture, sometimes two purifications by flash column chromatography are needed.

3. For the linker based on γ-aminobutyric acid approximately 13% of butyrolactam was formed as a side product.

**(1r,2R,3R,4s,5S,6S)-2,3,4,5,6-pentafluorocyclohexyl 3-(3,4,5-tris(tetradecyloxy)benzamido)propanoate - (FCH-C2-A)**

Prepared according to general procedure. Yield: 64%

**$^{1}$H NMR** (400 MHz, $CDCl_3$, 298 K) δ 6.96 (s, 2H), 6.62 (s, br, 1H), 5.57 – 4.25 (m, 6H), 3.98 (q, *J* = 6.1 Hz, 6H), 3.77 (q, *J* = 8.0 Hz, 2H), 2.83 (t, *J* = 5.7 Hz, 1H), 1.84 – 1.62 (m, 6H), 1.21 – 1.40 (m, 8H), 1.40 – 1.17 (m, 59H), 0.88 (t, *J* = 6.7 Hz, 9H).

**$^{13}$C NMR** (151 MHz, $CDCl_3$, 298 K) δ 171.0, 167.6, 153.2, 135.9, 128.4, 125.7, 107.8, 106.6, 105.6, 87.1, 85.7, 73.6, 69.4, 67.8, 67.6, 32.1, 30.5, 29.9, 29.9, 29.9, 29.8, 29.8, 29.8, 29.8, 29.6, 29.5, 26.2, 22.8, 14.2.

**$^{19}$F NMR** (376 MHz, $CDCl_3$, 298 K) mixture conformers δ -209.69 (br, s, 2F), -210.44, -211.42, -214.82 (br, s, 2F), -215.52, -216.83 (br, s, 1F).

**HR-MS** (MALDI) - m/z = Cal. For $C_{58}H_{100}F_5NO_6Na$: 1024.7363, observed: 1024.7363 $[M+Na]^+$.

**(1r,2R,3R,4s,5S,6S)-2,3,4,5,6-pentafluorocyclohexyl 5-(3,4,5-tris(tetradecyloxy)benzamido)pentanoate - (FCH-C4-A)**

Prepared according to general procedure. Yield: 71%

**$^{1}$H NMR** (400 MHz, $CDCl_3$, 298 K) δ 6.94 (s, 2H), 6.11 (t, *J* = 5.9 Hz, 1H), 5.51 – 4.30 (m, 6H), 3.99 (m, 6H), 3.46 (q, *J* = 6.6 Hz, 2H), 2.57 (t, *J* = 7.1 Hz, 2H), 1.89 – 1.62 (m, 11H), 1.53 – 1.40 (m, 8H), 1.40 – 1.06 (m, 63H), 0.88 (t, *J* = 6.7 Hz, 5H).

**$^{13}$C NMR** (151 MHz, $CDCl_3$, 298 K) δ 172.6, 167.7, 153.3, 141.3, 129.7, 105.8, 73.7, 69.6, 39.7, 39.6, 33.6, 32.1, 30.5, 29.9, 29.9, 29.9, 29.8, 29.8, 29.8, 29.8, 29.6, 29.5, 29.5, 29.1, 26.2, 22.8, 22.2, 14.3.

**$^{19}$F NMR** (376 MHz, $CDCl_3$, 298 K) mixture conformers δ -209.64 (br, s, 2F), -215.30 (br, s, 2F), -216.84 (br, s, 1F).

**HR-MS** (MALDI) - m/z = Cal. For $C_{60}H_{104}F_5NO_6Na$: 1052.7676, observed: 1052.7678 $[M+Na]^+$.

**(1r,2R,3R,4s,5S,6S)-2,3,4,5,6-pentafluorocyclohexyl 4-(3,4,5-tris(tetradecyloxy)phenylthioamido)butanoate - (FCH-C3-TA)**

Synthesis of **FCH-C3-TA** was carried out using Lawesson´s reagent. To a solution of **FCH-C3-A** (0.098 mmol, 1 eq) in toluene (50 ml) was added Lawesson´s reagent (0.5 eq.) and refluxed for 2 hr (TLC monitoring). The solvents were evaporated and column chromatography (dry loading) was performed using 3:2 petroleum ether, ethyl acetate eluent system to obtain the **FCH-C3-TA** in 80% yield.

**$^1$H NMR** (600 MHz, acetone-$d_6$, 298 K) δ 9.35 (t, $J$ = 5.6 Hz, 1H), 7.20 (s, 2H), 5.58 – 4.98 (m, 6H), 4.02 (t, $J$ = 6.2 Hz, 4H), 3.99 (t, $J$ = 6.4 Hz, 2H), 3.91 (q, $J$ = 6.7 Hz, 2H), 2.12 (p, $J$ = 7.3 Hz, 2H), 1.80 (dt, $J$ = 14.8, 6.4 Hz, 4H), 1.76 – 1.70 (m, 2H), 1.57 – 1.49 (m, 7H), 1.43 – 1.24 (m, 62H), 0.91 – 0.86 (t, 9H).

**$^{13}$C NMR** (151 MHz, acetone-$d_6$, 298 K) δ 199.1, 172.7, 153.5, 141.7, 137.7, 107.2, 87.4, 86.6, 73.9, 69.9, 67.7, 67.6, 46.4, 32.8, 32.8, 32.2, 31.4, 30.7, 30.7, 30.7, 30.7, 30.6, 30.6, 30.6, 30.6, 30.5, 30.5, 27.2, 27.1, 24.1, 23.5.

**$^{19}$F NMR** (376 MHz, acetone-$d_6$, 298 K) mixture conformers δ -209.68 (br, s, 2F), -210.38, -211.52, -215.84 (br, s, 2F), -215.84, -216.83 (br, s, 1F).

**HR-MS** (MALDI) - m/z = Cal. For $C_{59}H_{103}F_5NO_5S$: 1032.7472, observed: 1032.7466 $[M+H]^+$.

**(1r,2R,3R,4s,5S,6S)-2,3,4,5,6-pentafluorocyclohexyl 3,4,5-tris(tetradecyloxy)benzoate - (FCH-E)**

In a 10 mL round-bottom flask pentafluorophenyl ester (**S2**) (600 mg, 0,65 mmol, 2.5 equiv.) and **S3** (50 mg, 0.26 mmol, 1 equiv.) were dissolved in 3 mL of dry DMF and 3 mL of triethylamine. Reaction mixture was stirred at 120 °C until **S2** was consumed (ca. 2 – 3 hours). Solvents were removed under reduced pressure and the crude material was purified by flash column chromatography (gradient elution CH2Cl2 to CH2Cl2/acetone 8:2) to afford **FCH-E** (Yield 96%) as a colourless solid.

**$^{1}$H NMR** (600 MHz, THF-$d_8$, 298 K) δ 7.37 (s, 2H), 5.47 – 4.59 (m, 6H), 4.01 (t, *J* = 6.3 Hz, 6H), 1.81 (p, *J* = 6.6 Hz, 4H), 1.52 (p, *J* = 7.9, 7.4 Hz, 6H), 1.43 – 1.24 (m, 61H), 0.89 (t, *J* = 6.8 Hz, 9H).

**$^{13}$C NMR** (151 MHz, THF-$d_8$, 298 K) δ 165.6, 154.0, 144.4, 109.4, 87.0, 73.8, 69.9, 68.0, 67.8, 67.7, 32.9, 31.4, 30.8, 30.8, 30.8, 30.7, 30.7, 30.7, 30.6, 30.4, 30.4, 30.4, 14.5.

**$^{19}$F NMR** (376 MHz, THF-$d_8$, 298 K) mixture conformers δ -211.44 (br, s, 2F), -212.16, -213.2, -216.08 (br, s, 2F), -217.52, -217.91 (br, s, 1F).

**HR-MS** (ESI -ve) - m/z = Cal. For $C_{57}H_{95}F_8O_7$: 1043.6950, observed: 1043.6947 $[M+CF_3COO]^-$.

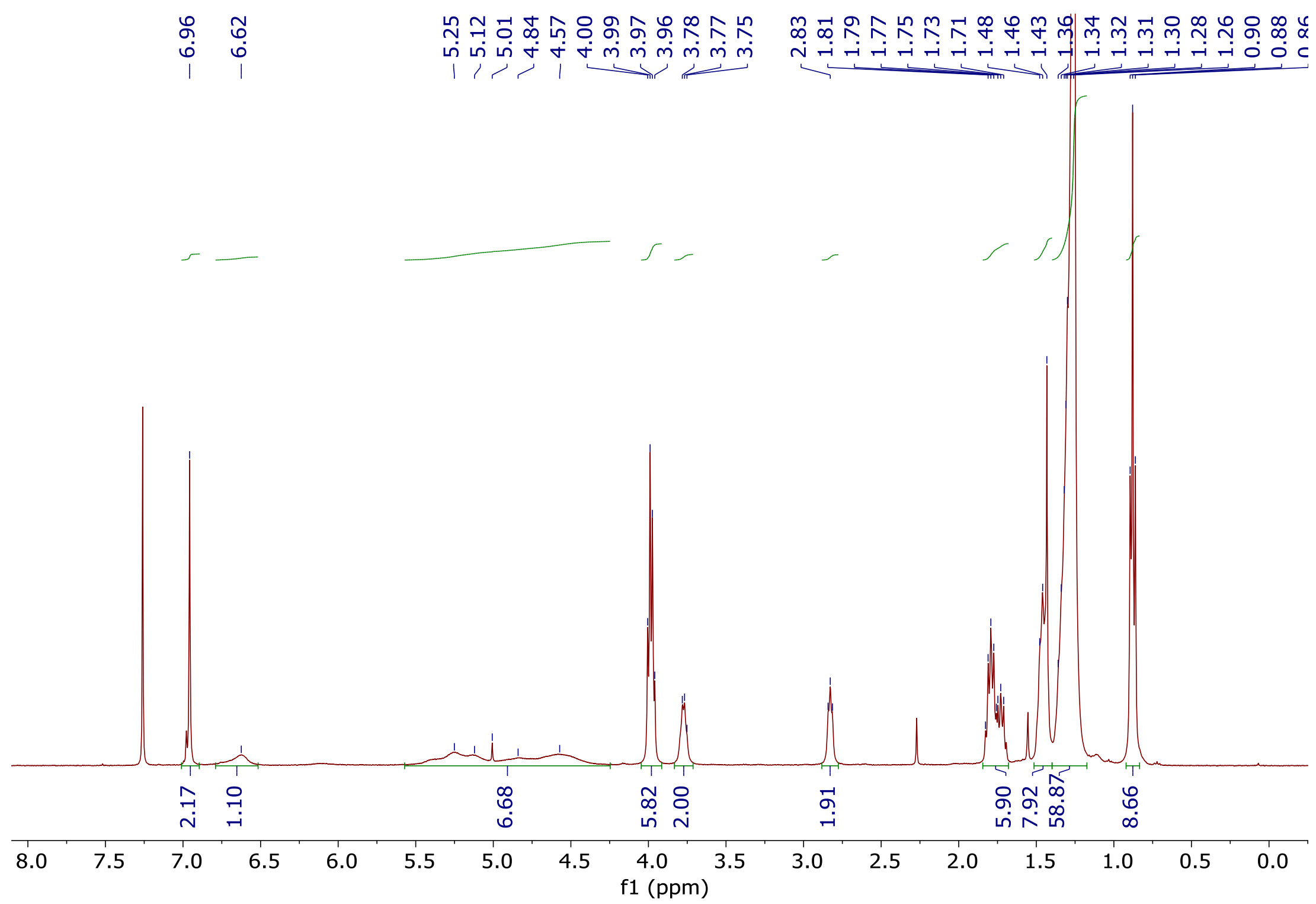

**Fig. SI.** $^{1}H$ NMR (400 MHz, $CDCl_3$, 298 K) of **FCH-C2-A**

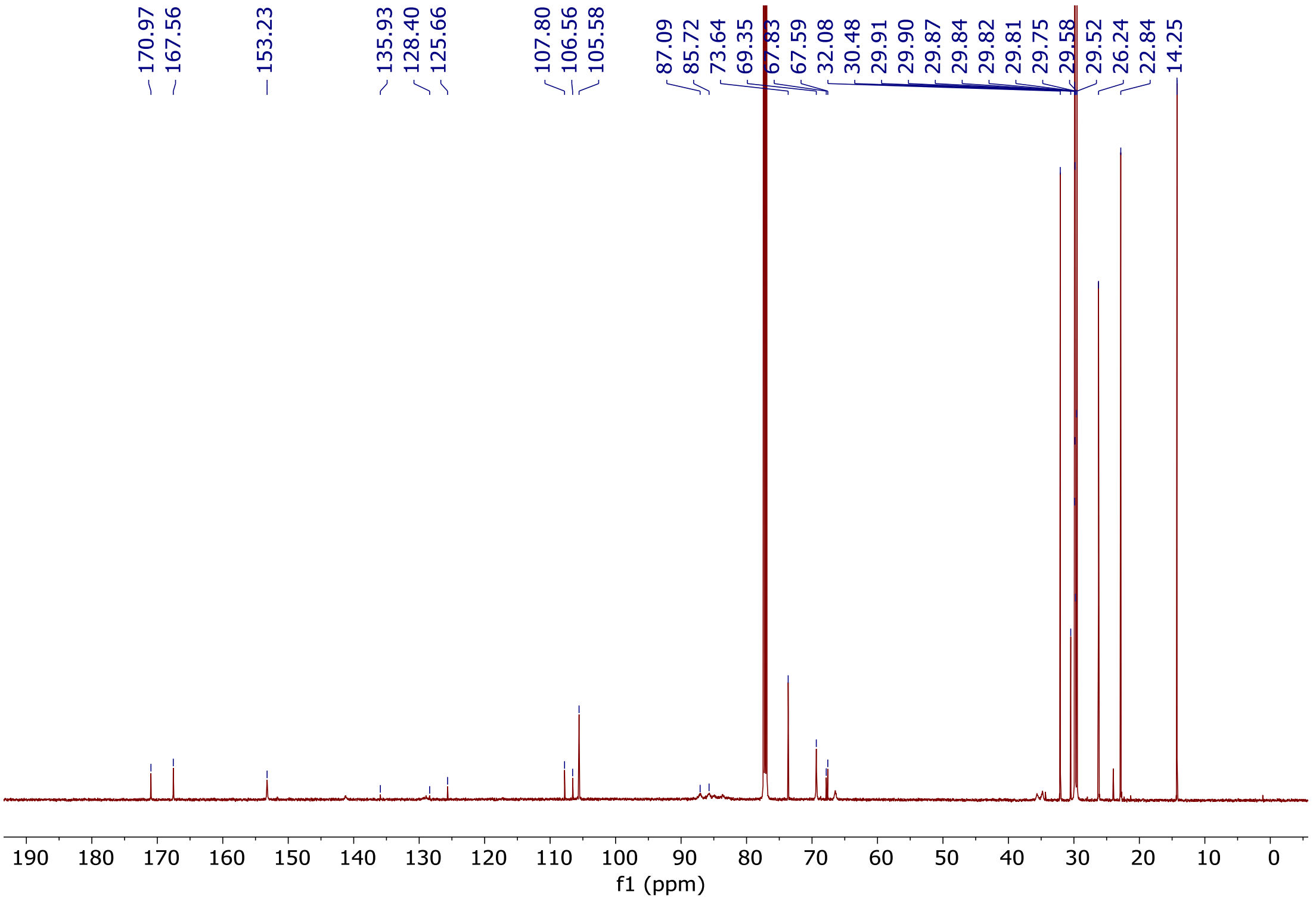

**Fig. SII.** $^{13}C$ NMR (151 MHz, $CDCl_3$, 298 K) of **FCH-C2-A**

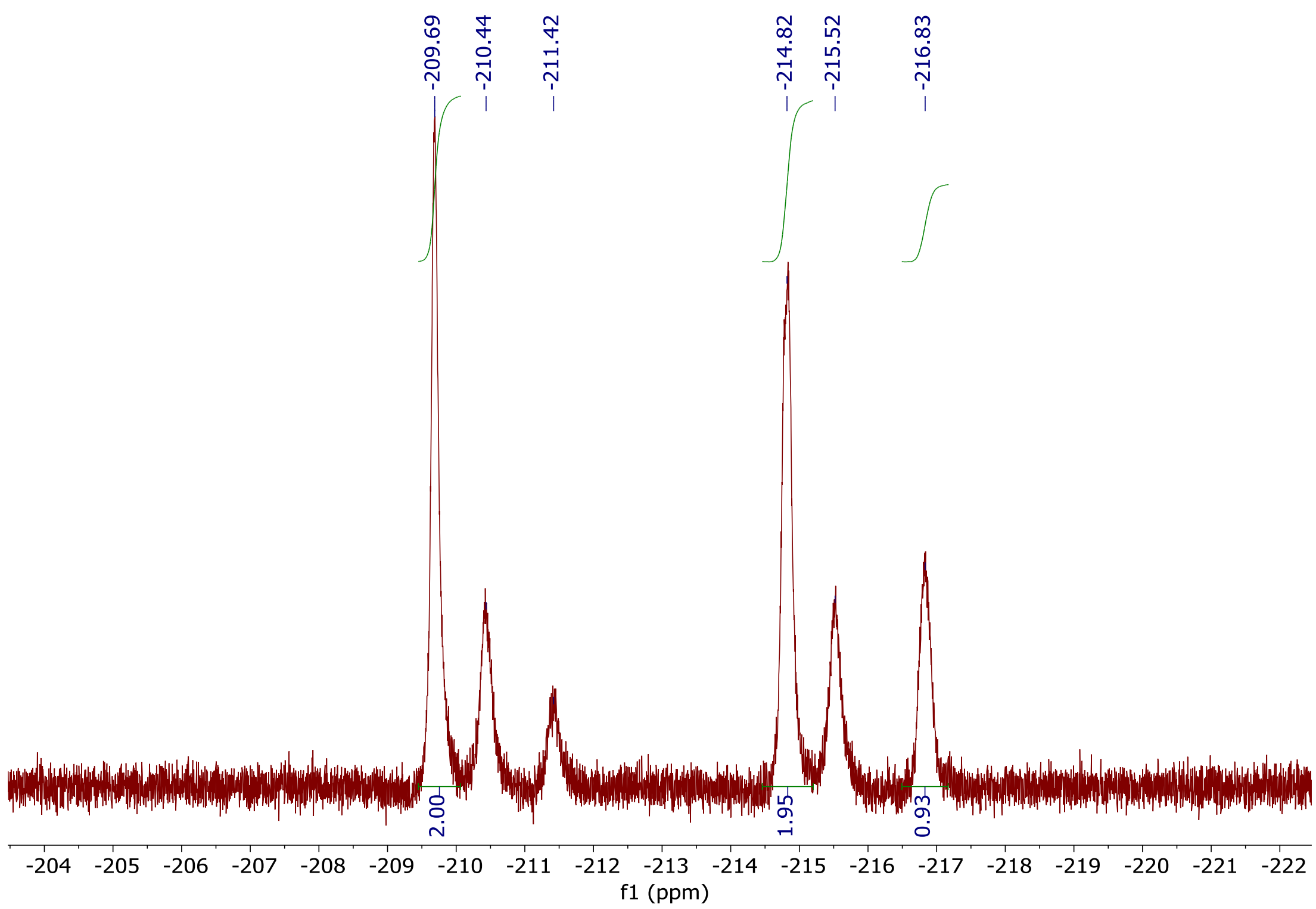

**Fig. SIII.** $^{19}$F NMR (376 MHz, $CDCl_3$, 298 K) of **FCH-C2-A**

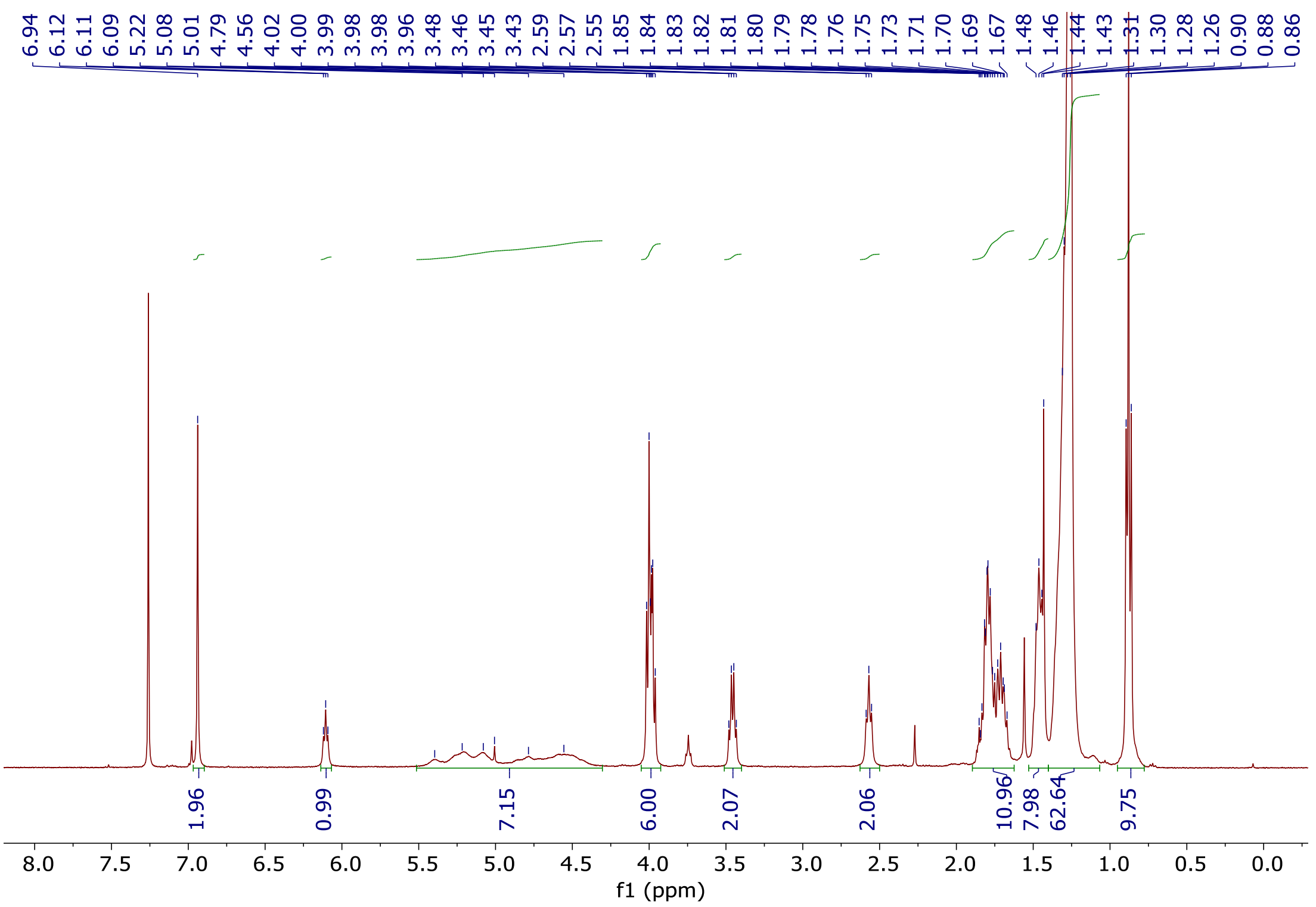

**Fig. SIV.** $^{1}$H NMR (400 MHz, $CDCl_3$, 298 K) of **FCH-C4-A**

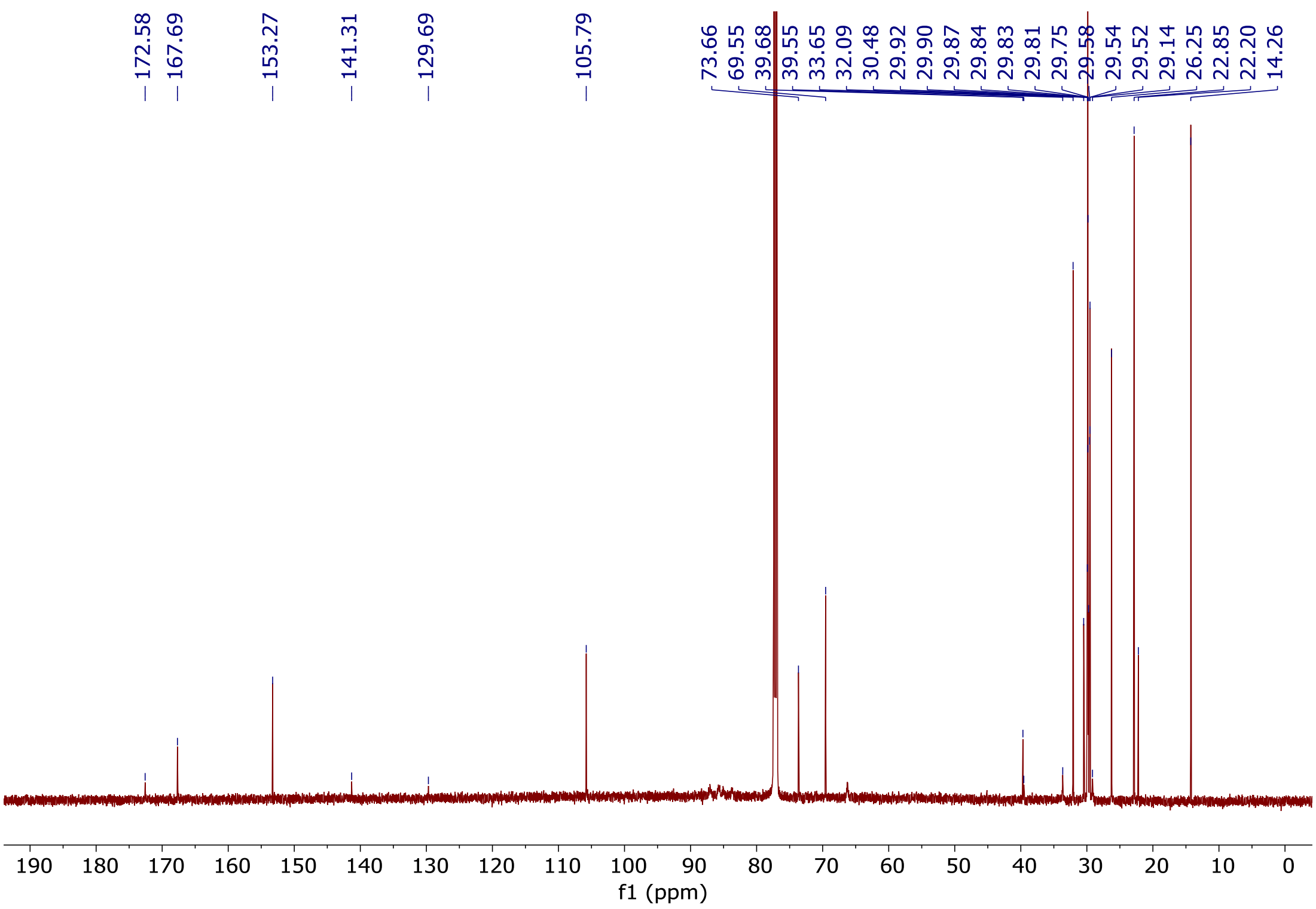

**Fig. SV.** $^{13}$C NMR (151 MHz, $CDCl_3$, 298 K) of **FCH-C4-A**

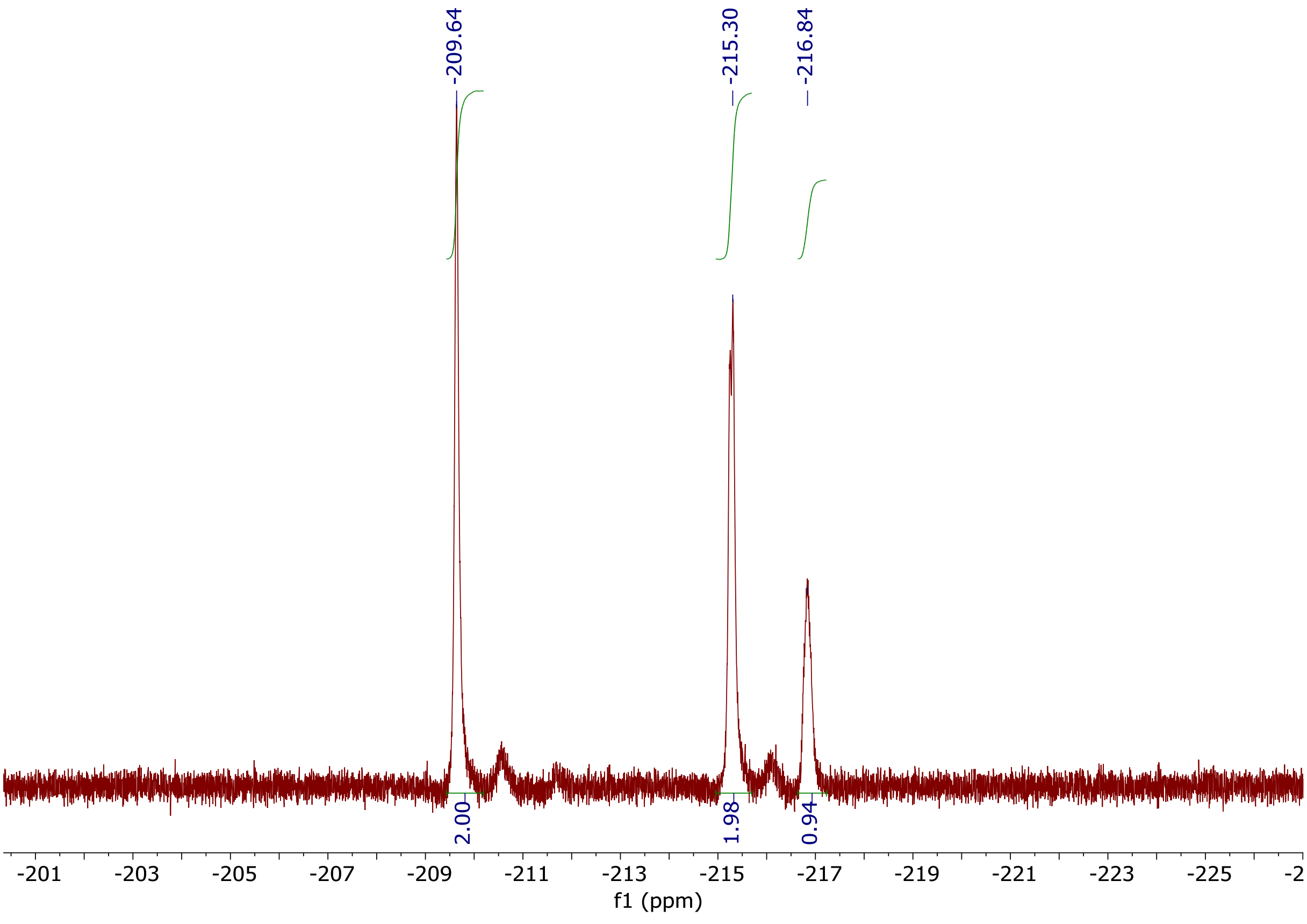

**Fig. SVI.** $^{19}$F NMR (376 MHz, $CDCl_3$, 298 K) of **FCH-C4-A**

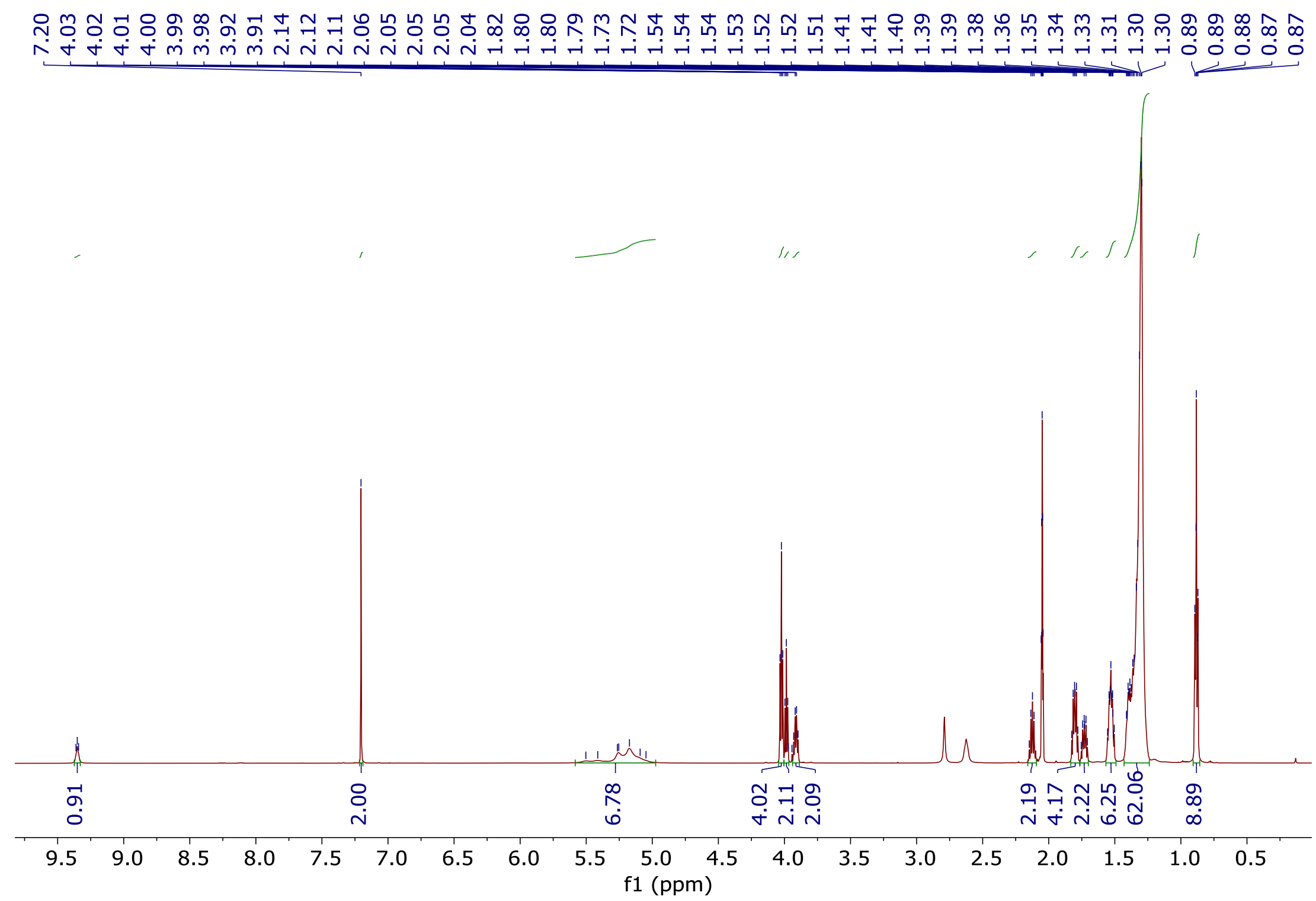

**Fig. SVII.** $^{1}H$ NMR (600 MHz, acetone-*d*$_6$, 298 K) of **FCH-C3-TA**

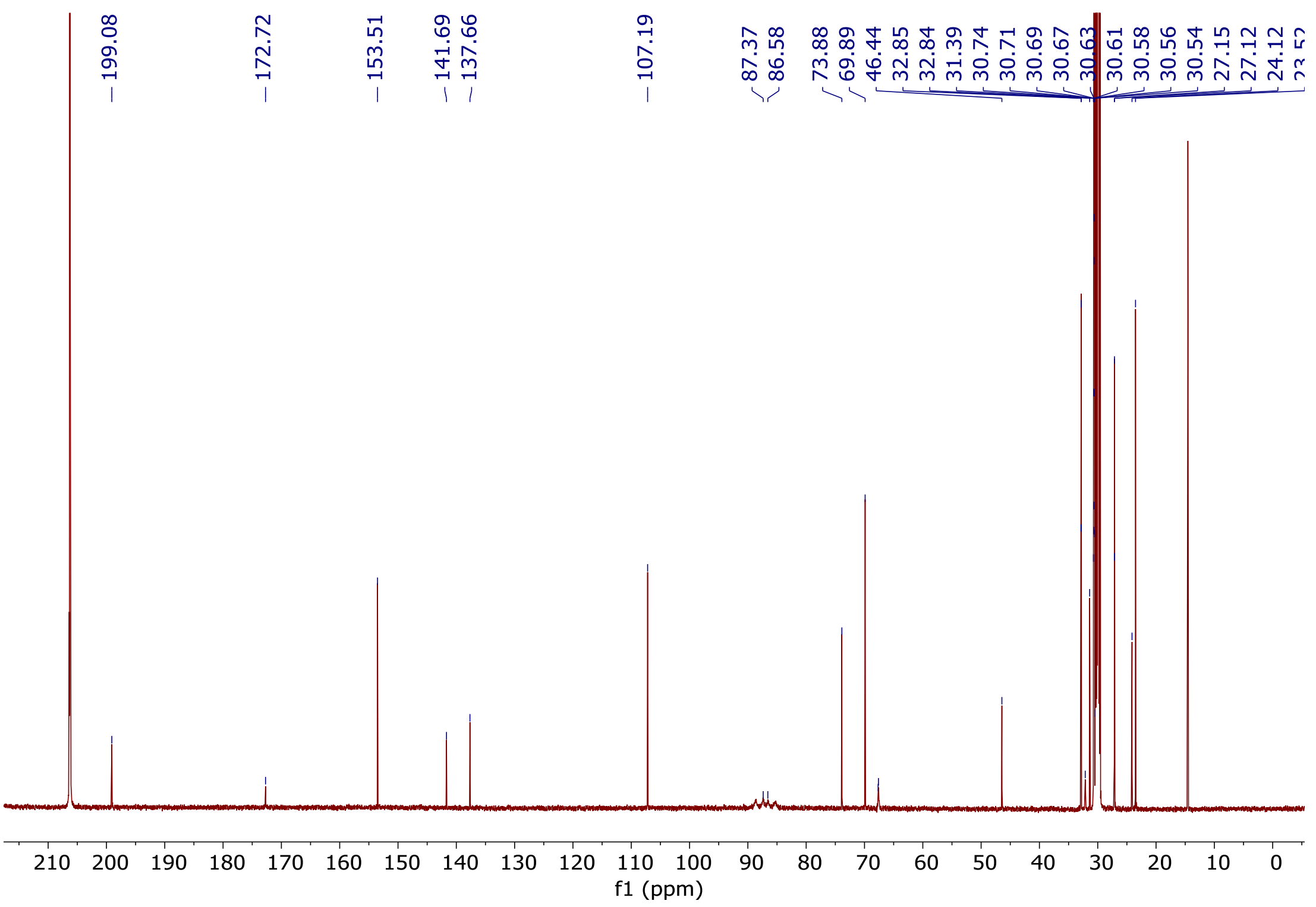

**Fig. SVIII.** $^{13}C$ NMR (151 MHz, acetone-*d*$_6$, 298 K) of **FCH-C3-TA**

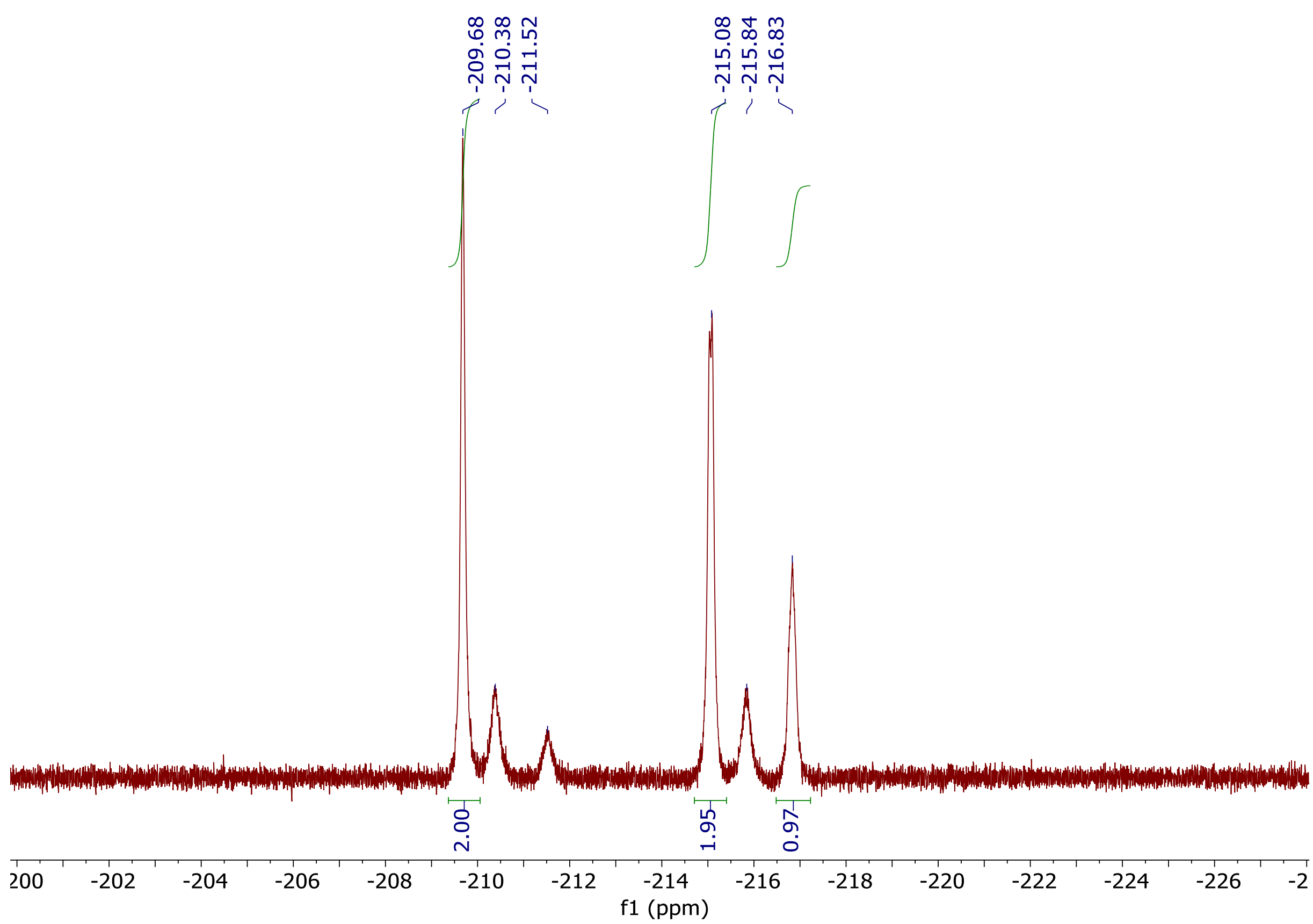

**Fig. SIX.** $^{19}$F NMR (376 MHz, acetone-$d_6$, 298 K) of **FCH-C3-TA**

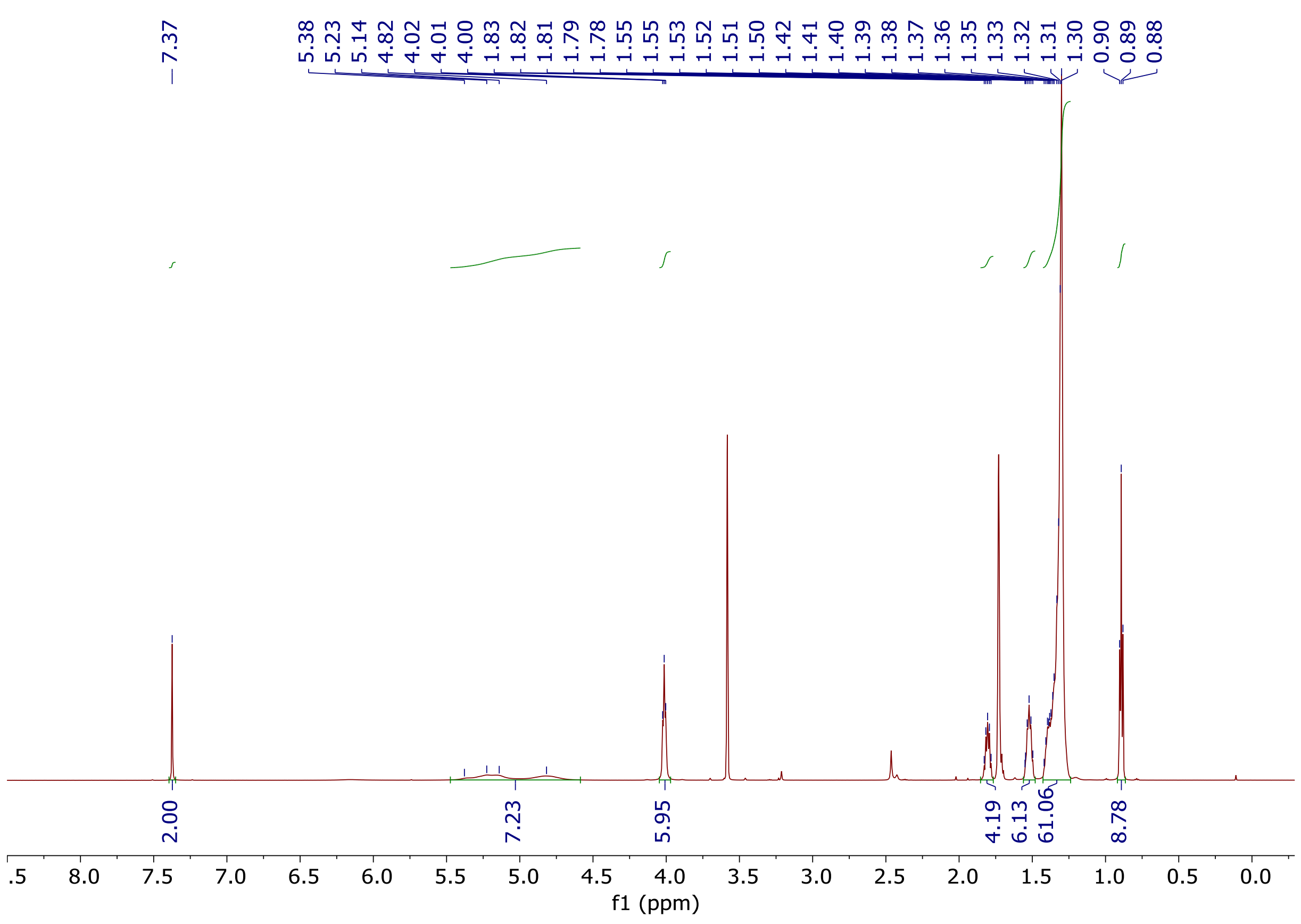

**Fig. SX.** $^{1}$H NMR (600 MHz, THF-$d_8$, 298 K) of **FCH-E**

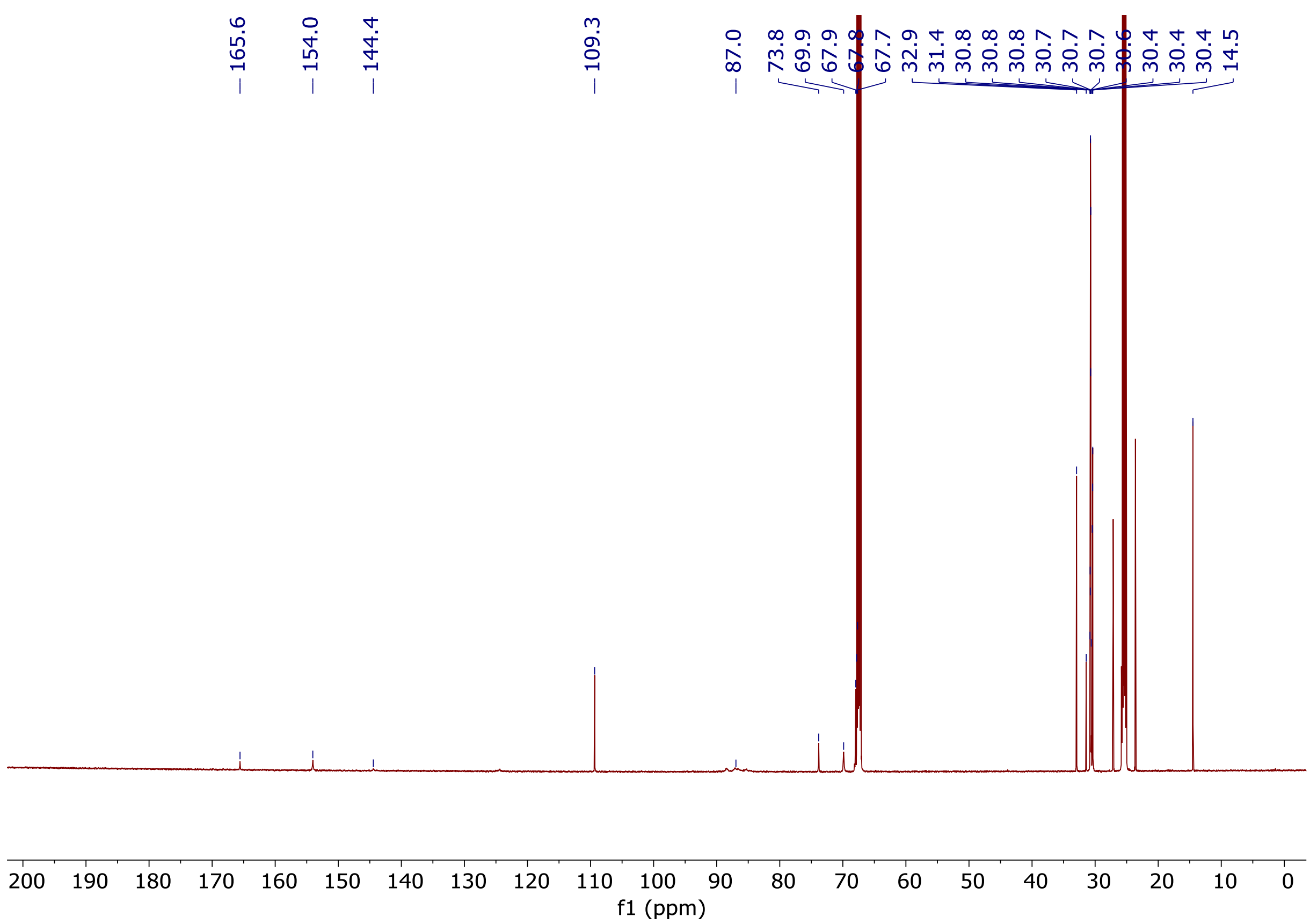

**Fig. SXI.** $^{13}C$ NMR (151 MHz, THF-$d_8$, 298 K) of **FCH-E**

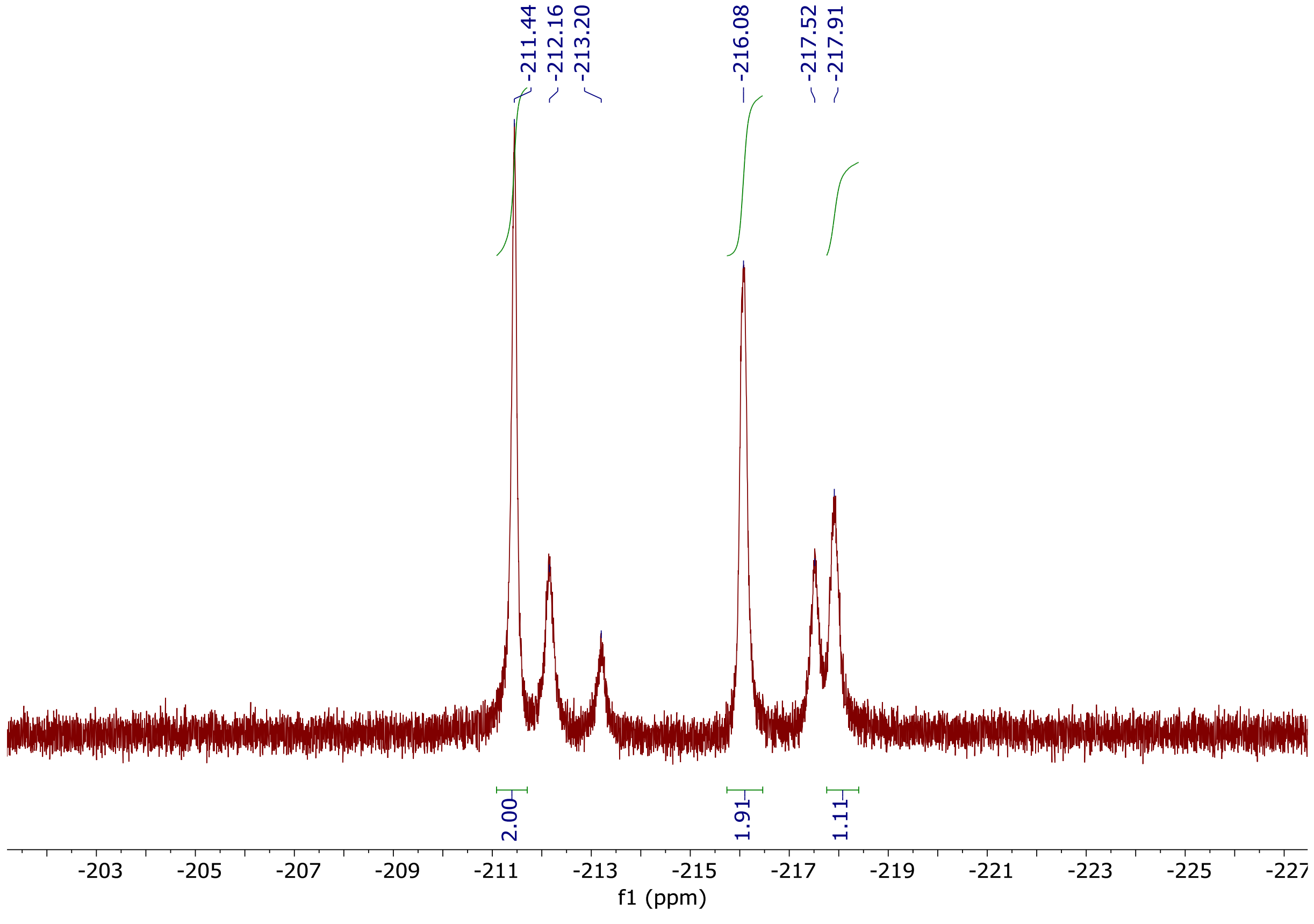

**Fig. SXII.** $^{19}F$ NMR (376 MHz, THF-$d_8$, 298 K) of **FCH-E**

### 1.2 Synthesis of molecular discotics:

*N*,*N'*,*N''*-Benzene-1,3,5-tricarboxamide **(BTA)** and *N*,*N'*,*N''*-Trialkylbenzene-1,3,5-tris(carbothioamide) **(BTTA)** were synthesized according to previously reports, respectively.[2,3]

*N1,N3,N5*-Tris(2-ethylhexyl)cyclohexane-1,3,5-tricarboxamide (**CTA**) was synthesized according to a previously reported procedure[4], starting from cis,cis-cyclohexane-1,3,5-tricarboxylic acid. Chiral (S)-3,7-dimethyloctylamine was obtained via a literature procedure.[5] *N1,N3,N5*-Tris((S)-3,7-dimethyloctyl)cyclohexane-1,3,5-tris(carbothioamide) (**CTTA**) was obtained from the corresponding trisamide by treatment with P2S5 using a previously reported procedure.[6]

### 1.3 Synthesis of TPA Trisamides:

*N*,*N*',*N*''-[Nitrilotri(2,1-phenylene)]tritridecanamide, *o*-TPA-N

*N*,*N*',*N*''-[Nitrilotri(4,1-phenylene)]tritridecanamide, *p*-TPA-N

4,4',4''-Nitrilotris(*N*-dodecylbenzamide), *p*-TPA-C

All solvents and reagents were purchased at reagent grade from commercial suppliers (Merck/Sigma-Aldrich, TCI, Thermo Fisher Scientific, Acros Organics, Honeywell, BLD Pharmatech) and used without additional purification. Thin layer chromatography was monitored on ALUGRAM aluminum plates from Macherey-Nagel, coated with 0.20 mm $SiO_2$, by irradiation with UV-light ($\lambda$=365 and 254nm). Flash column chromatography was carried out with $SiO_2$ from Macherey-Nagel (technical grade 60 M, pore size 60 Å, 40−63 µm particle size).

*Nuclear Magnetic Resonance* spectra were recorded at room temperature (295 K) on a Bruker Avance III 300, 400, 500, 600 or 700 at the Institute of Organic Chemistry (Heidelberg University). Proton broad band decoupling was applied for $^{13}C$ measurements. Deuterated solvents were used as purchased from Merck/Sigma-Aldrich and Deutero. Chemical shifts (reported in parts per million (ppm)) were referenced[7] to and 5.32 ppm ($CD_2Cl_2$) for $^{1}H$ and $\delta_C$ = 77.16 ppm ($CDCl_3$) and 53.84 ppm ($CD_2Cl_2$) for $^{13}C$ and interpreted with MestReNova Version 14.1.2-25024. Apparent signal multiplicity is reported as s (singlet), d (doublet), dd (doublet of doublets), t (triplet), or m (multiplet).

*Infrared Spectroscopy:* A JASCO FT/IR-6000 FT-IR spectrometer was operated in ATR mode to record infrared spectra. The respective transmission spectra are baseline corrected, the respective peaks are stated in $cm^{-1}$ and labeled according to the following abbreviations: s (strong), m (medium), w (weak), and br (broad).

*Melting Point:* The melting point was determined on a Büchi M-560 melting point apparatus in open capillaries.

*Mass Analysis:* Mass spectra were received from the facility of Heidelberg University and recorded on a Bruker timsTOFfleX (matrix assisted laser desorption ionization (MALDI)) instrument.

The compounds **S4**[8], **S5**[9,10], **S6**[11,12], **S7**[8,9], **S8**[9], **S9**[13] and **S10**[13] were synthesized according to literature procedures.

$K_2CO_3$, DMSO, 120 °C, 6 d (59%) → **S4**; Pd/C (10w%), $NH_2NH_2 \cdot H_2O$, EtOH, 60 °C, 2.5 h (71%) → **S5**; **S6** ($ClC(O)C_{12}H_{25}$), $NEt_3$, $CH_2Cl_2$, 0 °C to rt, 12 h (40%) → ***o*-TPA-N**

$K_2CO_3$, DMSO, 120 °C, 3 d (96%) → **S7**; Pd/C (10w%), $NH_2NH_2 \cdot H_2O$, EtOH, 60 °C, 4.5 h (73%) → **S8**; **S6**, $NEt_3$, $CH_2Cl_2$, 0 °C to rt, 17 h (73%) → ***p*-TPA-N**

R = $NHC(O)C_{12}H_{25}$

CsF, DMF, 120–140 °C, 3 d (50%) → **S9**; KOH, $H_2O$, EtOH, 120 °C, 10 d (75%) → **S10**; 1) oxalyl chloride, DMF, THF, 0 °C to rt, 3 h; 2) dodecan-1-amine, $NEt_3$, $CH_2Cl_2$, 0 °C to rt, 16 h (65%) → ***p*-TPA-C**

R = $C(O)NHC_{12}H_{25}$

**Scheme SII.** Synthetic route towards TPA-based trisamides ***o*-TPA-N**, ***p*-TPA-N** and ***p*-TPA-C**.

**_N_,_N_',_N_''-[Nitrilotri(2,1-phenylene)]tritridecanamide (_o_-TPA-N)**

$C_{57}H_{90}N_4O_3$
879.37 g/mol

Reaction conditions were adapted from literature procedure to the synthesis of ***o*-TPA-N**.[14] A solution of tridecanoyl chloride (**S6**) (1.19 g, 1.30 mL, 5.11 mmol) was prepared in dry $CH_2Cl_2$ (10 mL) under $N_2$ atmosphere and cooled to 0 °C. *N*1,*N*1-Bis(2-aminophenyl)benzene-1,2-diamine (**S5**) (450 mg, 1.55 mmol) was then suspended in dry $CH_2Cl_2$ (25 mL) under light exclusion and $N_2$ atmosphere and added dropwise to the acyl chloride. After the addition of $NEt_3$ (518 mg, 709 mL, 5.11 mmol), the solution was stirred at rt for 12 h. The reaction was treated with water (20 mL) and the organic phase was washed with sat. aq. $NaHCO_3$ (20 mL) and sat. aq. NaCl (20 mL). Purification by column chromatography ($SiO_2$ deactivated with $NEt_3$, PE/$CH_2Cl_2$ 1:1, PE/$CH_2Cl_2$/MeOH 20:20:1) and washing with MeOH afforded compound ***o*-TPA-N** (540 mg, 614 mmol, 40%) as colorless solid.

M.p. = 89 °C.

$R_f$ = 0.56 ($SiO_2$, PE/$CH_2Cl_2$/MeOH 20:20:1).

$^1$H NMR (700 MHz, $CD_2Cl_2$): $\delta$ 8.27–8.14 (br. m, 3H), 7.73 (s, 3H), 7.12–7.09 (m, 3H), 7.04–7.02 (m, 3H), 6.77 (d, $J$ = 7.6 Hz, 3H), 1.99–1.95 (m, 6H), 1.36–1.13 (m, 60H), 0.88 (t, $J$ = 7.0 Hz, 9H) ppm.

$^{13}$C NMR (176 MHz, $CD_2Cl_2$): $\delta$ 172.0, 138.6, 132.2, 126.1, 125.3, 124.8, 37.2, 32.4, 30.14, 30.11, 29.9, 29.80, 29.78, 29.6, 25.7, 23.1, 14.3 ppm. Two signals coincident or not observed.

IR (FT-ATR): $\tilde{\nu}$ 3338 (w), 3230 (w), 3123 (w), 3046 (w), 2917 (s), 2849 (s), 1696 (m), 1655 (m), 1648 (m), 1593 (m), 1523 (s), 1449 (s), 1297 (m), 1266 (m), 754 (s) $cm^{-1}$.

MALDI HR-MS (DCTB): *m/z* calcd for $C_{57}H_{90}N_4O_3$ $[M]^+$ 878.7007, found 878.7021, *m/z* calcd for $C_{57}H_{90}N_4NaO_3$ $[M+Na]^+$ 901.6905, found 901.6913, *m/z* calcd for $C_{57}H_{90}KN_4O_3$ $[M+K]^+$ 917.6645, found 917.6659, *m/z* calcd for $C_{114}H_{180}N_8NaO_6$ $[2M+Na]^+$ 1780.3918, found 1780.3956.

Elemental analysis calcd (%) for $C_{57}H_{90}N_4O_3$: C 77.85, H 10.32, N 6.37, O 5.46; found: C 77.94, H 10.19, N 6.61.

***N*,*N*',*N*''-[Nitrilotri(4,1-phenylene)]tritridecanamide (*p*-TPA-N)**

$C_{57}H_{90}N_4O_3$
879.37 g/mol

Reaction conditions were adapted from literature procedure to the synthesis of ***p*-TPA-N**.[14] Tridecanoyl chloride (1.06 g, 1.16 mL, 4.55 mmol) was dissolved in dry $CH_2Cl_2$ (24 mL) under $N_2$ atmosphere at 0 °C. Small portions of $N^1$,$N^1$-bis(4-aminophenyl)benzene-1,4-diamine (**S8**) (400 mg, 1.38 mmol) and $NEt_3$ (460 mg, 630 mL, 4.55 mmol) were added to the reaction mixture over a period of 10 min. The solution was allowed to warm to rt and stirred for 17 h. Water (100 mL) was added and the suspension was sonicated for 10 min. The solid was then filtered and washed with $CH_2Cl_2$ (200 mL) to afford compound ***p*-TPA-N** (884 mg, 1.01 mmol, 73%) as a colorless solid.

M.p. = 225 °C.

$R_f$ = 0.53 ($SiO_2$, $CH_2Cl_2$/MeOH 20:1).

$^1$H NMR (700 MHz, THF-$d_8$): $\delta$ 8.83 (s, 3H), 7.47 (d, *J* = 8.4 Hz, 6H), 6.89 (d, *J* = 8.4 Hz, 6H), 2.24 (t, *J* = 7.4 Hz, 6H), 1.65 (t, *J* = 7.3 Hz, 6H), 1.33–1.29 (m, 54H), 0.88 (t, *J* = 6.8 Hz, 9H) ppm.

$^{13}$C NMR (176 MHz, THF-$d_8$): $\delta$ 170.9, 144.2, 135.8, 124.6, 120.7, 37.7, 32.9, 30.63, 30.60, 30.55, 30.49, 30.3, 26.5, 25.8, 25.6, 23.6, 14.4 ppm.

IR (FT-ATR): $\tilde{\nu}$ 3290 (m), 3188 (w), 3044 (w), 2917 (s), 2850 (s), 1654 (s), 1651 (s), 1599 (m), 1505 (s), 1465 (s), 1405 (m), 1315 (m), 1270 (m), 1244 (m), 817 (m), 720 (m) cm$^{-1}$.

MALDI HR-MS (DCTB): *m/z* calcd for $C_{57}H_{90}N_4O_3$ $[M]^+$ 878.7007, found 878.7019, *m/z* calcd for $C_{57}H_{90}N_4NaO_3$ $[M+Na]^+$ 901.6905, found 901.6925, *m/z* calcd for $C_{57}H_{90}KN_4O_3$ $[M+K]^+$ 917.6645, found 917.6663, *m/z* calcd for $C_{114}H_{180}N_8O_6$ $[2M]^+$ 1757.4020, found 1757.4081, *m/z* calcd for $C_{114}H_{181}N_8O_6$ $[2M+H]^+$ 1758.4099, found 1758.4122.

Elemental analysis calcd (%) for $C_{57}H_{90}N_4O_3$: C 77.85, H 10.32, N 6.37, O 5.46; found: C 77.51, H 10.60, N 6.49.

This data is in accordance with literature.[15]

**4,4',4''-Nitrilotris(*N*-dodecylbenzamide) (*p*-TPA-C)**

$C_{57}H_{90}N_4O_3$
879.37 g/mol

Compound ***p*-TPA-C** was synthesized according to a modified literature procedure.[16] Compound **S10** (100 mg, 265 mmol) was dissolved in dry THF (3.6 mL) under $N_2$ atmosphere. A solution of DMF (775 mg, 10.6 mmol) in dry THF (800 mL) was added and the mixture was cooled to 0 °C. Oxalyl chloride (89.7 mL, 135 mg, 1.06 mmol) was added dropwise, and the acyl chloride was stirred at rt for 2.5 h. The solvent was removed under reduced pressure, before the acyl chloride was dissolved in dry $CH_2Cl_2$ (2.8 mL). A suspension of dodecyl amine (162 mg, 875 mmol) and $NEt_3$ (129 mL, 93.9 mg, 928 mmol) in dry $CH_2Cl_2$ (2.0 mL) was added at 0 °C, before the reaction mixture was stirred at rt for 16 h. Subsequently, the precipitate was filtered, washed with water (1 mL), diluted aq. HCl (0.2 M, 0.5 mL), water (10 mL), EtOH (5 mL), PE (2 mL), and purified by column chromatography ($SiO_2$, toluene/MeOH 9:1). Compound ***p*-TPA-C** (152 mg, 172 mmol, 65%) was isolated as a colorless solid.

M.p. = 201-202 °C.

$R_f$ = 0.42 ($SiO_2$, toluene/MeOH 9:1).

$^1$H NMR (600 MHz, THF-$d_8$): $\delta$ 7.75 (d, *J* = 8.7 Hz, 6H), 7.50 (t, *J* = 5.7 Hz, 3H), 7.08–7.06 (m, 6H), 3.33 (q, *J* = 6.7 Hz, 6H), 1.58–1.54 (m, 6H), 1.39–1.28 (m, 54H), 0.88 (t, *J* = 7.0 Hz, 9H) ppm.

$^{13}$C NMR (151 MHz, THF-$d_8$): $\delta$ 166.1, 150.1, 131.4, 129.4, 124.3, 40.5, 32.9, 30.9, 30.64, 30.62, 30.60, 30.4, 30.3, 28.0, 23.6, 14.4 ppm. *One signal coincident or not observed.*

IR (FT-ATR): $\tilde{\nu}$ 3293 (br., m), 2920 (s), 2850 (s), 1629 (s), 1599 (s), 1538 (s), 1496 (s), 1466 (m), 1310 (s), 1276 (s), 1179 (m), 840 (m), 766 (m), 688 (m) cm$^{-1}$.

MALDI HR-MS (DCTB): *m/z* calcd for $C_{57}H_{90}N_4O_3$ $[M]^+$ 878.7007, found 878.7006.

Elemental analysis calcd (%) for $C_{57}H_{90}N_4O_3$: C 77.85, H 10.32, N 6.37, O 5.46; found: C 77.57, H 8.77, N 6.44.

This data is in accordance with literature.[16]

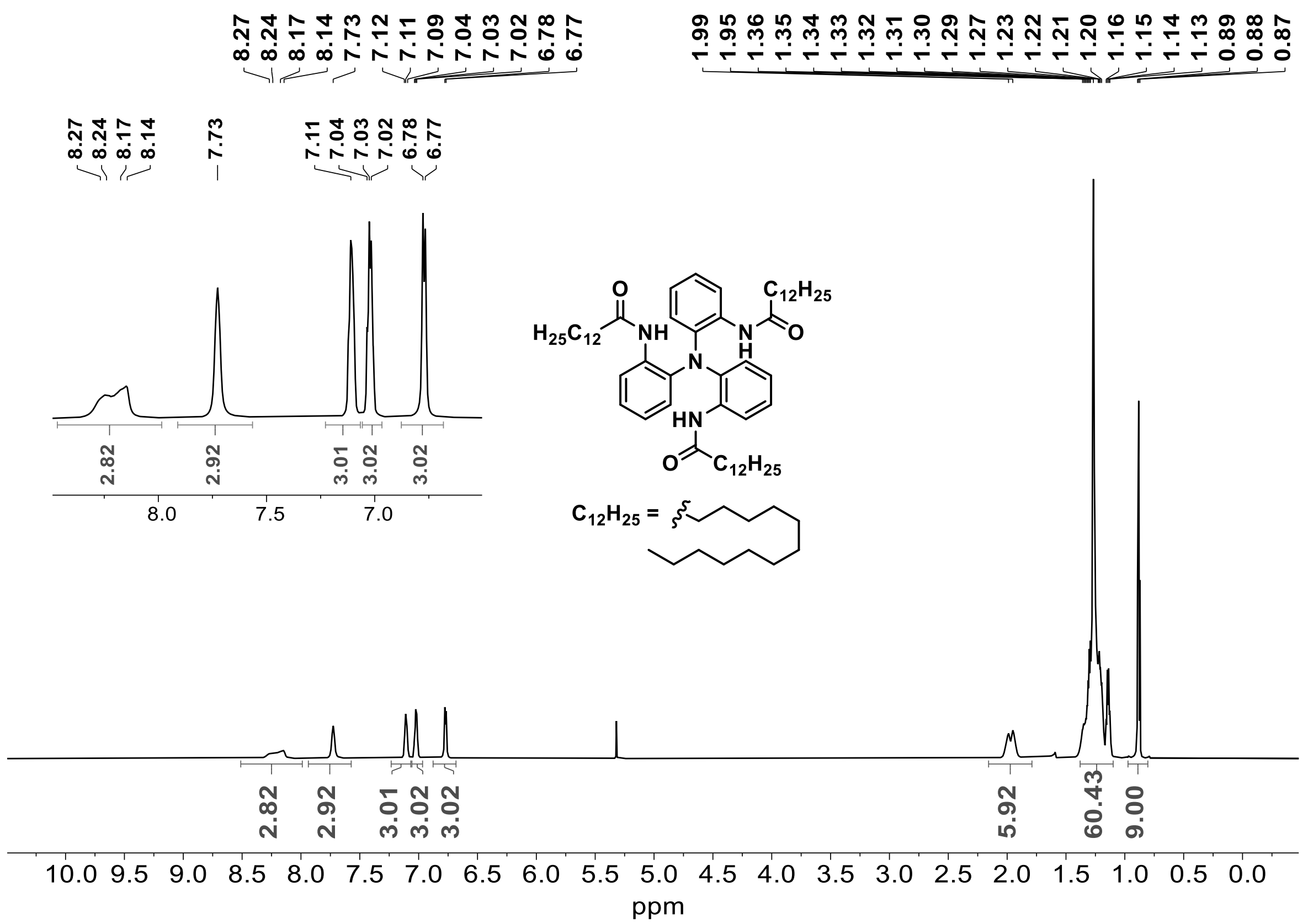

**Fig. SXIII.** $^{1}H$ NMR spectrum of compound ***o*-TPA-N** (700 MHz, $CD_2Cl_2$).

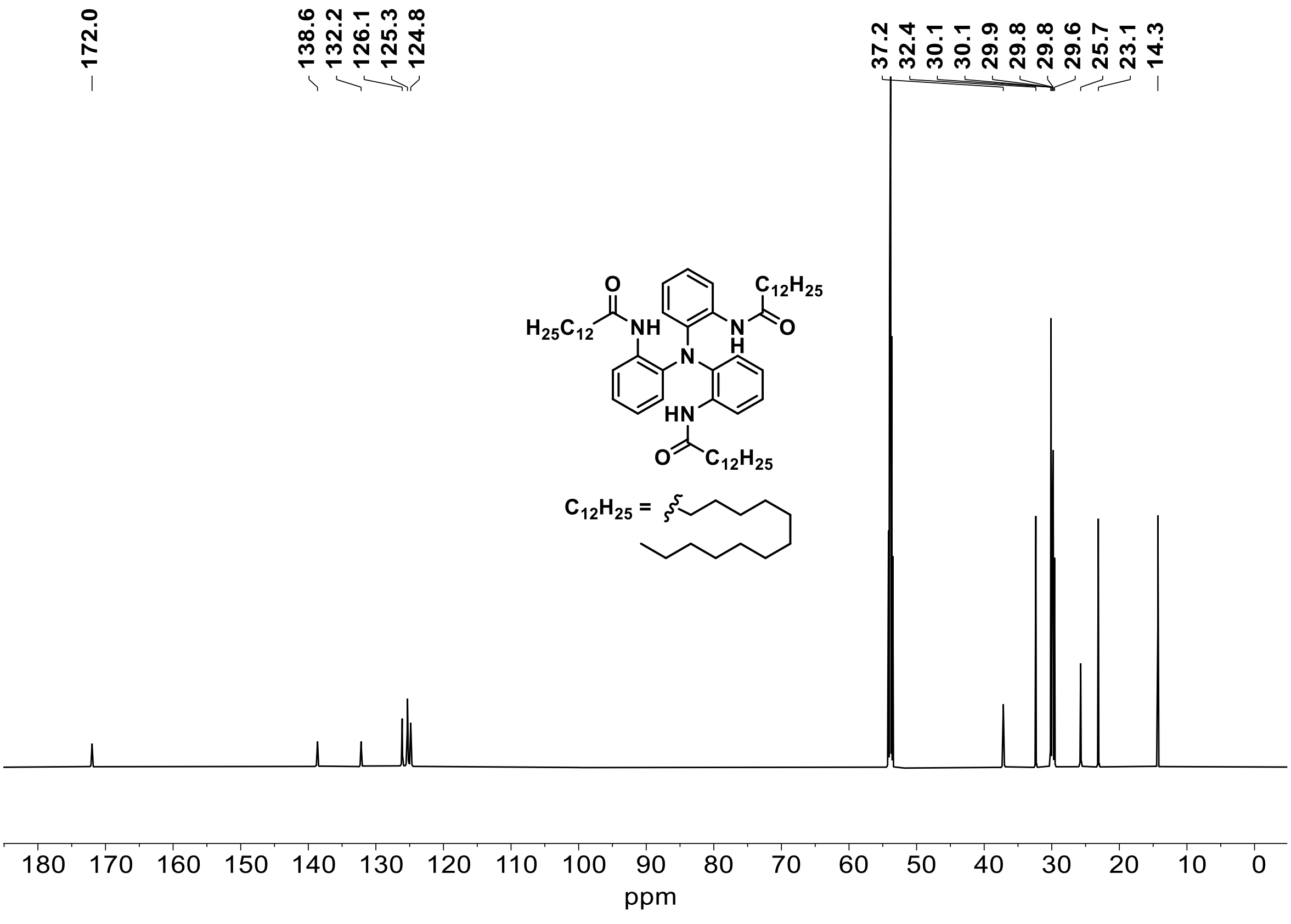

**Fig. SXIV.** $^{13}C$ NMR spectrum of compound ***o*-TPA-N** (176 MHz, $CD_2Cl_2$).

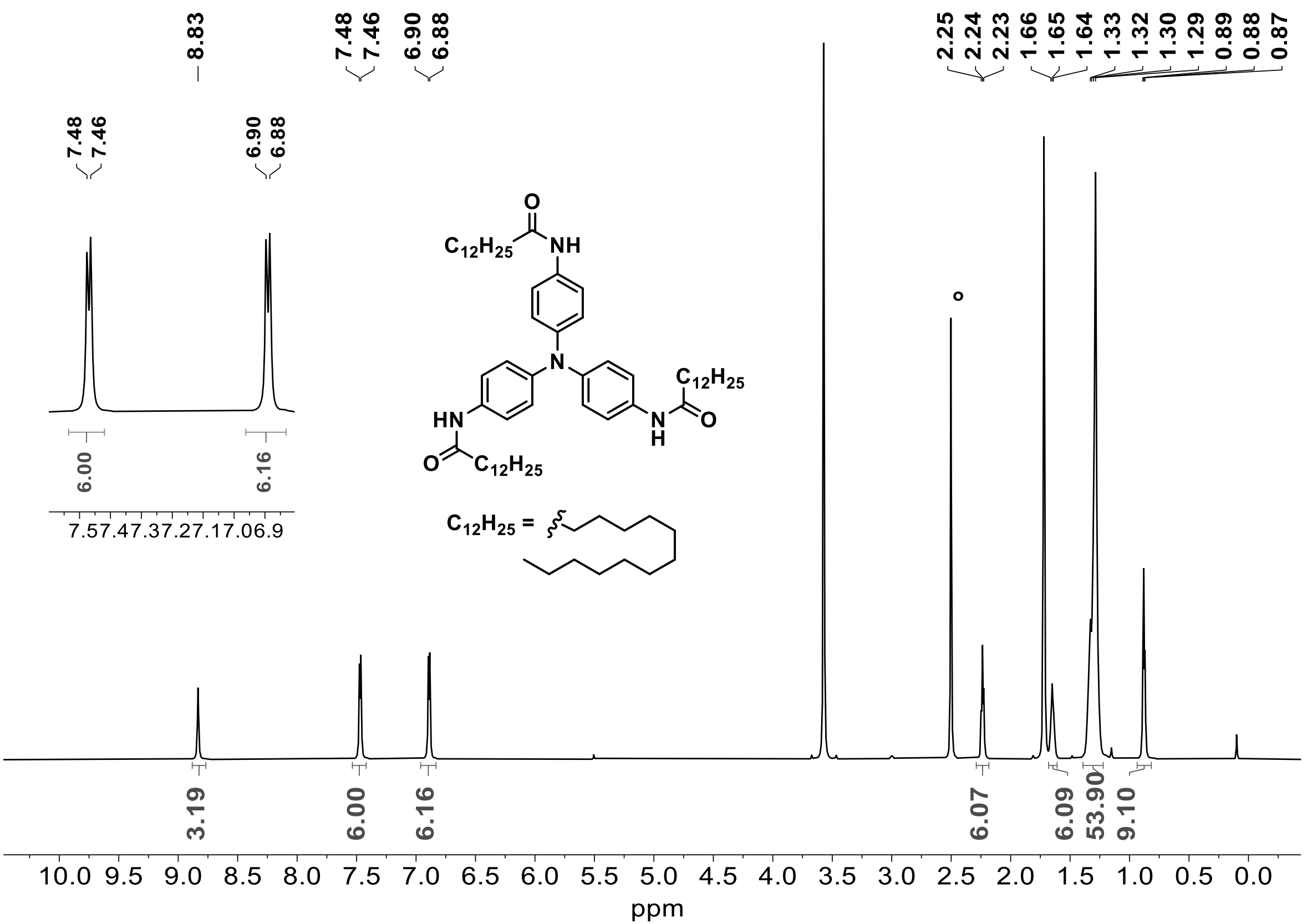

**Fig. SXV.** $^{1}$H NMR spectrum of compound ***p*-TPA-N** (700 MHz, $C_4D_8O$); °water.

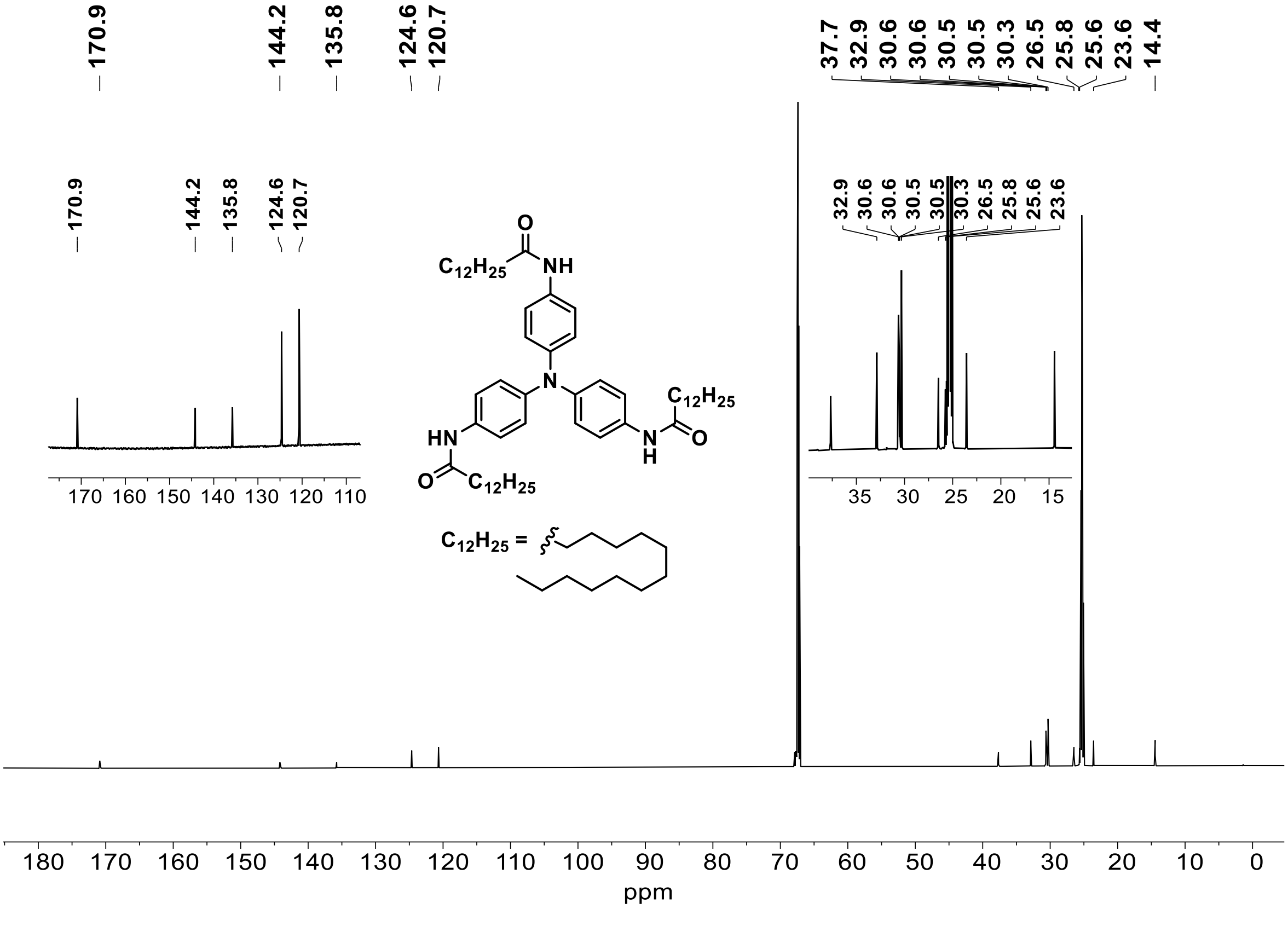

**Fig. SXVI.** $^{13}$C NMR spectrum of compound ***p*-TPA-N** (176 MHz, $C_4D_8O$).

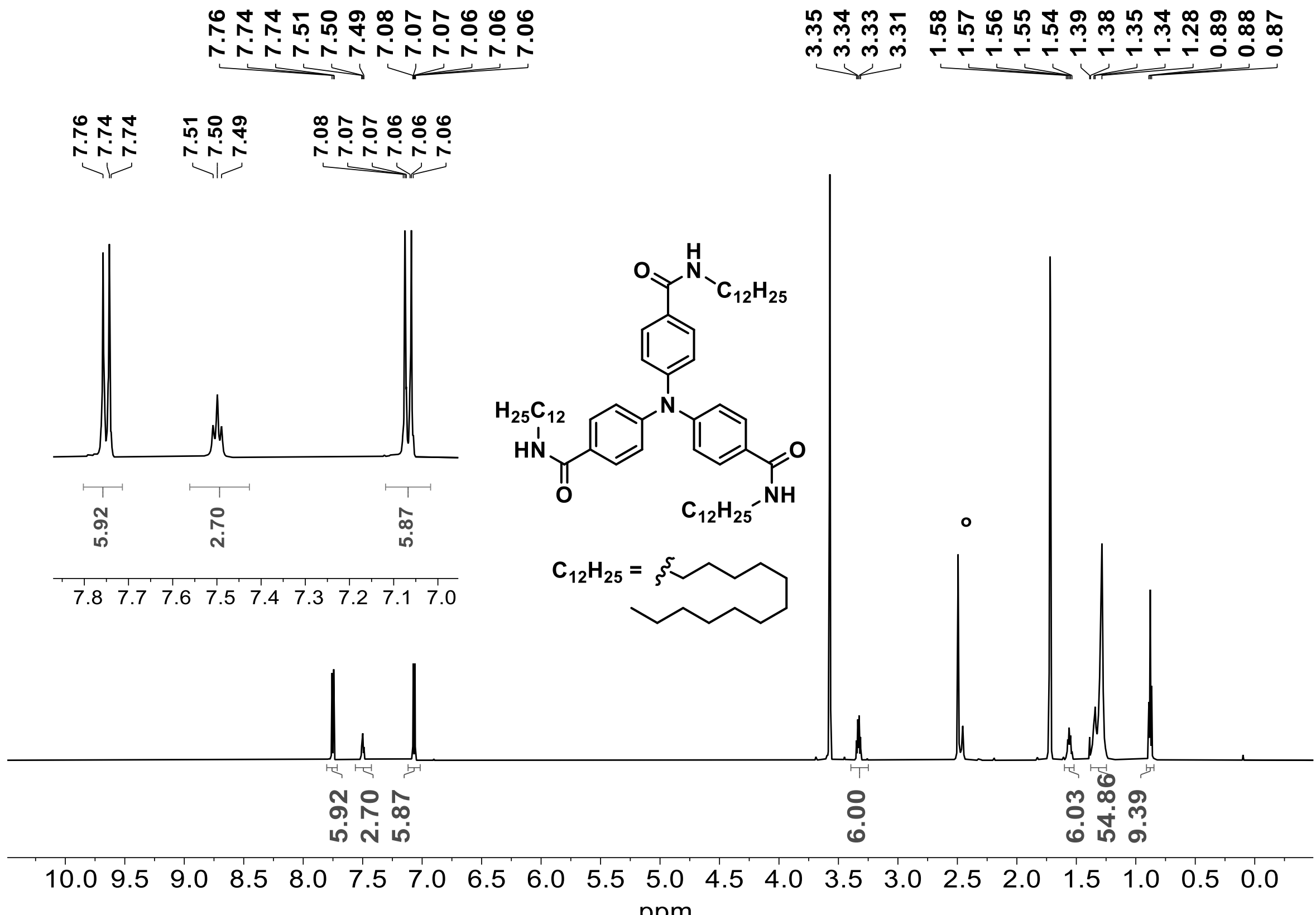

**Fig. SXVII.** $^1$H NMR spectrum of compound ***p*-TPA-C** (600 MHz, $C_4D_8O$); °water.

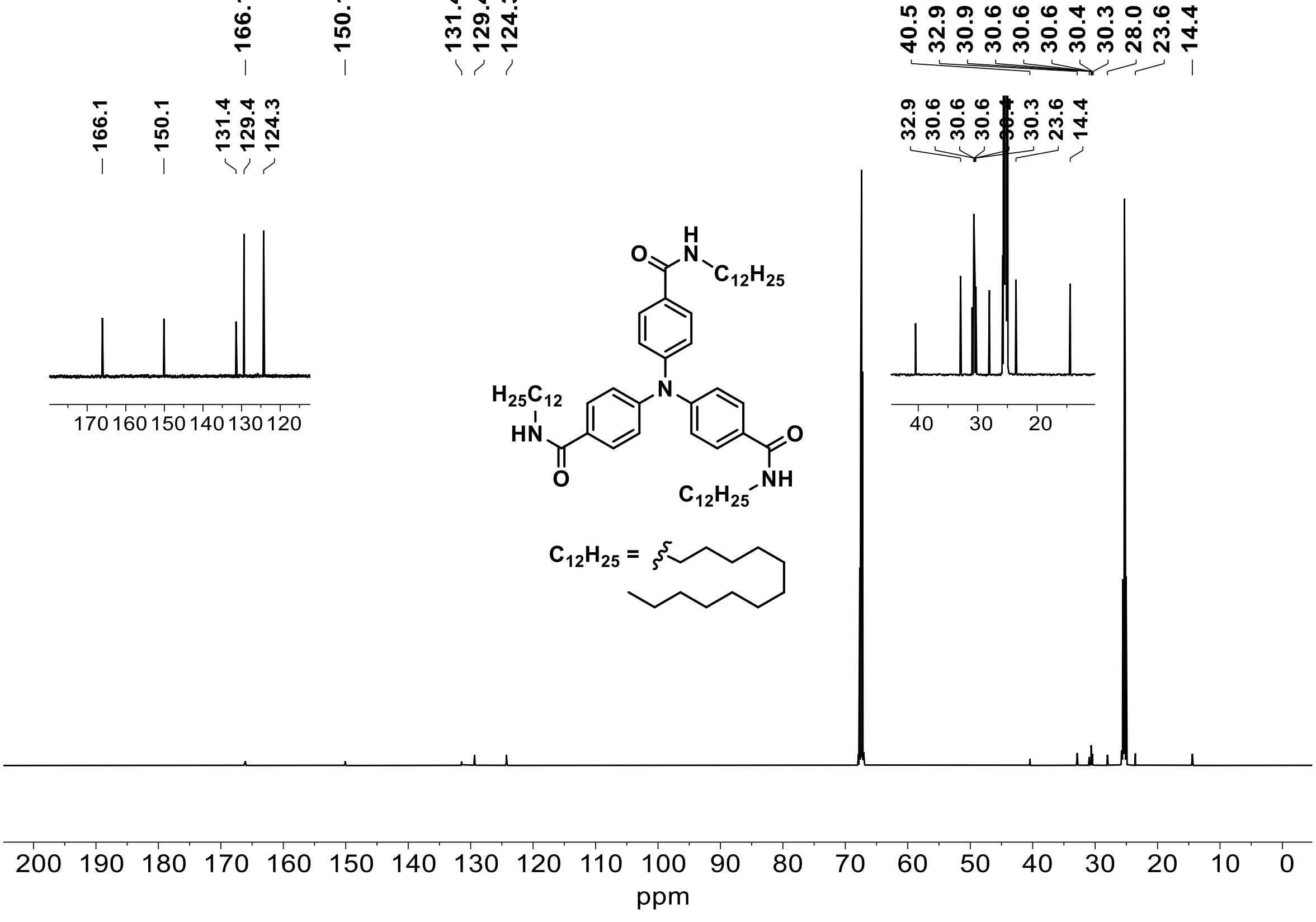

**Fig. SXVIII.** $^{13}$C NMR spectrum of compound ***p*-TPA-C** (151 MHz, $C_4D_8O$).

## 2 – Sample fabrication and methods

*Substrate preparation and characterization:* The thin film characterization was done on interdigitated electrodes (IDEs) which were either supplied by MicruX Technologies IDE or made by ourselves. For the former, the 180 pairs electrode pairs consist of 150 nm gold on top of 50 nm titanium deposited on a glass substrate. The total channel width amounts to $\sim 0.8\mathrm{m}$. The latter IDEs are produced by a UV photolithography process inside a cleanroom and consist of a typically 25-30 nm gold on top of 3 nm chromium evaporated on a glass substrate. The total channel width amounts to $\sim 4\mathrm{m}$. In both cases, the electrodes were 5 µm wide with a gap of 5 µm between the electrodes.

Prior to active layer deposition, all substrates were cleaned the same way. The substrates were first mechanically washed with soap and water. Then, they were spaced out and put in succession into water, acetone and isopropanol for chemical cleaning. For each solution, the substrates were cleaned for 10 min at room temperature using an ultrasonic bath. Finally, the substrates were blown dry with nitrogen.

Despite the extensive cleaning, empty IDEs showed a finite background conductivity as shown in Fig. S5. Since background currents varied slightly between samples, measurement data in the main text are not corrected for background. The relevant currents discussed herein exceed the background by at least an order of magnitude.

*Film deposition:* For reasons of material conservation, all samples were drop-casted from solution. The **FCH-Cn-As**, **FCH-E** and **FCH-C3-TA** were additionally heated over the melting point and cooled back down before any measurement was taken. This was done to improved film homogeneity but was not necessary to observe the conductivity effects. The **FCH-Cn-As**, **FCH-E**, **FCH-C3-TA, CTA** and **CTTA** were dissolved in tetrahydrofuran (THF) with concentrations between 10 to 20 mg/mL and drop-casted on glass substrates patterned with IDEs. To ensure full electrode coverage, 2-5 times 4-5 µL were drop-casted as needed. **BTA-C10** and **BTTA-C12** were solved in pure chloroform. The 10 mg/mL solution was heated at 40 °C for 15 min for improved solubility before drop casting 2 times 5 µL on the electrodes. The samples were then heated at 70 °C for 15 min to produce more homogenous films. All **TPA**s were drop casted from a 10 mg/mL solution onto the electrodes. ***o*-TPA-N** was solved in an equal amount of THF an DMC. For both ***p*-TPA-N** and ***p*-TPA-C**, pure THF was used as the solvent. In all cases, the amount of deposited solution was 2 times 5 µL. No additional temperature treatment was used during the deposition process.

All samples had full electrode coverage with inhomogeneous films with film thicknesses typically ranging from 1 to 3 µm.

*Electrical characterization:* Both the $I(V)$ and $IV(t)$ measurements were acquired via a Keithley 2636B System SourceMeter. The samples were fixed on a sample plate and contacted with needles within a Janis vacuum probe station with a base pressure of $\sim 10^{-6}\mathrm{mbar}$. The temperature regulation was done via cooling with liquid nitrogen by applying a constant flow from a nitrogen dewar and adjusting heating of the sample plate via a LakeShore 336 temperature controller. All measurements were carried out under high vacuum conditions unless explicitly stated otherwise. Although this was not pursued in detail here, we found that the observed conductivity is stable in air during measurements, but reduces over time when the sample is left untouched in air for days, cf. Fig. S10.

The capacitance-voltage (*CV*) and double wave method (DWM) measurements were taken using an aixACCT Systems Research Line DBLI together with a Linkam HFS600E-PB4 probe stage. For the *CV* measurements, a DC-bias is swept back and forth with a superimposed small AC-bias $\left(1V_{\mathrm{p\text{-}p}}\right)$ to measure the capacitance. Depending on the applied DC-bias, the permanent dipole moments become highly susceptible to external perturbation leading to a peak in dielectric permittivity and therefore capacitance. The double wave method (DWM) was used for ferroelectric characterization.[17]

*Surface characterization:* Atomic force microscopy (AFM) images were obtained using a Bruker MultiMode 8-HR AFM operated in tapping mode. Kelvin probe force microscopy (KPFM) images were taken with an Oxford Instruments Jupiter XR Asylum Research AFM in sideband mode together with a Linkam HFS600E-PB4 probe stage for temperature control and device contacting.

*Structural analysis:* The XRD measurements were carried out using a Rigaku SmartLab setup (Cu source with a 0.154 nm wavelength). The GIWAXS measurements were done in 2D mode with an incident angle of 0.30° and an exposure time of 3600 s.

*Differential Scanning Calorimetry*: DSC was conducted on Mettler Toledo DSC 2 STARe system under nitrogen atmosphere (heating and cooling was carried out at a rate of 5 °C/min).

*Elemental analysis (purity determination)*: For TXRF analysis, a S2 Picofox (Bruker Nano GmbH, Berlin, Germany) equipped with high efficiency module and Mo X ray tube was used. Spectra were recorded at 50 kV and 600 µA over a live time of 1000 s. During measurement the system was purged with nitrogen gas (N2 5.0, mti Industriegase AG, Neu-Ulm, Germany).

*UV-Vis spectroscopy*: UV-Vis absorption spectra for the **FCH**s were recorded on a Nanodrop OneC (Thermo Fisher Scientific) in tetrahydrofuran (THF) at room temperature.

*Cyclic Voltammetry*: For the **FCH** molecules, cyclic voltammetry was recorded on a Bio-Logic SAS (SP-150) instrument. A 2 mM solution of each analyte solution was prepared for cyclic voltammetry measurements in dry THF solvent. The supporting electrolyte used was 0.1 M n-butylammonium hexafluorophosphate $(nBu_4N)(PF_6)$, and the experiments were conducted under argon atmosphere at 25°C. A glassy carbon electrode served as the working electrode, a platinum wire as the counter electrode, and the reference electrode was Ag/Ag+. The scan rate was 100 mV/s. The potentials were referenced to the ferrocene/ferrocenium ($Fc/Fc^+$) couple.

*Calculation of HOMO through Cyclic Voltametry*[18–20]: Ferrocene was used as external standard. It shows two peaks at 0.60 and 0.44 V, cf. Fig. SXIX, hence $E_{1/2}$ (ferrocene) is equal to 0.52 V, which is used to calculate $E_{HOMO}$ as:

$$E_{HOMO}= -\left[\left(E_{ox}\text{-}E_{1/2(ferrocene)}\right)+4.8\right]$$

$E_{LUMO}$ is then calculated by $E_{LUMO}$=($E_{HOMO}$-$E_g$) where the optical bandgap is calculated from[21]

$$E_g = \frac{hc}{\lambda_{onset}} = \frac{1240}{\lambda_{onset}\ (nm)}\ eV$$

with $E_g$ the band gap; $h$ Planck's constant; $c$ the velocity of light and $\lambda_{onset}$ the wavelength of the onset of absorption, cf. Fig. SXX.

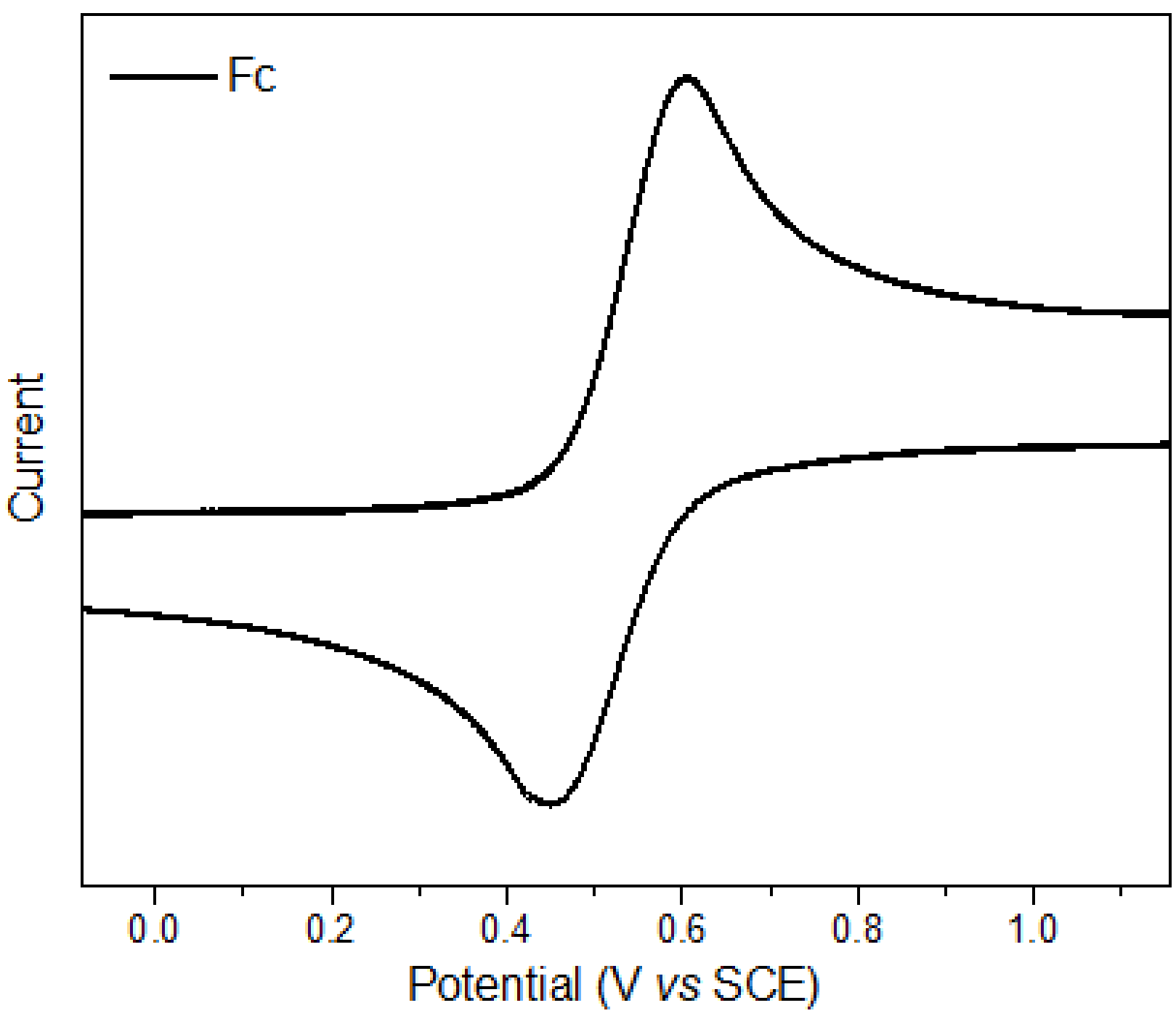

**Fig. SXIX.** Cyclic voltammogram of ferrocene at a scan rate of 100 mV/s (0.1 M tetrabutylammonium hexafluorophosphate in anhydrous THF) (electrodes: working- glassy carbon; counter- Pt wire; reference- Ag/Ag$^+$).

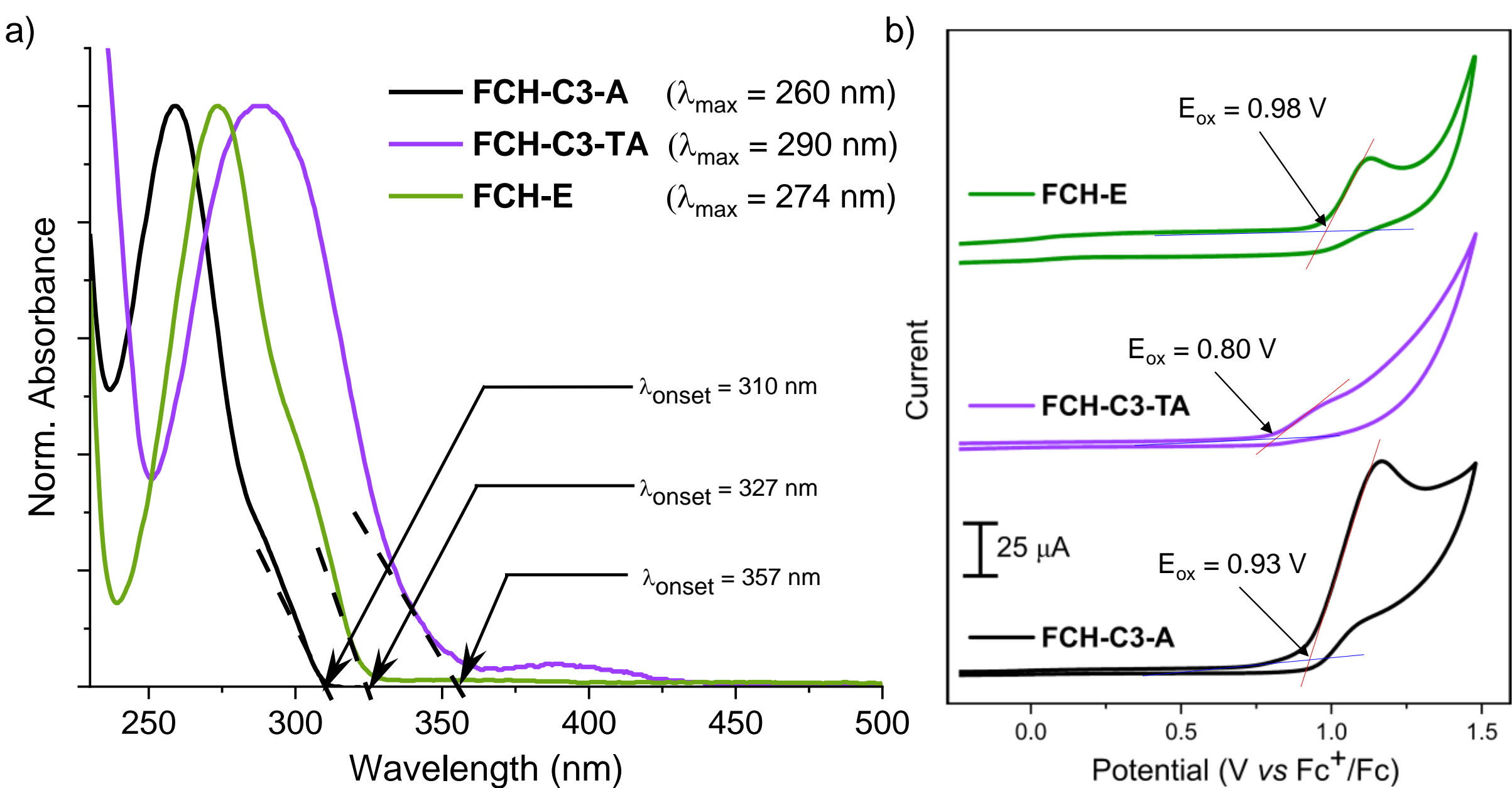

**Fig. SXX.** a) Normalized UV–Vis absorption spectra of **FCA** compounds in THF at room temperature. b) Cyclic voltammograms at a scan rate of 100 mV/s (0.1 M tetrabutylammonium hexafluorophosphate in anhydrous THF) (electrodes: working- glassy carbon; counter- Pt wire; reference- $Ag/Ag^{+}$).

Although the cyclic voltammograms in Fig. SXX show largely non-reversible oxidation peaks for the **FCA** compounds, it is important to realize that the relevant timescale of a CV experiment is very different from that in a conductivity experiment. In CV in the experiments in Fig. SXX b, the positive charge persists on the molecule for several seconds, dictated by the scan rate, which is what leads to chemical degradation. In contrast, typical hopping rates are orders-of-magnitude faster. Estimating a hopping rate $\tau$ from $\mu \cong qa_{typ}^2/6k_BT\tau$, with $a_{typ}$ a typical hopping length scale (~1 nm) and $k_BT/q$ the thermal energy (~25 meV), we find for the **FCA** compounds (mobility $\mu$ ~$10^{-6}$ $cm^2/Vs$) $\tau \cong 10^{-7}$ s, that is, seven orders of magnitude shorter.

For the **TPA** molecules, cyclic voltammetry was recorded on a BASi cell stand employing the Epsilon Eclipse software from BASi at room temperature with a scan rate of 50 mV $s^{-1}$. The three-electrode setup included a glassy carbon working electrode of 3 mm diameter, a platinum wire counter electrode, and a Ag+/Ag (3 M NaCl solution) quasi-reference electrode, controlled by means of a potentiostat. The electrolyte solution (0.1 M in either dry THF or dry CH2Cl2) was prepared with $(n\text{-}Bu_4N)(PF_6)$ and deoxygenated with a nitrogen flow for 20 min prior to the analysis. The ferrocene/ferrocenium (Fc/Fc+) redox couple was used for referencing.

HOMOs were estimated according to $E_{HOMO}$=-$(E_{ox,1}$+4.8$)$ $eV$ where $E_{ox,1}$ denotes the first oxidation potentials as experimentally determined by CV in dry THF (vs. $Fc/Fc^{+}$), vide supra and Fig. SXXI.

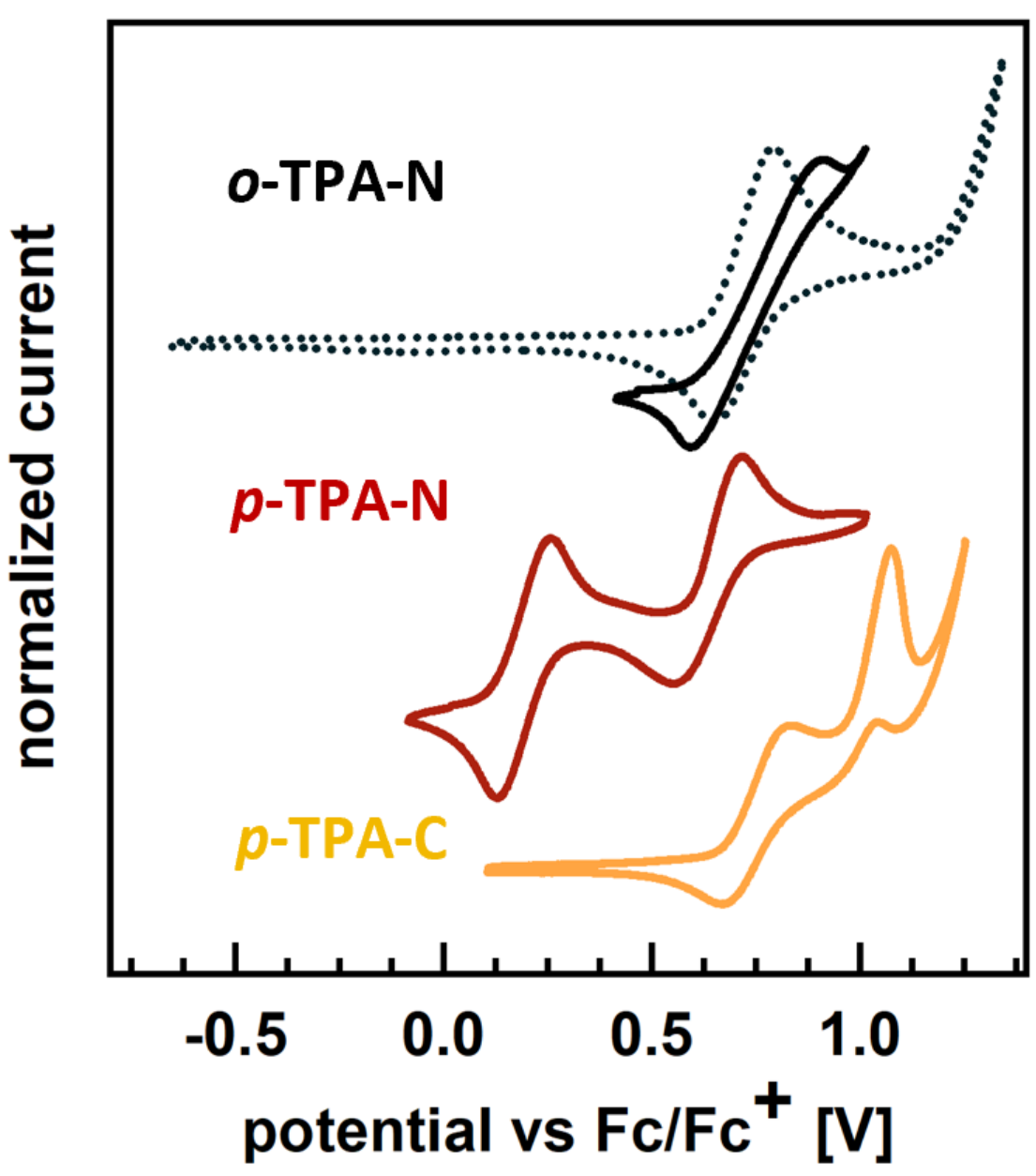

**Fig. SXXI.** Cyclic voltammetry data for **TPA** compounds in dry THF (solid lines) or dry $CH_2Cl_2$ (dotted line) at room temperature and referenced to $Fc/Fc^+$ (scan rate 50 $mVs^{-1}$)

## 3 – Alternative explanations for conductivity

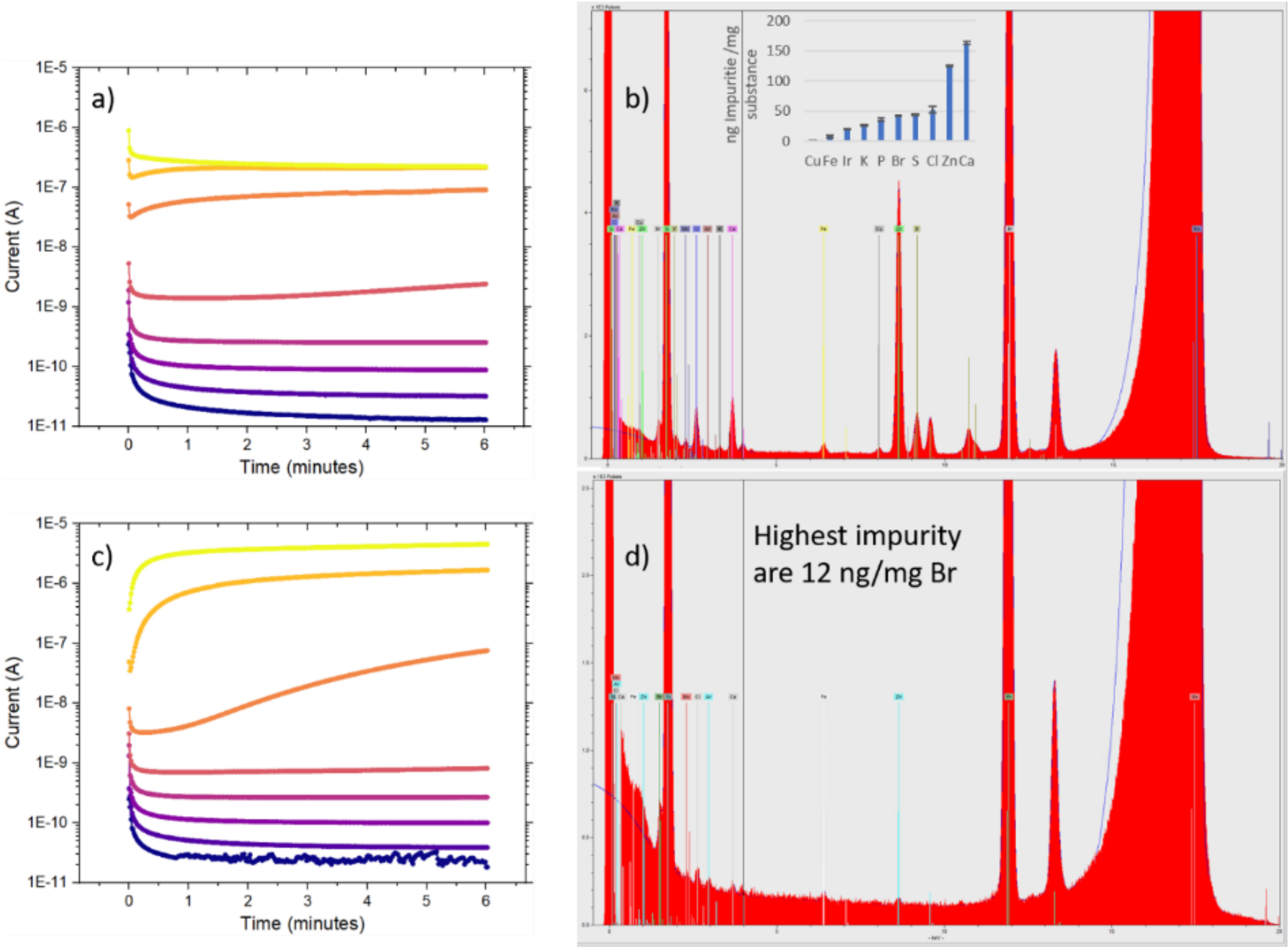

**Figure S1.** Measured current vs. time at constant applied voltage for increasing temperature steps for **FCH-C3-A** samples of a batch with high impurity content (a) and low impurity content (c). b) and d) X-ray fluorescence data depicting the impurity concentration of the corresponding material batch.

We remark that the conductivity of the low impurity content batch (Fig. S1b, d) is equal or higher than that of the high impurity content batch (panels a, c), ruling out that the observed conductivity is an impurity-related artefact. However, since most of the identified elements will be present in the form of ions (e.g. Br as bromide anion), this does explain the small transient currents at short times after applying the voltage step in Fig. 2a.

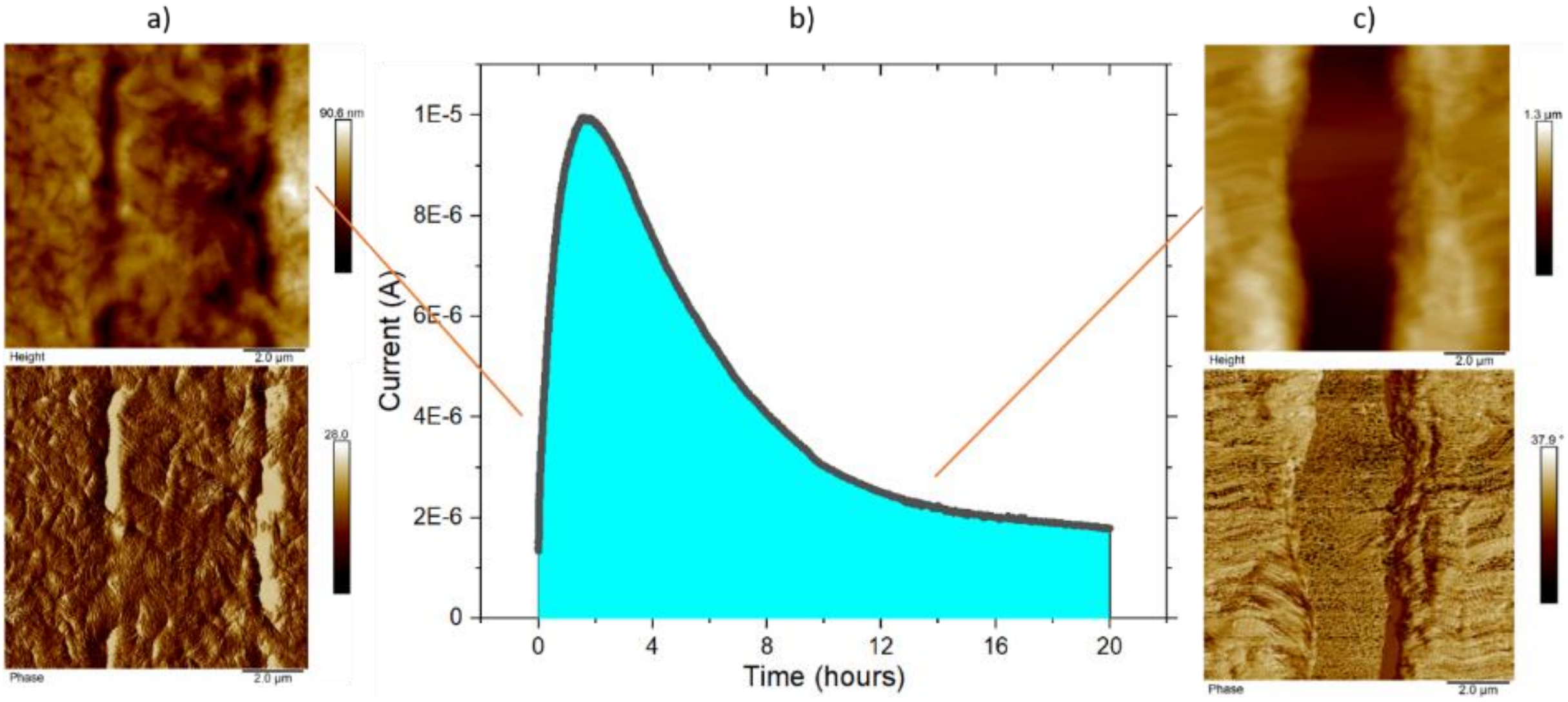

**Figure S2.** Longtime current measurement of an **FCH-C3-A** device under constant applied field of 20 V/µm at 370 K shown in (b). The integrated (cyan) area shows the charge extracted at the electrodes over 20 hours of measurement. The peak shape can be explained by an increase in long-range order of the material (compare Fig. S6a), followed by material transport on top of the electrodes. At the beginning, this leads to an enhanced conductivity as film alignment increases, while the successive decay in current is caused by material being pushed out of the active region onto the electrodes. This effect slows down considerably over time. a) and c) show AFM images (top – height and bottom – phase) taken before and after long time field annealing, showing the unaligned film (cf. Fig. 2d in the main text) and the material piling up on the electrodes, respectively.

The charge obtained by integrating the longtime current measurement amounts to 307 mC. To rule out effects like electric double layer formation or degradation, we calculate the maximally possible number of charged monomers. That is, if every single monomer in the film would be ionized. Assuming a solvent concentration of 20 mg/mL and the film being formed by drop-casting $5\times5\mu\mathrm{l}=25\mu\mathrm{l}$ we have an upper limit of 500 µg material in our film. With a molecular weight of the **FCH-C3-A** of 1003 u or $1.67\cdot10^{-15}\mu\mathrm{g}$ the film is made up of a total of $\sim3\cdot10^{17}$ molecules. Assuming every single one of them is singly ionized, a total charge of 48.1 mC is obtained, which is exceeded by a factor of more than six by the measured charge in Fig. S2. The reversibility of the conductivity shown in Fig. S10a further confirms the absence of significant degradation processes.

To calculate the upper limit of possible charge originating from double layer formation, we assume a high dielectric constant of 10. The total electrode surface area including the top and the sides amounts to 4.32 $\mathrm{mm}^2$. With $C=\epsilon A/d$ and a lower double layer thickness of 0.4 nm, we obtain a double layer capacitance of 383 nF, corresponding to a charge of 38.3 µC when applying 100 V. Again, this is negligibly small compared to the integrated charge measured in Fig. S2 and excludes double layer formation as possible origin of the conductivity.

Proton conductivity can be excluded, as we observe conductivity in the C3-amide at temperatures well below its melting point, while in high vacuum and at temperatures above the evaporation temperature of $H_2O$.[22]

Finally, we should rule out the possibility that, rather than involving the states in the intrinsic molecules, transport takes place in an impurity band that roughly aligns with the contact Fermi level at around -4.5 eV. For some form of hopping transport to be possible, this requires the distance between the impurities to be at most a few nm. Taking 3 nm as an upper limit (using the Mott condition for the formation of a true impurity band gives an order of magnitude shorter distances) gives a lower limit for the required impurity density of $\sim 4\times10^{25}$ $m^{-3}$. That is an unrealistically high value, namely more than 1% of the molecular density. Moreover, this would have to be an impurity that has an energy level around -4.5 eV, which would therefore have to be a relatively large molecular compound. From the synthesis procedure, any significant concentrations of such compounds can be ruled out. Furthermost importantly, the X-ray fluorescence spectroscopy data in Fig. S1d rule out any impurity densities of this magnitude. Taking the measured concentration of the most abundant impurity, $\sim$12 ng/mg Br, and the molecular weight of 1003u, an impurity concentration of $\sim$0.015% is obtained, two orders of magnitude below what would be needed to enable a plausible hopping distance. As an alternative scenario, one could imagine free charge generation by thermal excitation over the 3 to 4 eV gap between the impurity level and the LUMO. Around room temperature, such a gap is effectively unsurmountable. Moreover, the fact that we measure space charge limited behavior indicates that the free charge density is small, i.e., negligible at finite fields.

## 4 – Estimation of energy level shift by interfacial dipoles

Starting from

$$\Delta\varphi = \frac{P_r}{\varepsilon_0 \varepsilon_r} d_{typ}$$

with $\varepsilon_0\varepsilon_r$ as the dielectric constant of the active layer and $d_{typ}$ as a typical thickness of the dipolar layer. The polarization $P_{\mathrm{r}} = p_{\mathrm{mol}}/V_{\mathrm{mol}}$ becomes, for a molecular dipole moment of the fluorinated cyclohexane ring $p_{mol}$=6.2Debye and a molecular volume $V_{mol} \approx 0.4\mathrm{nm}^3$, $P_r \approx 52\mathrm{mC/m^2}$. Using $\varepsilon_r \approx 2$ and $d_{\mathrm{typ}} \approx 0.4\mathrm{nm}$, we arrive at $\Delta\varphi \approx 1.2\mathrm{eV}$. Additionally, accounting for the $\sim 3.7\mathrm{Debye}$ amide dipole moment in the same molecular volume then gives a total estimated energy level shift $\Delta\varphi$ due to interfacial dipoles by $\sim 1.9\mathrm{eV}$ for the **FCH-C3-A** amide. A similar calculation for **BTA-C10**, which has a measured remnant polarization of $\sim 45\mathrm{mC/m^2}$,[23] leads to $\Delta\varphi \approx 1\mathrm{eV}$.

## 5 – Computational details

Quantum chemical calculations were performed using the Gaussian 16 program package.[24] Long alkyl chains were simplified to methyl groups to reduce computational cost. Ground state geometry optimizations were performed by employing the B3LYP[25–28] functional, the 6-311G(d,p)[29,30] basis set and Grimmes D3 dispersion correction[31] with BJ-damping.[32] Thereby, ultra-tight convergence criteria of the respective computational method were used. Frequency calculations at the same level of theory were employed to verify the geometries as local minima possessing no imaginary frequencies.

Fig. S3 shows the calculated HOMO energy levels and corresponding electron wavefunction for all compounds presented in the main text. On the **FCH-C3-A** and **FCH-E** molecules, the HOMO level is localized predominantly on the phenyl group, whereas it shifts to the thio-amide group for the **FCH-C3-TA** molecule, in line with the significant shift in the HOMO energy upon replacing the amide oxygen by a sulfur in the latter molecule. The fact that **FCH-C3-A** and **FCH-C3-TA** do show pronounced conductivity and **FCH-E** does not, suggests that the wavefunction residing on the π-system, or not, is not essential for the observed conductivity. This picture is corroborated by the **BT(T)A** and **CT(T)A** compounds. While **BTA** and **CTA**, and especially **BTTA-C12** and **CTTA** show similar conductivity, the **CTA** compounds lack a π-system altogether and accordingly show negligible wavefunction amplitude on the central carbon ring. As expected for small molecular semiconductors, the HOMO wavefunction of ***o*-TPA-N**, ***p*-TPA-N** and ***p*-TPA-C** is delocalized over the entire C3-symmetric π-conjugated system with little amplitude on the amide groups.

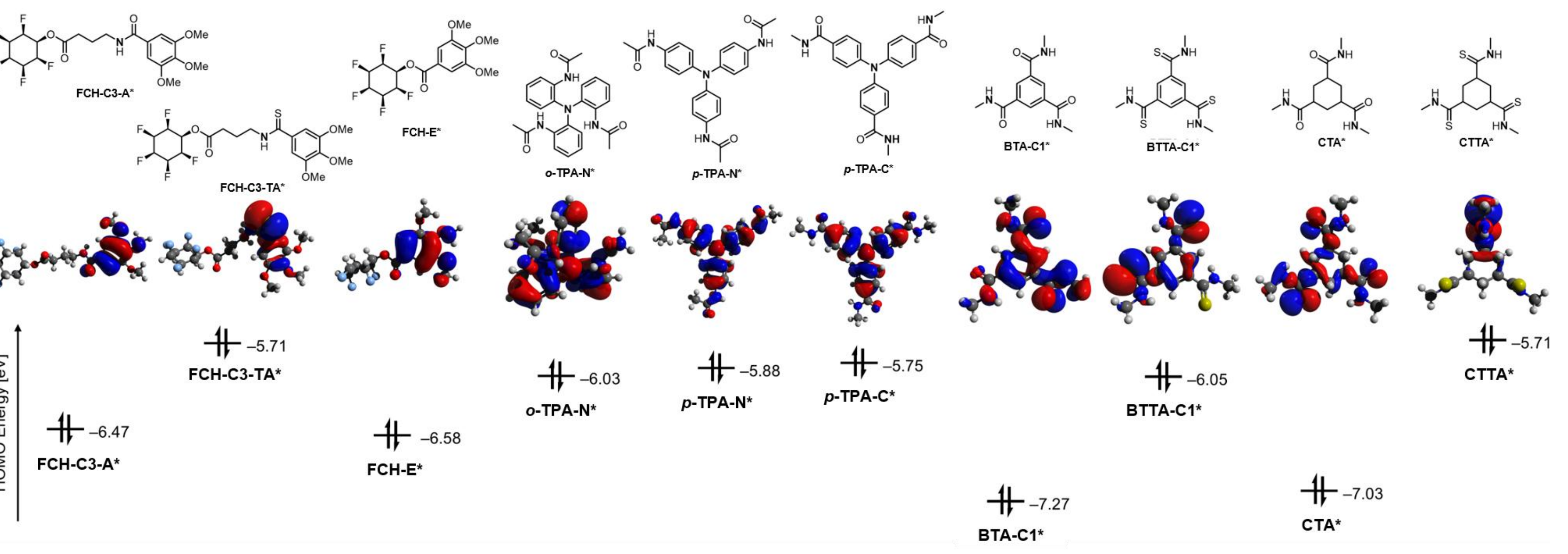

**Figure S3.** HOMO energy levels and electron orbitals calculated via DFT for all presented molecules.

## 6 – Additional measurements and analysis

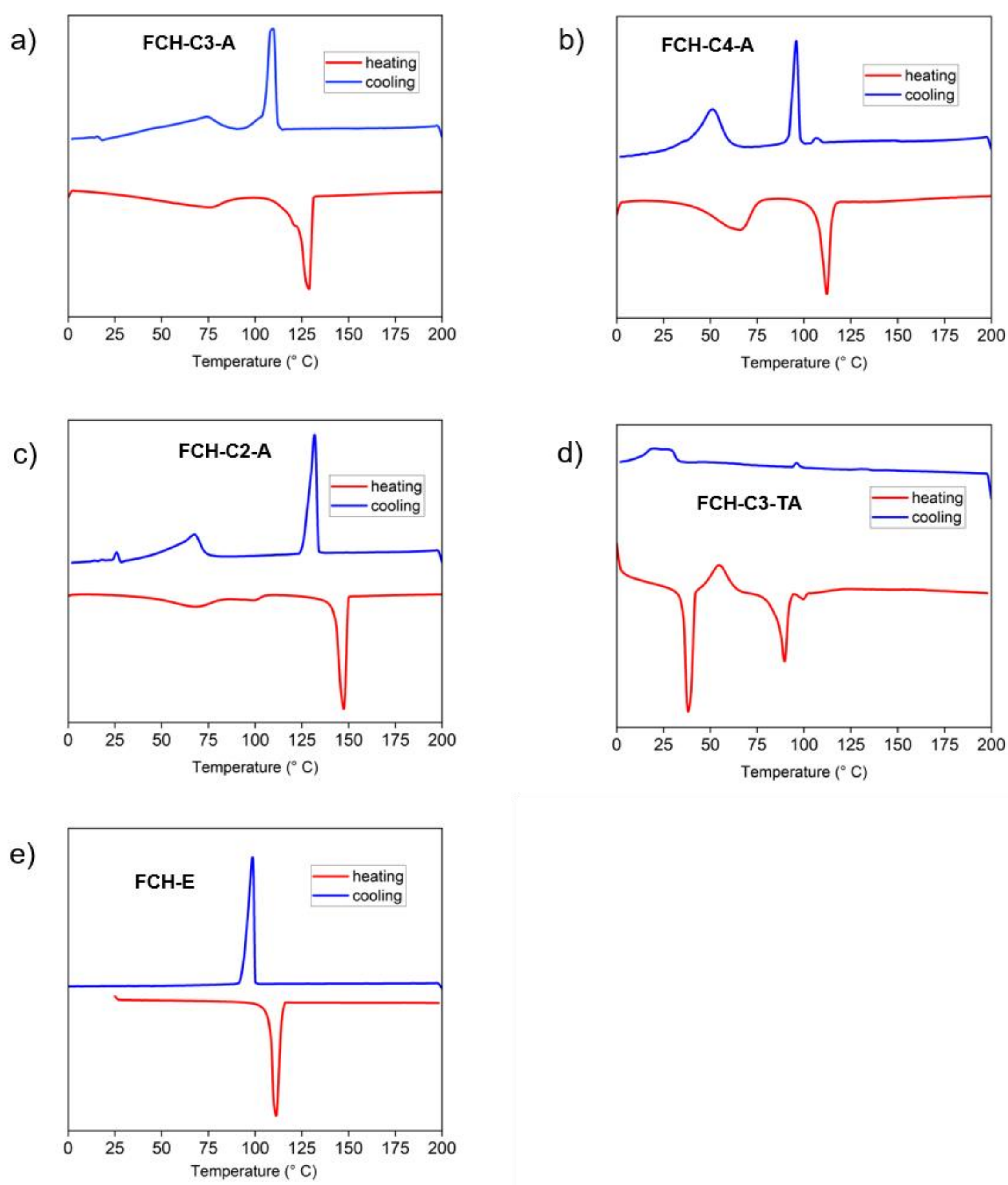

**Figure S4**. DSC curves of the **FCH** molecules studied in this work, a) **FCH-C3-A**, b) **FCH-C2-A**, c) **FCH-C4-A**, d) **FCH-C3-TA**, e) **FCH-E**. Heating, cooling at 5 °C/min. The large peaks originate from the melting and respective freezing point of the materials.

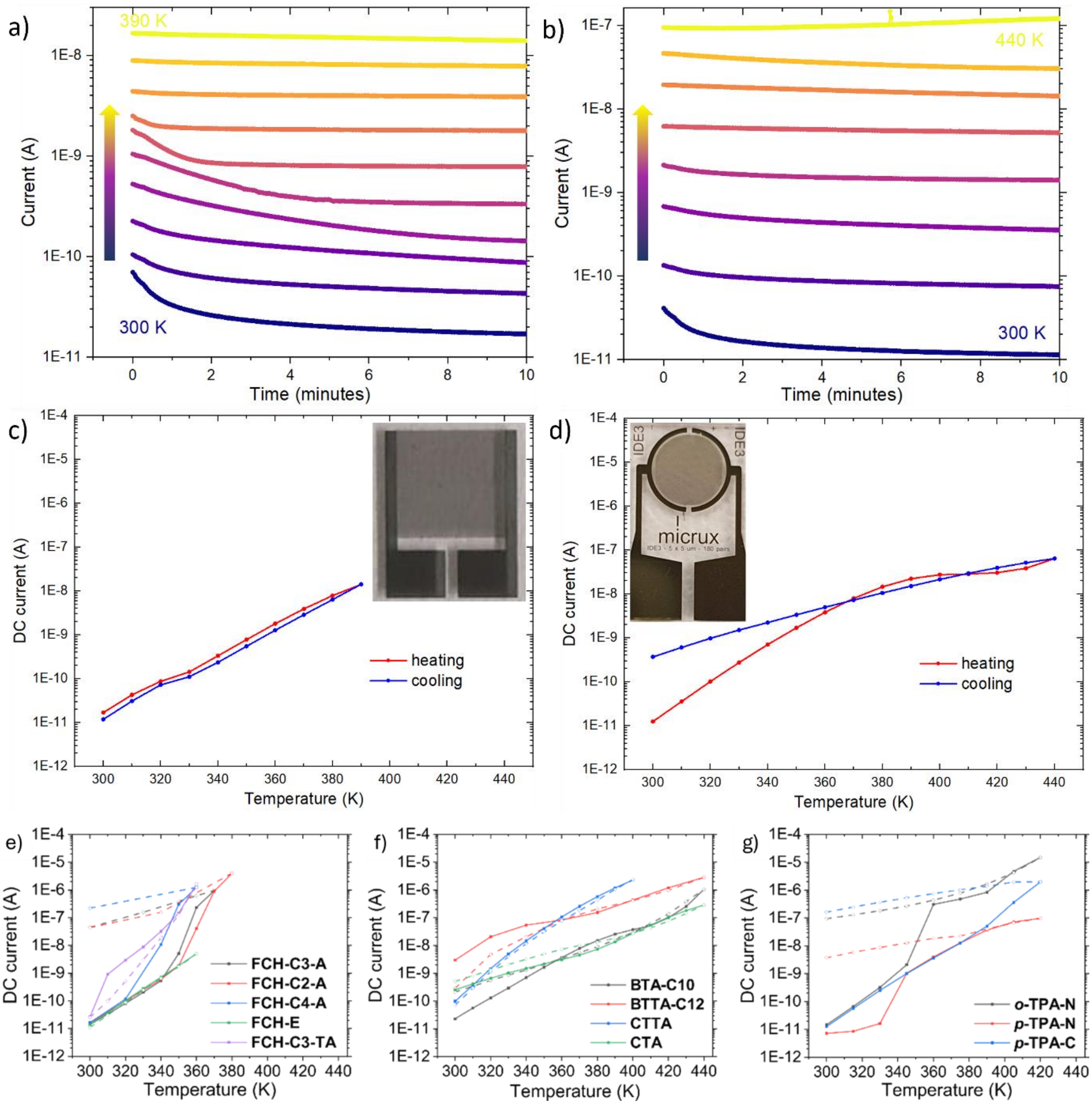

**Figure S5**. Background DC current extracted from current over time measurements at varying temperatures in the relevant ranges for a) empty homemade interdigitated electrode (IDE) substrates and b) empty commercially available Micrux IDEs. c) and d) show the corresponding equilibrium current plotted against temperature. Pictures of the corresponding IDE are inset in c) and d). Substrates were chemically cleaned and afterwards measured in high vacuum conditions. Measured DC current of all materials was compared to the relevant background measurement and evaluated accordingly. For comparison, the data of Fig. 3 in the main text are plotted as absolute currents in panels e-g).

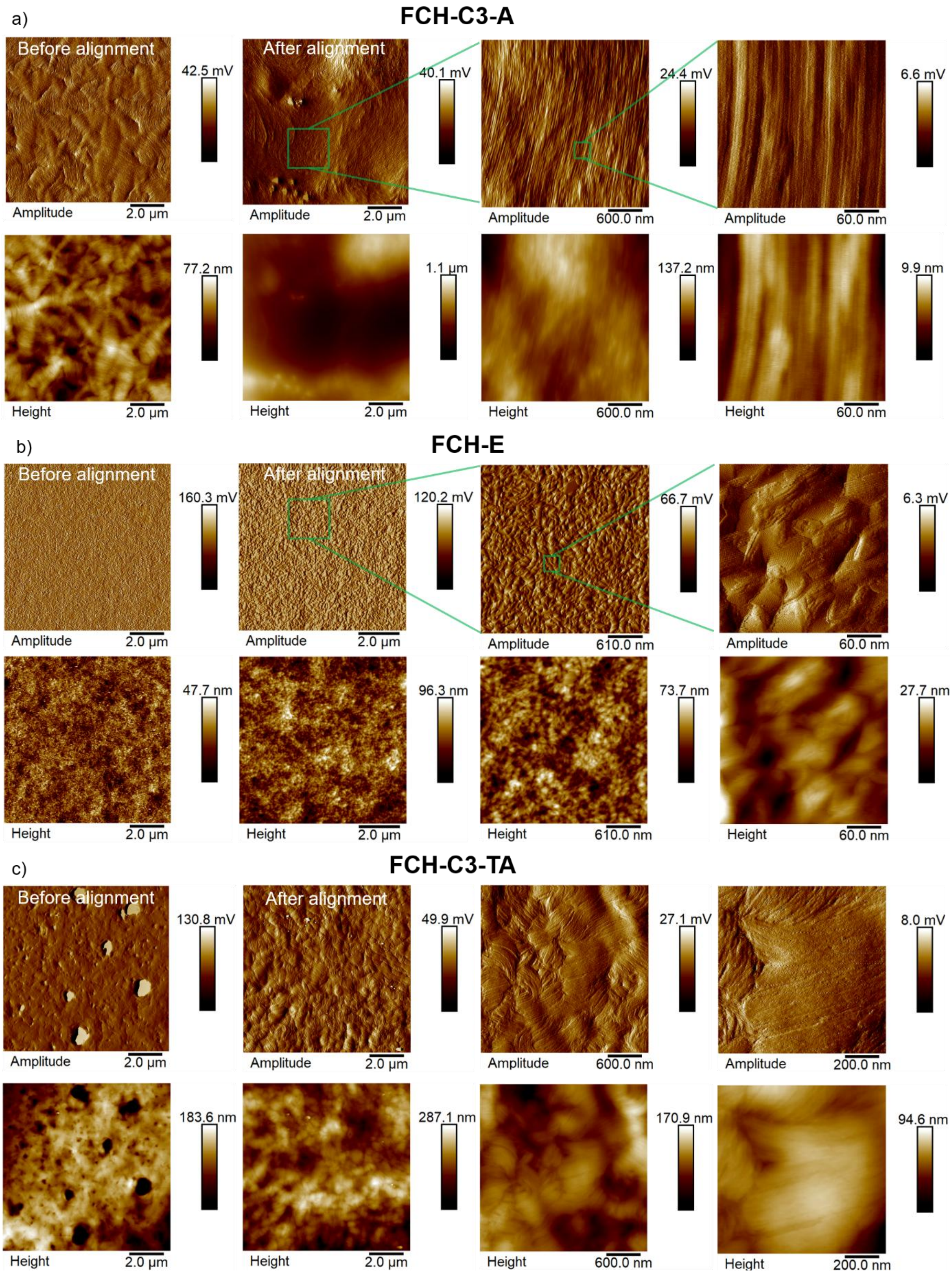

**Figure S6.** AFM amplitude and corresponding topographic images of drop-casted and molten films before and after field annealing, for a) **FCH-C3-A**, b) **FCH-E** and **FCH-C3-TA** c).

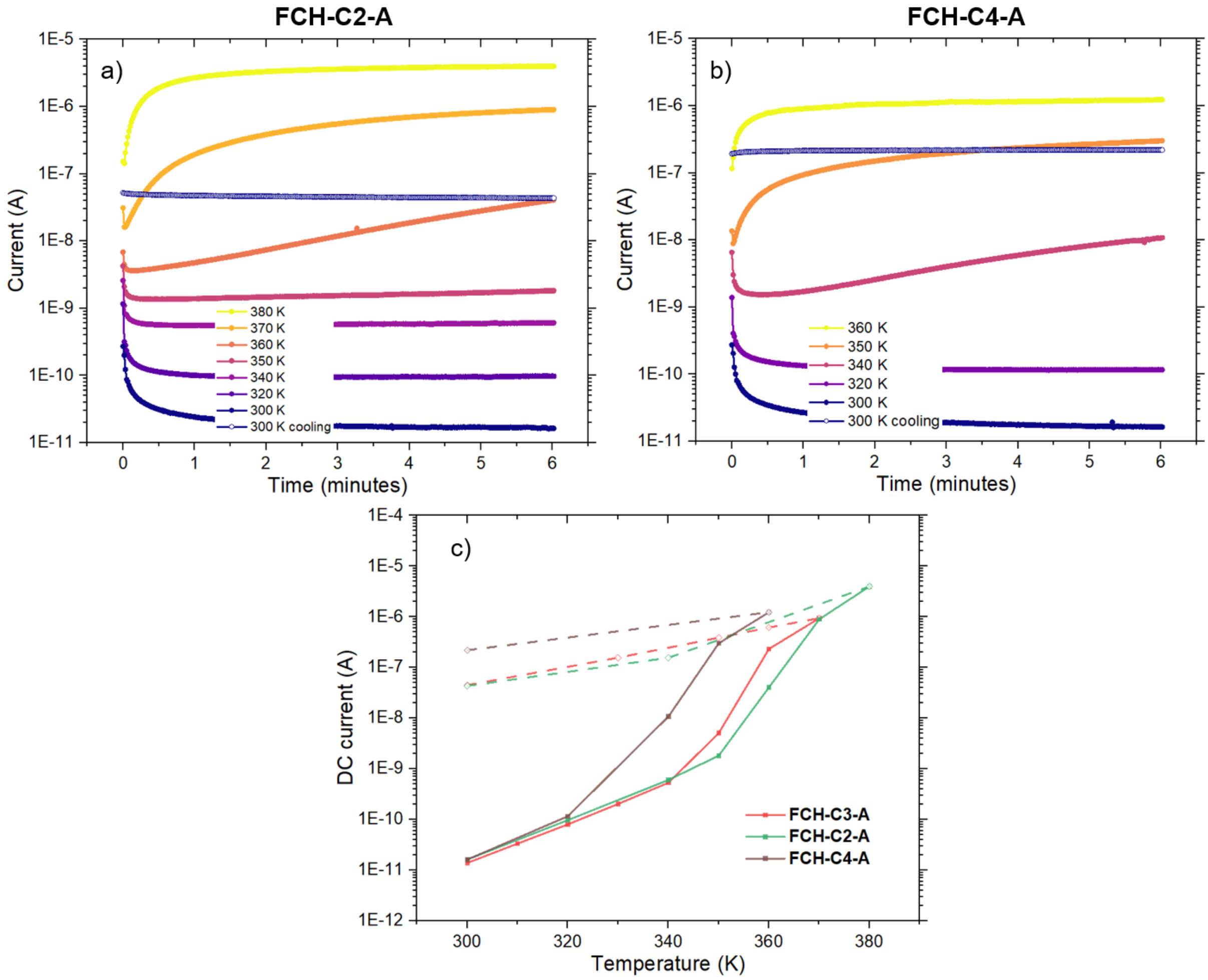

**Figure S7.** Current measurements at constant applied voltage at increasing (closed symbols) and decreasing (open symbols) temperatures for **FCH-C2-A** a) and **FCH-C4-A** b). Both materials exhibit analogous behavior to the **FCH-C3-A**, with the onset of increasing conductivity shifted relatively to the corresponding melting point of the materials; compare DSC traces in Fig. S4. This trend becomes clearer when looking at c), where the current at six minutes taken from $IV(t)$ measurements is plotted against temperature for all **FCH-Cn-A** materials. Solid lines represent increasing temperatures, dashed lines decreasing temperatures.

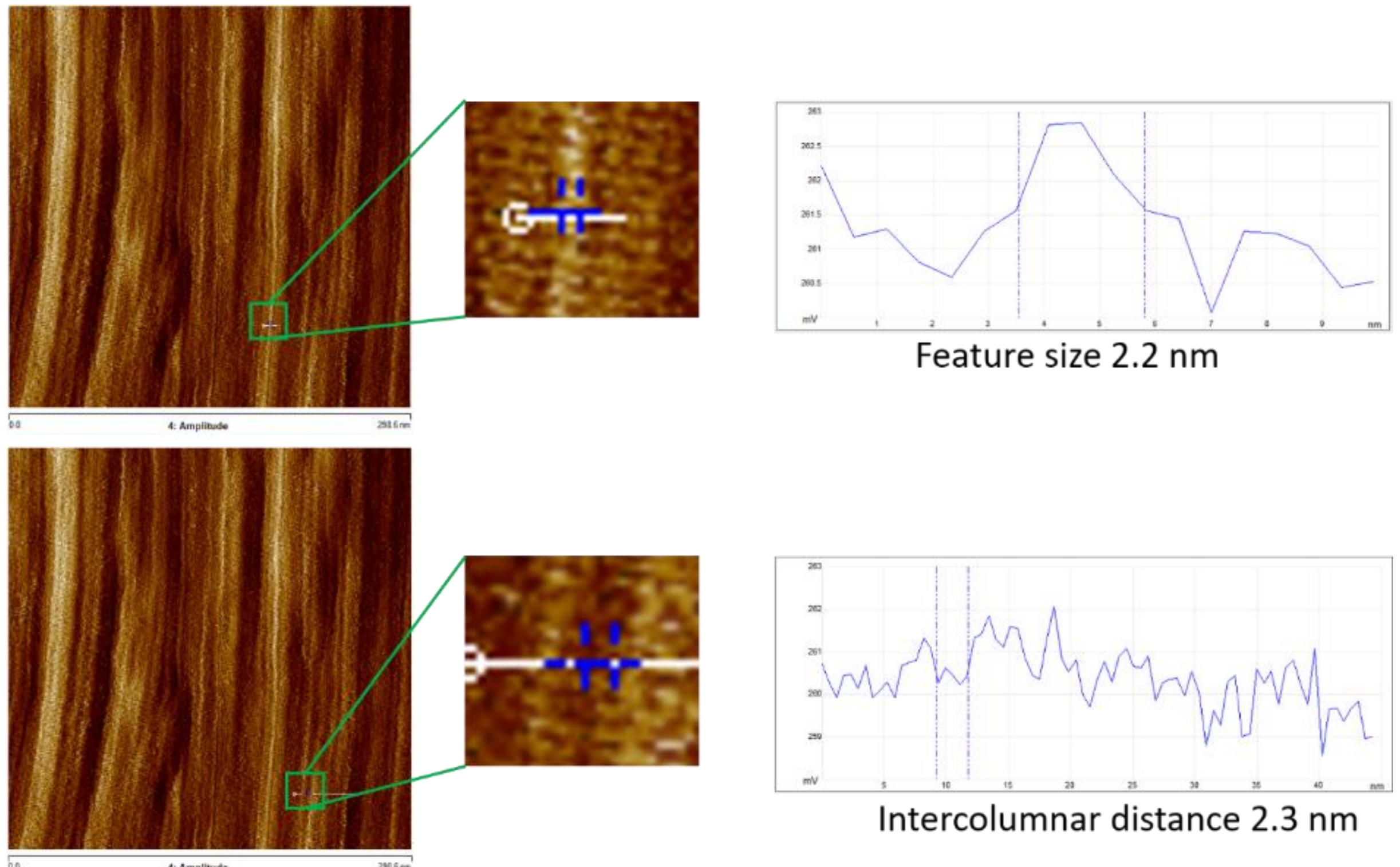

**Figure S8.** High resolution AFM images of **FCH-C3-A** after field annealing.

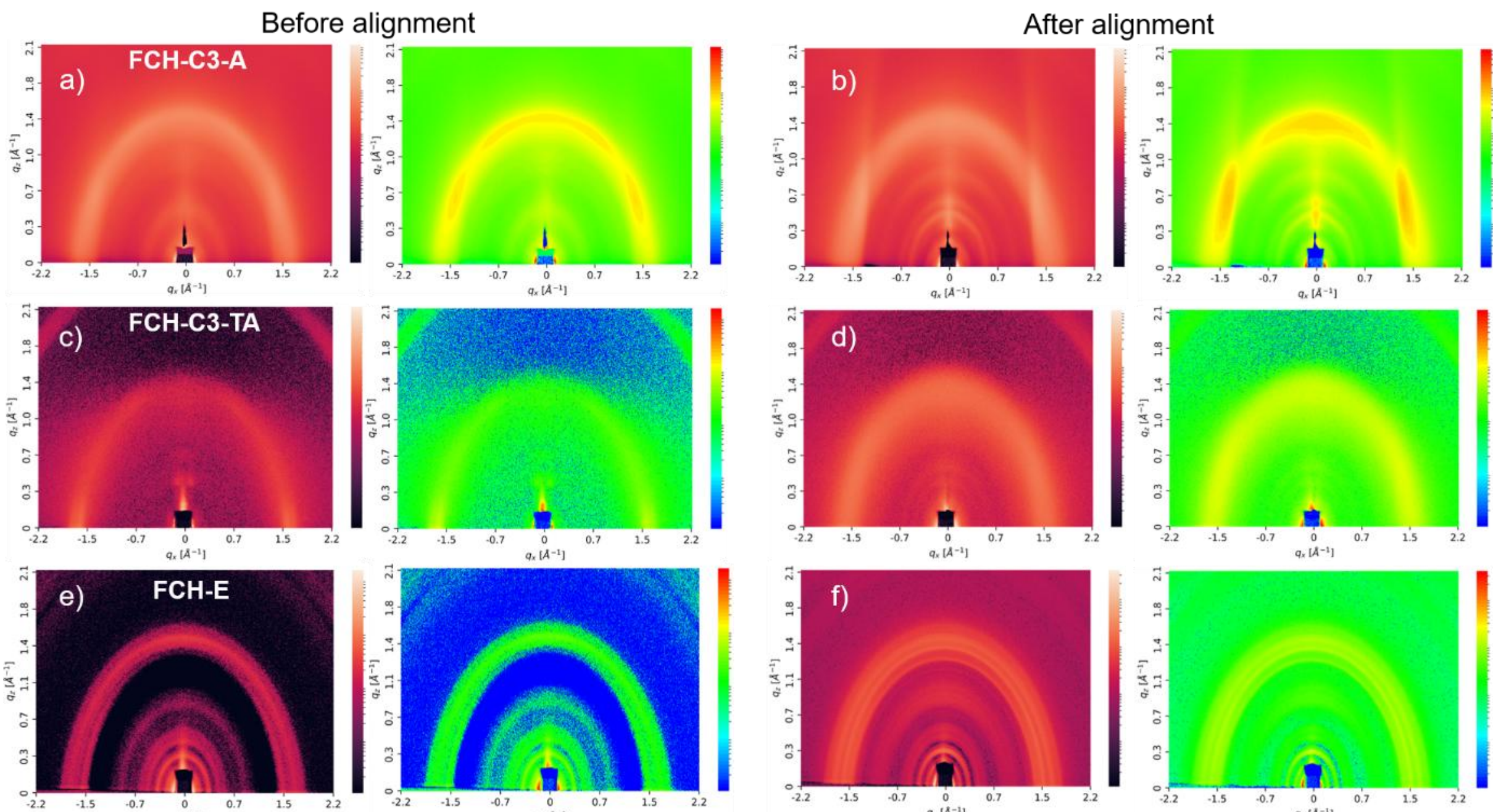

**Figure S9.** 2D GIWAXS measurements of the **FCH-C3-A** (a-b), **FCH-C3-TA** (c-d) and **FCH-E** (e-f) molecules. Spectra both before (a, c, e) and after (b, d, f) the alignment process are shown in both standard and high-contrast coloring. The spectra were taken perpendicular to the electrodes and converted to *q*-space.

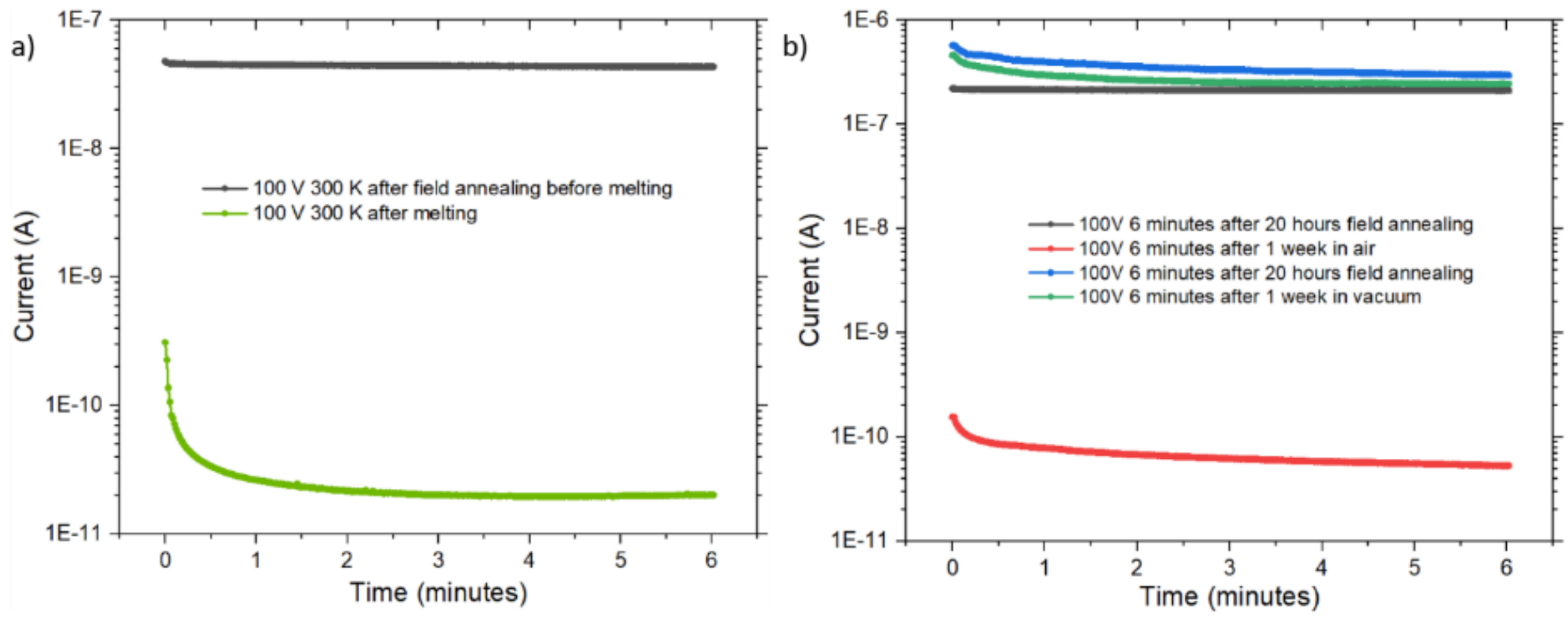

**Figure S10.** a) Current measurements of an **FCH-C3-A** device at constant applied voltage over time. a) For a device after field annealing (black) and after melting the same sample and remeasuring it under the same conditions (green) of 20 V/µm and 300 K. Field annealing the device again will bring it back to its starting condition of enhanced conductivity. b) Stability of the permanently enhanced conductivity at room temperature with the black line as starting point, taken after long time field annealing. After leaving the device one week at ambient air the conductivity reduces as shown by the red line. It has to be noted that the film alignment persists, so the reduction in conductivity is hypothesized to be caused by the formation of oxygen or water traps.[33,34] Consecutive field annealing in vacuum restores the original state (blue). Leaving the sample one week in vacuum leads to only a very minor reduction in conductivity (green), further supporting this hypothesis.

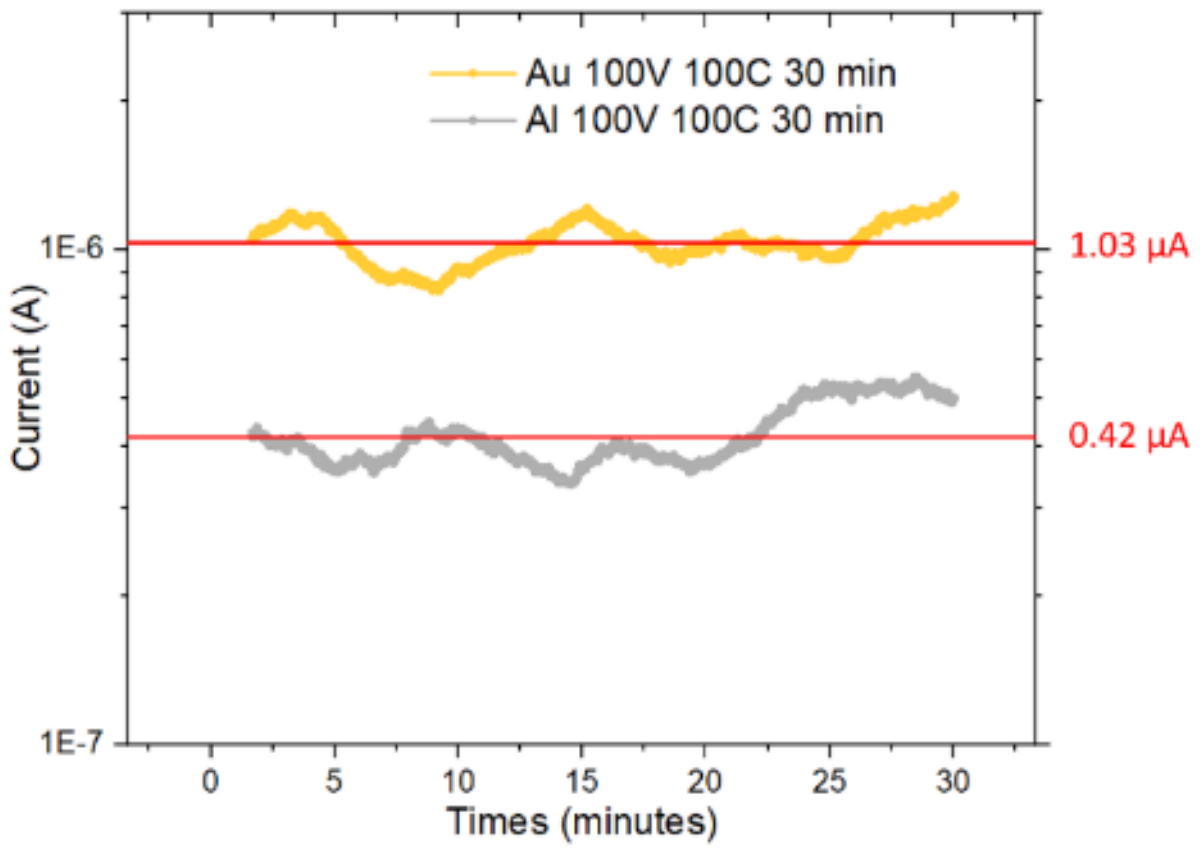

**Figure S11.** Measured current of **FCH-C3-A** devices under 20 V/µm applied field and 100 °C in air. Devices are identical except the electrode material, with the golden colored line measured on gold electrodes and the grey colored line measured on aluminum electrodes. Current fluctuations originate from 2-3 °C temperature fluctuations caused by measuring in air and PID controls. The more than a factor of two increase in current on gold electrodes is attributed to the lower injection barrier.

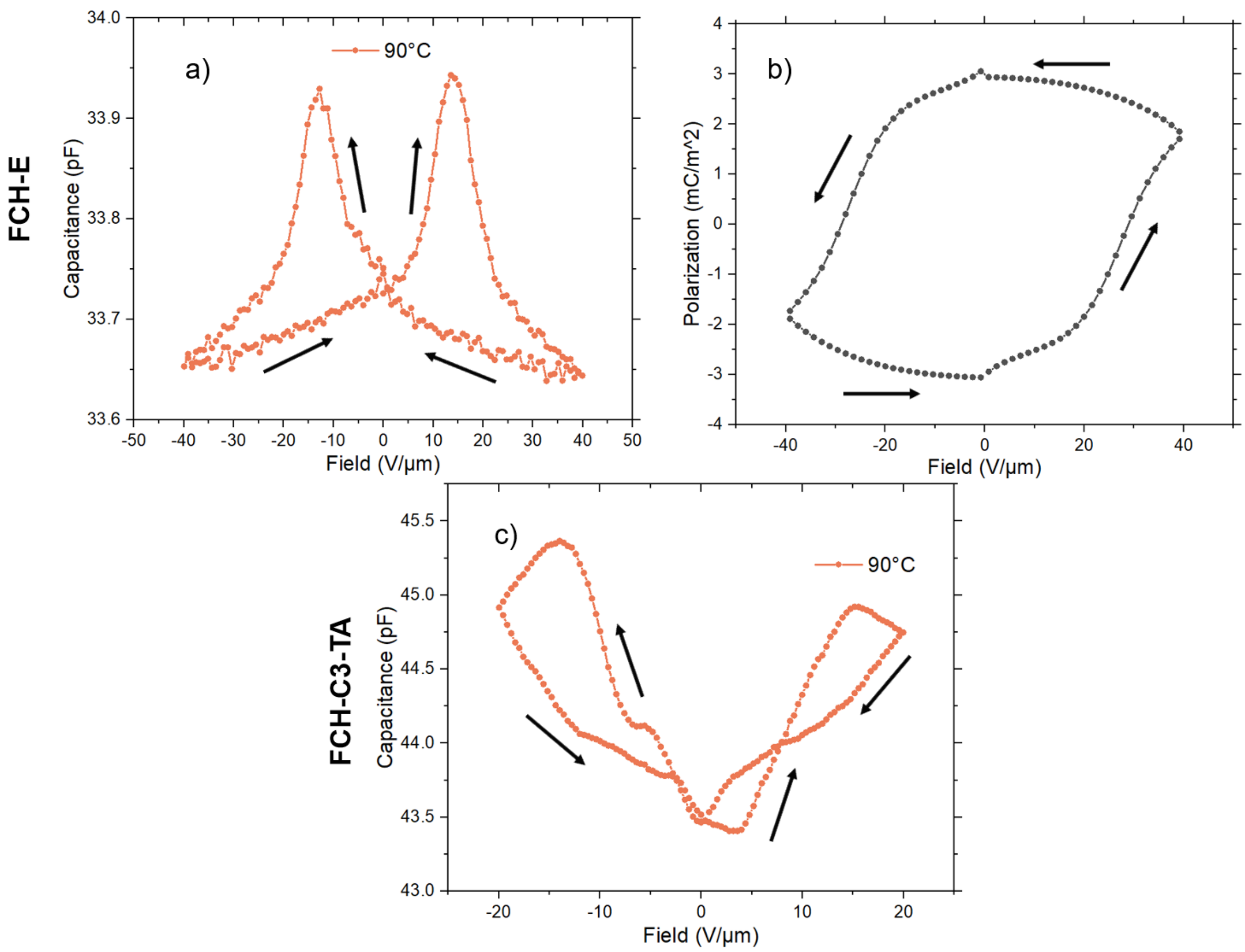

**Figure S12.** Capacitance-voltage sweeps with a superimposed AC bias of a) **FCH-E** (10 Hz, 1 V/µm peak to peak field) and c) **FCH-C3-TA** (1 kHz, 0.2 V/µm peak to peak field). c) is corrected linearly for background currents. Both exhibit an imperfect but characteristic butterfly shape of ferroelectric materials, caused by heightened dipolar responsivity near the coercive field. This demonstrates dipolar activity in both materials. b) A polarization hysteresis loop measured at 10 mHz 90 °C, typical for ferroelectric materials.

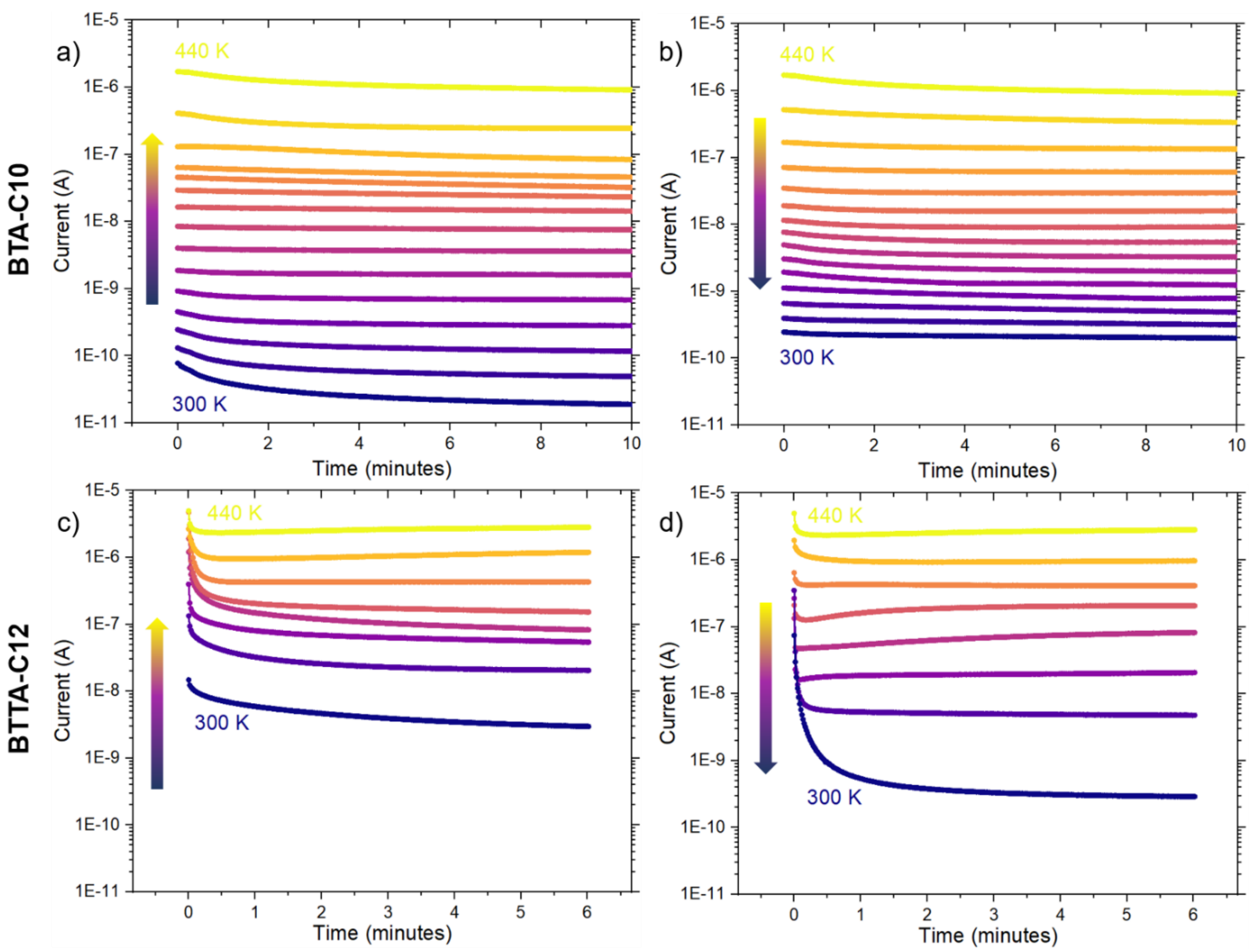

**Figure S13.** Measured current over time at constant applied voltage for increasing and decreasing temperatures, for **BTA-C10** in (a) and (b) as well as **BTTA-C12** in (c) and (d). The temperature steps are 10 K in (a) and (b) and 20 K in (c) and (d).

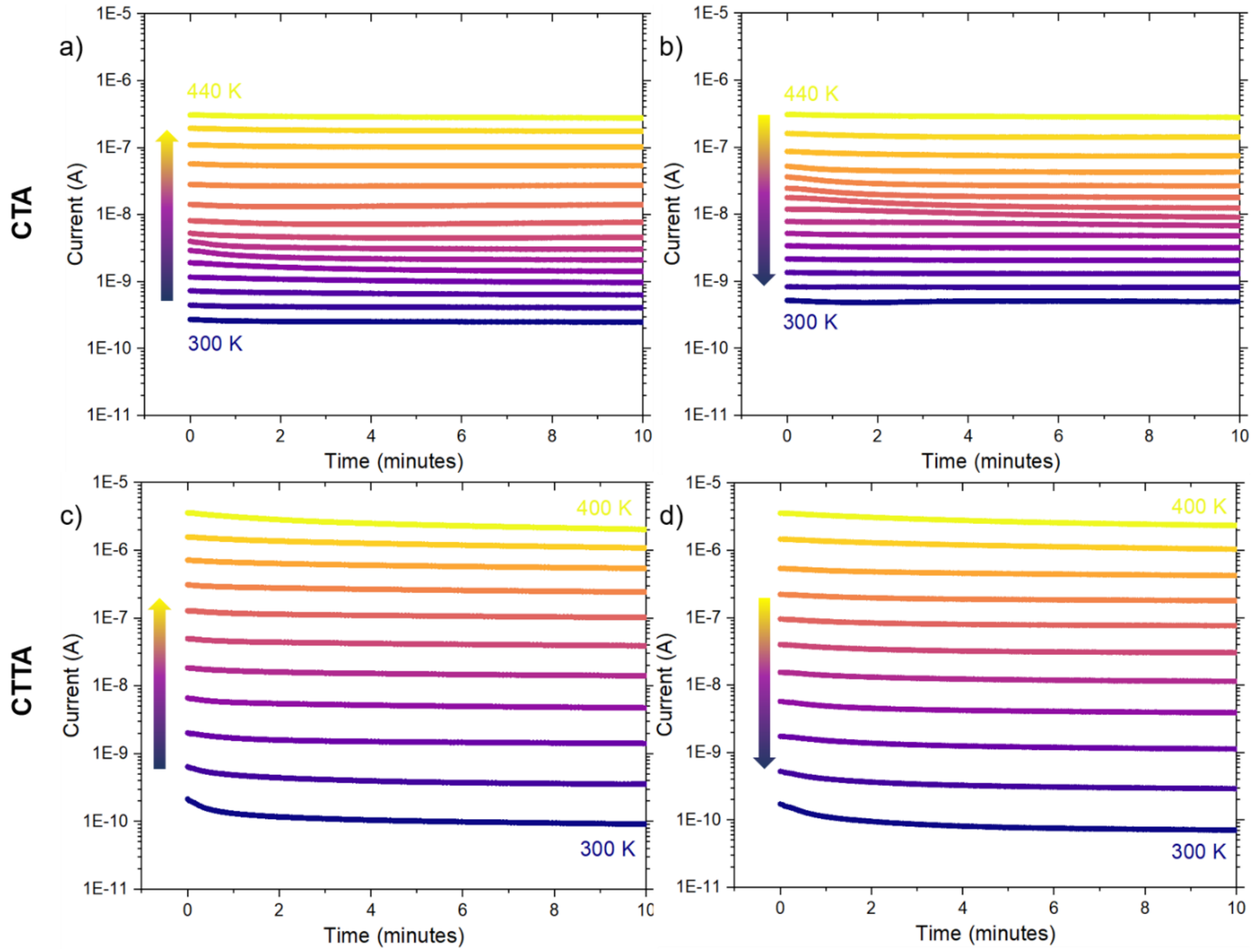

**Figure S14.** Measured current over time at constant applied voltage for increasing and decreasing temperatures, for **CTA** in (a) and (b) as well as **CTTA** in (c) and (d). The temperature steps are 10 K.

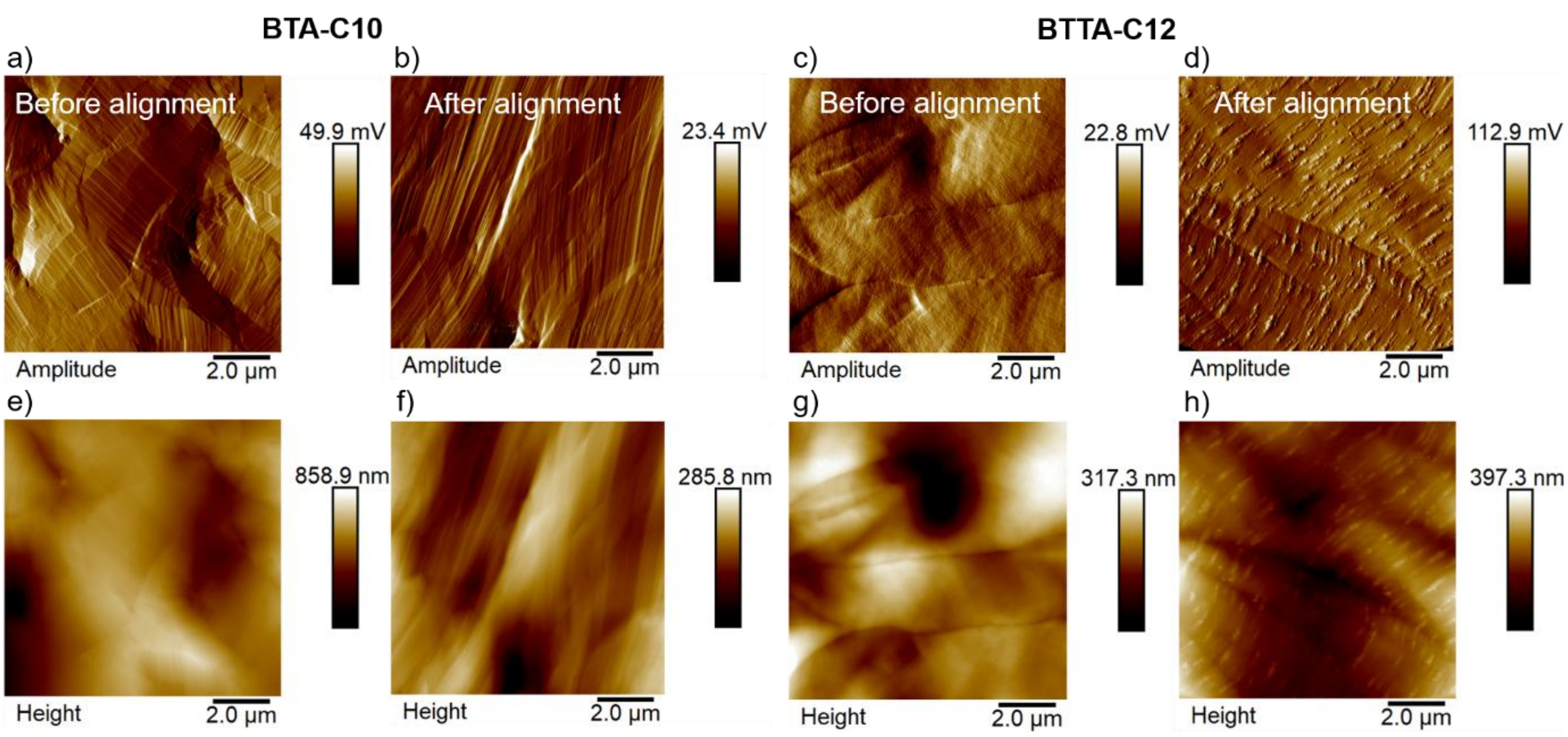

**Figure S15.** AFM morphology images of the **BTA-C10** (a, b, e, f) and **BTTA-C12** (c, d, g, h) molecules before and after the alignment process. All measurements were taken with same resolution of $10x10\mu m$. Here, the typical morphologies measured for the corresponding films are presented. The occurrence of blob-like surface features in panel (d) might indicate partial

material degradation at the highest temperatures, which are used in analogy to **BTA-C10** but not required for the observation of significant conductivity. GIWAXS data (Fig. S16) supports this observation, as the scatter reflexes for the field annealed sample are slightly broadened compared to before.

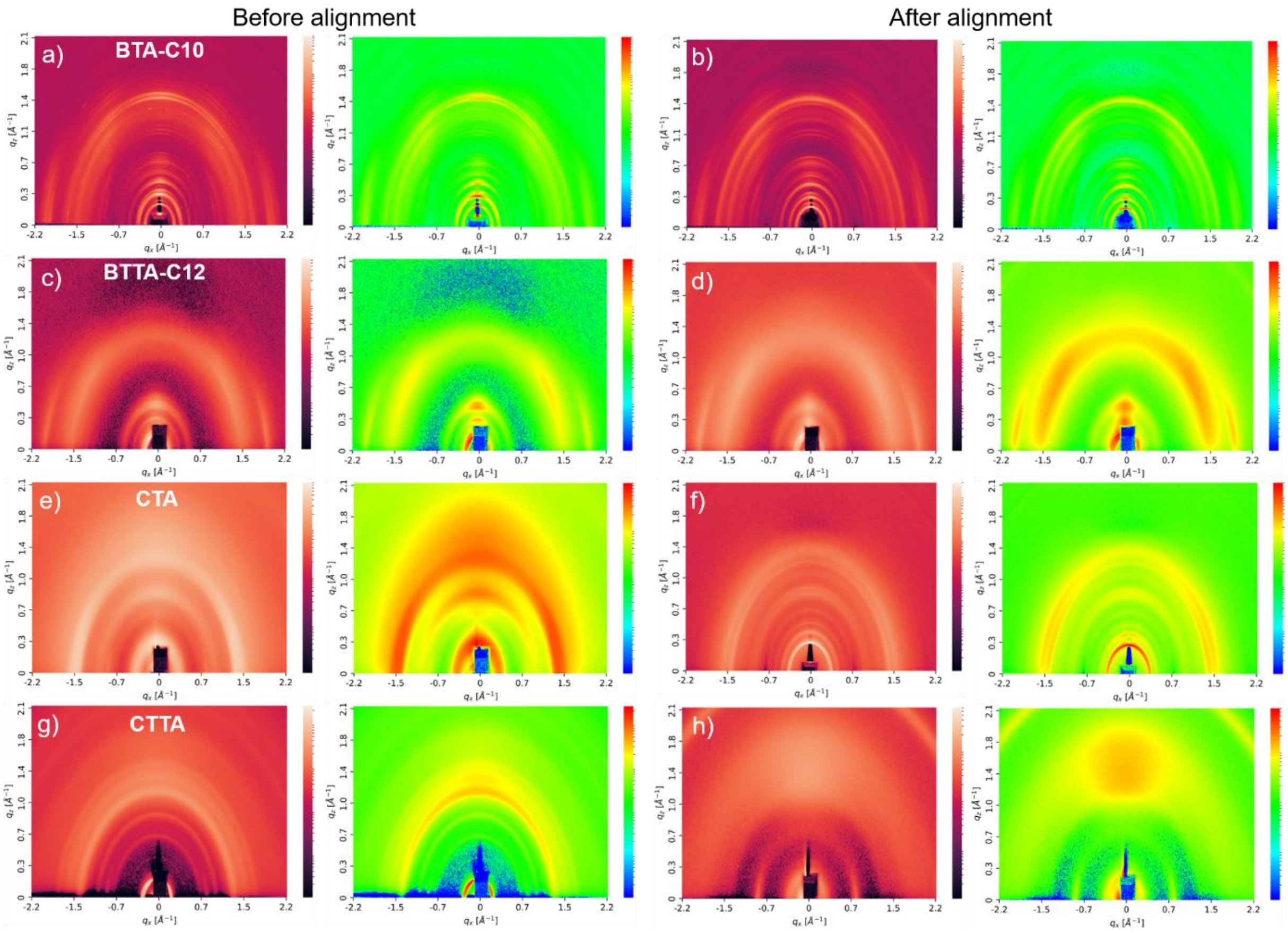

**Figure S16.** 2D GIWAXS measurements of the **BTA-C10** (a-b), **BTTA-C12** (c-d), **CTA** (e-f) and **CTTA** (g-h) molecules. Spectra both before (a, c, e, g) and after (b, d, f, h) the alignment process are shown in both standard and high-contrast coloring. The spectra were taken perpendicular to the electrodes and converted to *q*-space.

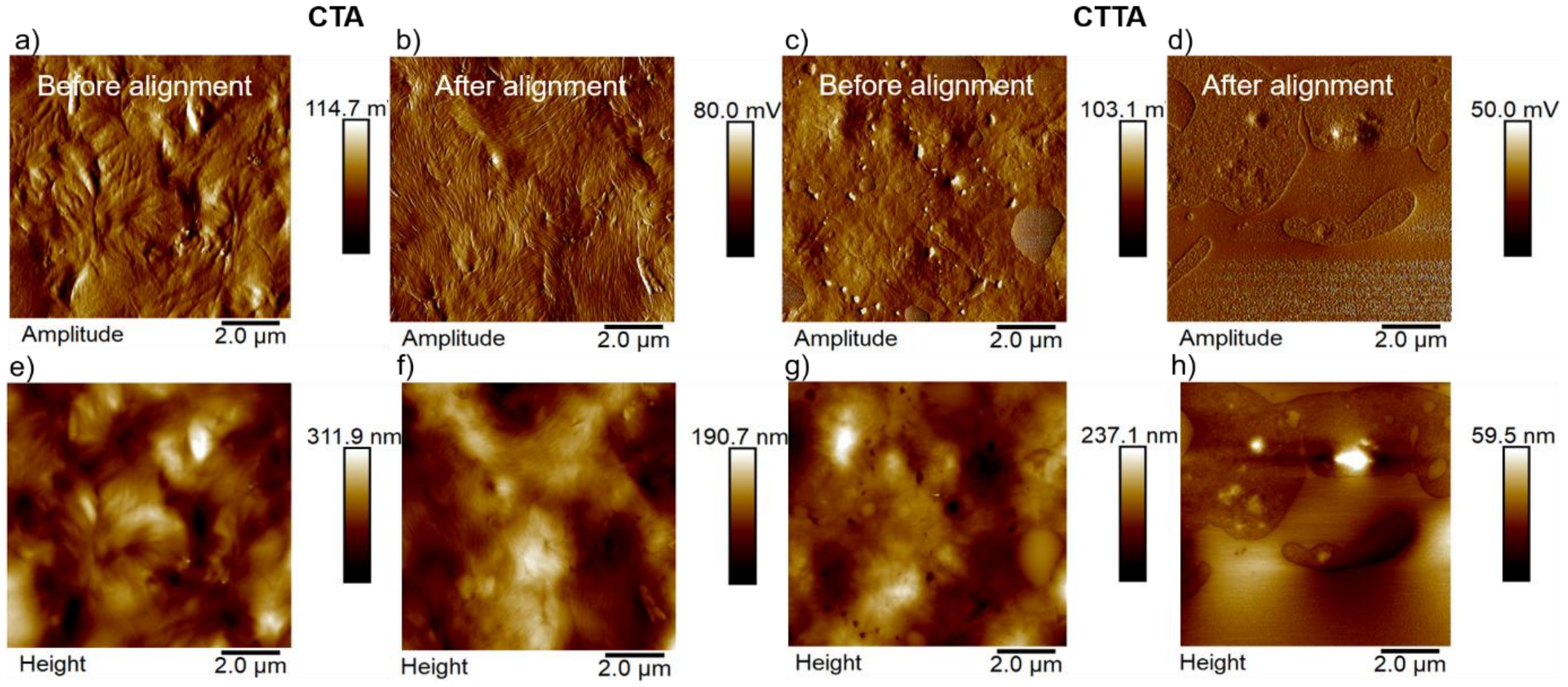

**Figure S17.** AFM morphology images of the **CTA** (a, b, e, f) and **CTTA** (c, d, g, h) molecules before and after the alignment process. All measurements were taken with the same resolution of $10\times10\mu\mathrm{m}$. Here, the typical morphologies measured for the corresponding films are presented.

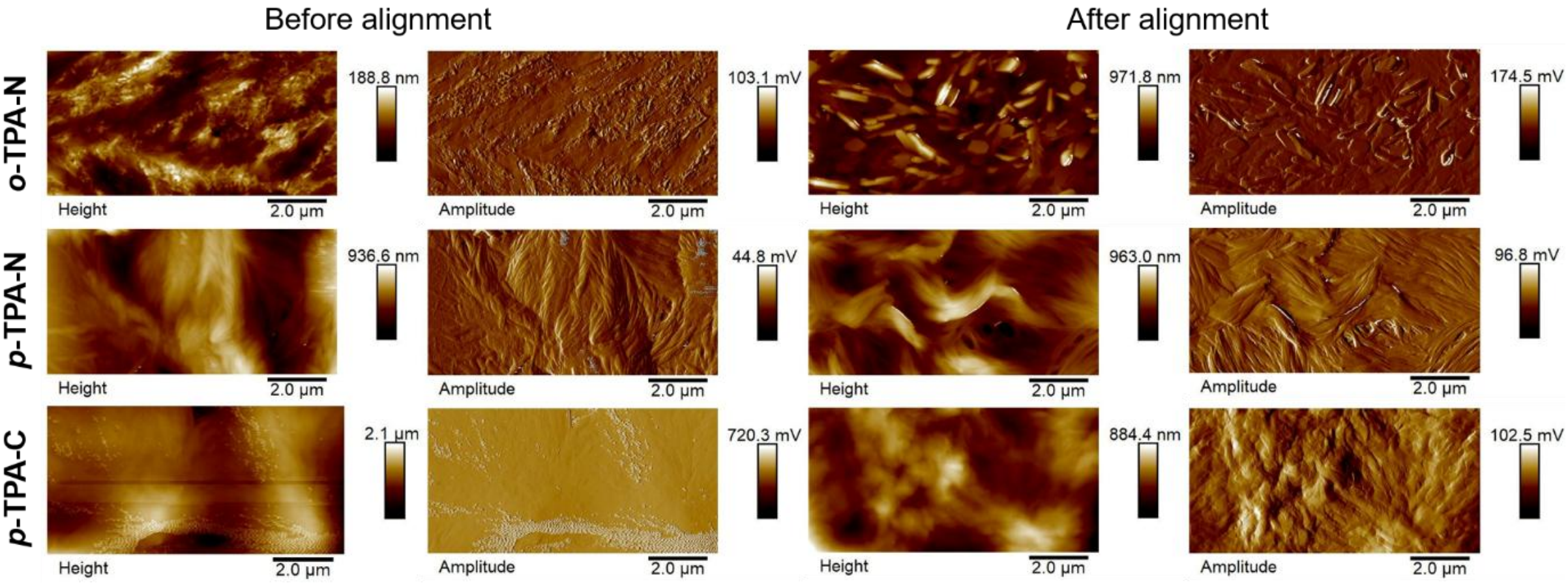

**Figure S18.** AFM morphologies of the ***o*-TPA-N**, ***p*-TPA-N** and ***p*-TPA-C** molecules before and after alignment process. All measurements were taken with the same resolution of $10\times5\mu\mathrm{m}$. Here, the typical morphologies measured for the corresponding films are presented.

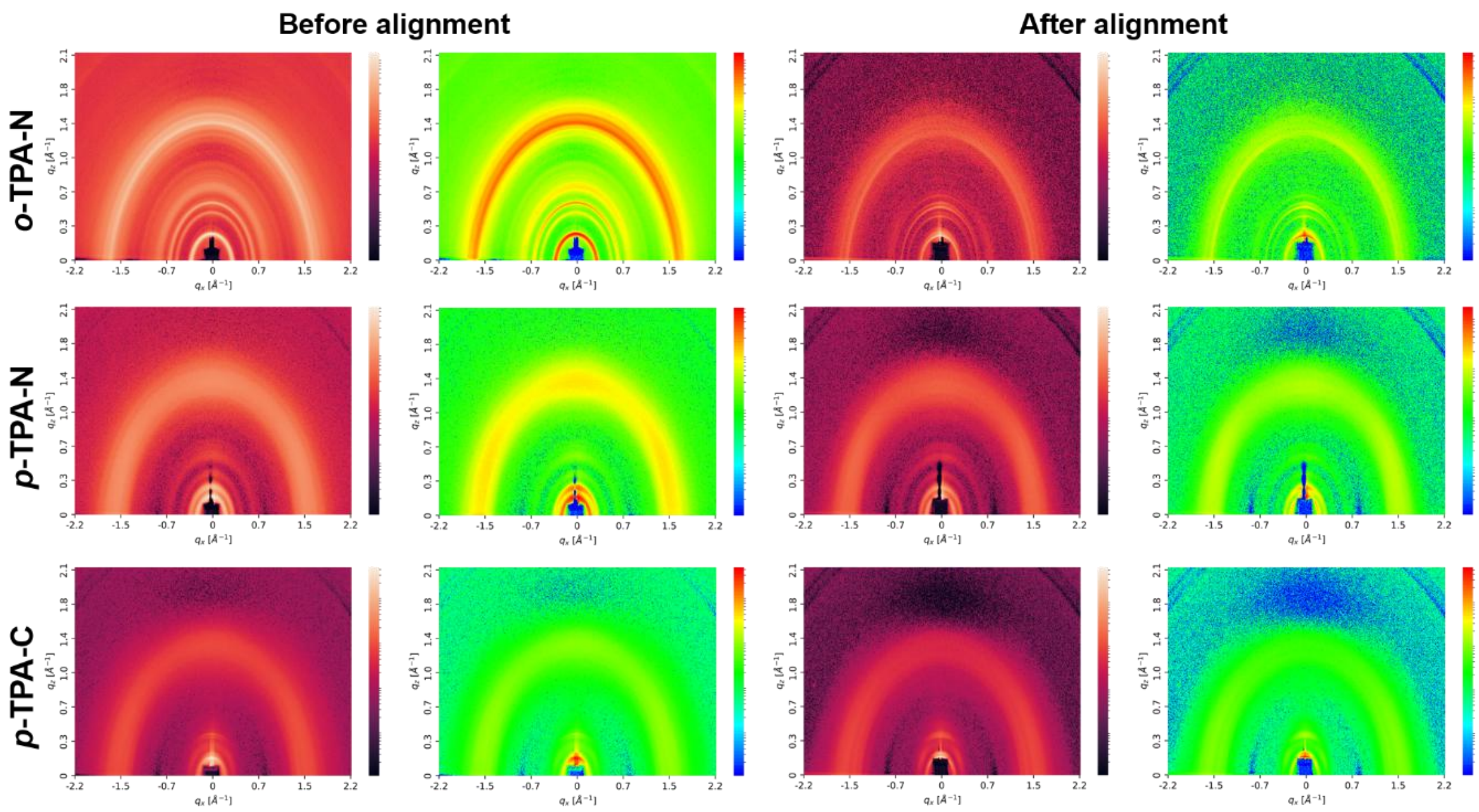

**Figure S19.** 2D GIWAXS measurements of the ***o*-TPA-N**, ***p*-TPA-N** and ***p*-TPA-C** molecules. Spectra both before and after the alignment process are shown in both standard and high-contrast coloring. The spectra were taken perpendicular to the electrodes and converted to *q*-space.

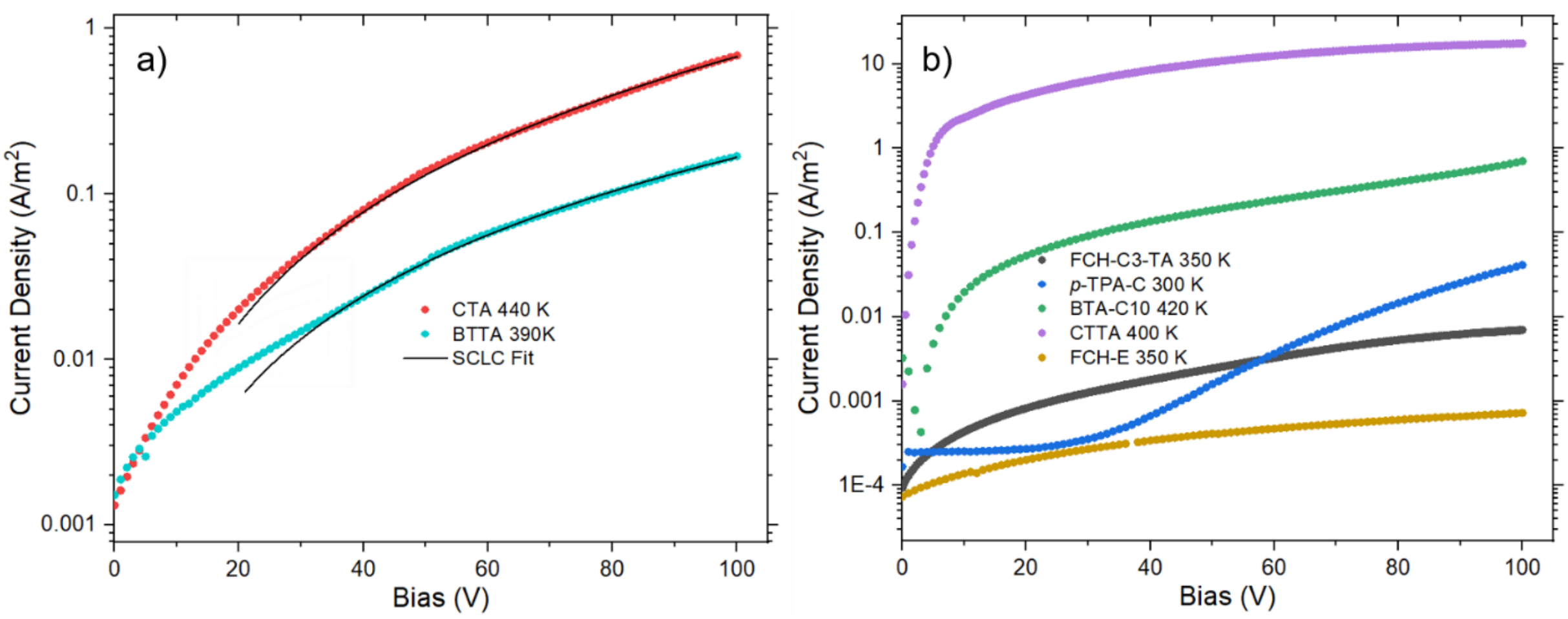

**Figure S20.** Current voltage characteristics for selected materials (symbols) fitted by the Murgatroyd SCLC model (black lines) (a) and the linear IV of **FCH-E** as well as further non-SCLC like IVs of the **FCH-C3-TA**, ***p*-TPA-C**, **BTA-C10** and **CTTA** (b).

Of the members of the **BTA** materials family that were tested, **CTA** and **BTTA-C12** exhibit SCLC-like behavior at elevated temperatures, as shown by the black lines in Fig. S20 and the perfectly reasonable values found for the fitting parameters in Table S2. Although BTA and **CTTA** do show a finite conductivity at elevated temperatures, cf. Fig. 5b, the functional form is not described well by known analytical models, which we attribute to non-ohmic contacts (**BTA**) and degradation (**CTTA**). The linear IV of the non-amide containing **FCH-E** can be seen in Fig. S20b, and is in line with the notion that the conductivity of this material does not or hardly exceed that of the substrate.

| | Cyclic voltammetry and UV-Vis | | | DFT calculations | | |
|---|---|---|---|---|---|---|
| Material | HOMO (eV) | LUMO (eV) | Bandgap (eV) | HOMO DFT (eV) | LUMO DFT (eV) | Bandgap (eV) |
| **FCH-C3-A** | -5.73 | -1.73 | 4.00 | -6.47 | -1.31 | 5.16 |
| **FCH-C3-TA** | -5.60 | -2.13 | 3.47 | -5.71 | -1.66 | 4.05 |
| **FCH-E** | -5.78 | -1.99 | 3.79 | -6.58 | –1.71 | 4.87 |
| ***o*-TPA-N** | -5.55 | | | -6.03 | -1.37 | 4.66 |
| ***p*-TPA-N** | -4.99 | | | -5.88 | –2.20 | 3.66 |
| ***p*-TPA-C** | -5.55 | | | -5.75 | -1.73 | 4.02 |

**Table S1.** Energy levels of the frontier orbitals of selected compounds, determined experimentally by a combination of cyclic voltammetry and UV-Vis absorption spectroscopy and theoretically by DFT calculations. Comparing the two methodologies shows that the DFT calculations provide very reasonable estimates, and in some cases lead to a slight overestimation of the injection barrier to the gold contacts ($E_{Fermi}$~ -4.5 eV) that needed to be overcome. Hence, a smaller, but still significant, modulation of the injection barrier by the interfacial dipoles is needed to enable efficient charge injection.

| **Material** | **Gaussian Disorder Fit σ (meV)** | **Murgatroyd SCLC Fit μ ($cm^{2}$*$V^{-1}$*$s^{-1}$)** | **Murgatroyd SCLC Fit γ ($cm^{1/2}$*$V^{-1/2}$)** |
|---|---|---|---|
| **FCH-C3-A** | 76 ± 2 | 7.9 ± 0.1 E-7 (300 K) | 2.99 ± 0.03 E-3 |
| **FCH-C4-A** | - | 2.7 ± 0.2 E-8 (300 K) | 5.7 ± 0.2 E-3 |
| ***o*-TPA-N** | 59 ± 1 | 1.6 ± 0.1 E-9 (300 K) | 5.7 ± 0.1 E-3 |
| ***p*-TPA-N** | 57 ± 3 | 2.4 ± 0.1 E-6 (300 K) | 1.1 ± 0.1 E-3 |
| **BTTA-C12** | - | 4.0 ± 0.1 E-7 (390 K) | 6.5 ± 0.6 E-4 |
| **CTA** | - | 8.7 ± 0.2 E-7 (440 K) | 2.2 ± 0.1 E-3 |

**Table S2.** Fit parameters from Fig. 6 and Fig. S20

## Supplementary references